\documentclass[a4paper,12pt]{article}
\pdfoutput=1
\usepackage{jheppub_X} 
\usepackage{hyperref}
\usepackage{lineno}
\usepackage{xcolor}
\usepackage{setspace}
\usepackage{subcaption}

\usepackage{cleveref}
\crefname{table}{Table}{Tables}
\crefname{equation}{Eq.}{Eqs.}
\crefname{appendix}{App.}{Apps.}
\crefname{section}{Sec.}{Secs.}
\crefname{figure}{Fig.}{Figs.}

\makeatletter
\g@addto@macro\bfseries{\boldmath}
\makeatother

\definecolor{colorTC}{rgb}{.2,.7,.2}

\newcommand{\s}{\hspace{0.8pt}}
\newcommand{\LD}{\Lambda_\text{D}} 
\newcommand{\aD}{\alpha_\text{D}} 

\preprint{CERN-TH-2023-194}

\title{Dark Sector Glueballs at the LHC}

\author[a]{Austin Batz,}
\author[b,c,a]{Timothy Cohen,}
\author[d]{David Curtin,}
\author[d]{Caleb Gemmell,}
\author[a]{and Graham D. Kribs\s}
\affiliation[a]{Institute for Fundamental Science and Department of Physics, \\
University of Oregon, Eugene, OR 97403, USA}
\affiliation[b]{Theoretical Physics Department, CERN, 1211 Geneva, Switzerland}
\affiliation[c]{Theoretical Particle Physics Laboratory, EPFL, 1015 Lausanne, Switzerland}
\affiliation[d]{Department of Physics, University of Toronto, Toronto, ON M5S 1A7, Canada}

\emailAdd{abatz@uoregon.edu}
\emailAdd{tim.cohen@cern.ch}
\emailAdd{dcurtin@physics.utoronto.ca}
\emailAdd{caleb.gemmell@mail.utoronto.ca}
\emailAdd{kribs@uoregon.edu}

\abstract{
We study confining dark sectors where the lightest hadrons are glueballs.  Such models can provide viable dark matter candidates and appear in some neutral naturalness scenarios.
In this work, we introduce a new phenomenological model of dark glueball hadronization inspired by the Lund string model.
This enables us to make the most physically-motivated predictions for dark glueball phenomenology at the LHC to date.
Our model approximately reproduces the expected thermal distribution of hadron species as an emergent consequence of hadronization dynamics.
The ability to predict the production of glueball states heavier than the lightest species significantly expands the reach of long-lived glueball searches in MATHUSLA compared to previous simplified estimates.
We also characterize regions of parameter space where emerging and/or semivisible jets could arise from pure-glue dark sectors, thereby providing new benchmark models that motivate searches for these signatures.
}

\begin{document}
\maketitle
\flushbottom

\section{Introduction}
There has been substantial work investigating the phenomenology of confining dark sectors, or so-called hidden valley models \cite{Strassler:2006im,Strassler:2006ri,Strassler:2006qa,Han:2007ae,Strassler:2008bv,Strassler:2008fv,Luo:2009kf,Cvetic:2012kj}. In a collider setting, dark partons are produced through some portal \cite{Patt:2006fw,March-Russell:2008lng,Delgado:2008rq,Krolikowski:2008qa,Hambye:2008bq,Falkowski:2009yz} to the Standard Model (SM) before showering and eventually hadronizing. These dark hadrons may decay entirely or partially into SM particles, resulting in various experimental signatures including semivisible jets \cite{Cohen:2015toa,Cohen:2017pzm,Beauchesne:2017yhh,Bernreuther:2020vhm,CMS:2021dzg,ATLAS:2023swa}, emerging jets \cite{Schwaller:2015gea,CMS:2018bvr,Treado:2748283,Archer-Smith:2021ntx,Carrasco:2023loy}, SUEPs \cite{Harnik:2008ax,Knapen:2016hky,Barron:2021btf,Lory:2022upc}, and more \cite{Baumgart:2009tn,Cheng:2015buv,Csaki:2015fba,Park:2017rfb,Kribs:2018ilo,Kribs:2018oad,Costantino:2020msc,Cohen:2020afv,Knapen:2021eip}, see \cite{Albouy:2022cin} for a recent review. A subset of these dark sector models have strong theoretical motivations as solutions to the ``little hierarchy problem,'' referred to as neutral naturalness models \cite{Chacko:2005vw,Burdman:2006tz,Cai:2008au,Poland:2008ev,Curtin:2015fna,Craig:2015pha,Cohen:2015gaa, Craig:2016kue,Cohen:2018mgv,Cheng:2018gvu}.  In these models, the Higgs mass is rendered calculable due to the inclusion of new particles that are neutral under SM QCD. Dark sectors also feature frequently in models of dark matter \cite{Hur:2007uz,Krolikowski:2008qa,Kribs:2009fy,Bai:2013xga,Appelquist:2015yfa,Appelquist:2015zfa,Antipin:2015xia,Freytsis:2016dgf,Kribs:2016cew,Mitridate:2017oky,Beauchesne:2018myj,Francis:2018xjd}.

In this work, we are interested in scenarios where the lightest dark hadronic states are glueballs.
Dark glueballs have been studied extensively, in the context of collider searches \cite{Curtin:2015fna,Chacko:2015fbc,Burdman:2018ehe,Kilic:2018sew}, models of dark matter \cite{Faraggi:2000pv,Boddy:2014yra,Boddy_2014_1,Garc_a_Garc_a_2015,Soni_2016,Yamanaka:2019aeq,Yamanaka:2019yek,Curtin:2022oec,Asadi:2022vkc}, and cosmology \cite{Forestell:2016qhc,Forestell:2017wov,Soni:2017nlm,Buen-Abad:2018mas,Jo:2020ggs,Carenza:2022pjd}.
They also appear in neutral naturalness models, such as fraternal twin Higgs (FTH)~\cite{Craig:2015pha} and folded supersymmetry (FSUSY)~\cite{Burdman:2006tz}, which invoke a $\mathbb{Z}_2$ symmetry between the Standard Model and a hidden sector, thereby introducing a dark $SU(3)$ gauge sector.  It is then possible that new massive fields charged under this dark $SU(3)$ are significantly heavier than the dark confinement scale, in which case dark glueballs are the only light hadrons. 

Previous collider studies have been limited by the absence of dedicated simulation tools for dark glueball production.
\textsc{Pythia}'s Hidden Valley module \cite{Carloni:2010tw,Carloni:2011kk} is the current state-of-the-art in simulating strongly-coupled dark sector hadronization, but it does not accommodate the qualitatively different pure-glue ($N_f = 0$) case. Recently, the Python module \texttt{GlueShower} \cite{Curtin:2022tou} was created to address this gap.  This package implemented a perturbative gluon shower and exploited only energy conservation and the large pure-glue mass-gap to 
parameterize the unknown details of glueball hadronization. While this enabled the first quantitative studies including the effects of  the shower and multiple glueball species~\cite{Curtin:2022oec}, the lack of any realistic hadronization dynamics resulted in very large uncertainties for exclusive quantities, like the production of specific glueball species, that can strongly influence collider signals. 

In this work, we present a more sophisticated glueball hadronization model, based on applying Lund string dynamics~\cite{Andersson:1983ia} to the pure-glue regime, and implement it in a customized version of \textsc{Pythia} 8 \cite{Bierlich:2022pfr}. This enables more theoretically robust collider studies with improved  uncertainties compared to the earlier \texttt{GlueShower} approach.\footnote{
Our code is also orders of magnitude faster than the Python-based \texttt{GlueShower}, greatly facilitating large-scale phenomenology studies and making the potential future incorporation of our algorithm  into a public release of \textsc{Pythia} significantly easier.
} 
We find that our hadronization algorithm dynamically realizes certain theoretically expected features, such as the thermal distribution of produced glueball species. These results suggest that the most important pure-glue hadronization dynamics may be captured by our approach. 

We apply this implementation of dark glueball hadronization to classify the phenomenology across the parameter space of two specific models, determining which regions could possibly be probed by future emerging or semivisible jet searches. 
We consider applications to long-lived particle (LLP) searches and semivisible/emerging jet searches.
In particular, we update the predicted signal  at the proposed MATHUSLA LLP detector~\cite{Chou:2016lxi,MATHUSLA:2018bqv, MATHUSLA:2020uve} of dark glueballs produced in exotic Higgs decays in \cref{section:MATHUSLA}. Compared to earlier estimates based on two-body Higgs decays \cite{Curtin:2015fna,Curtin:2018mvb}, our dark shower and hadronization dynamics lead to a dramatically expanded reach estimate, as the production of various glueball species with different lifetimes generates signals in different parts of parameter space. 

Semivisible jets and emerging jets have become targets for LHC searches, see~\cite{CMS:2021dzg,ATLAS:2023swa} and \cite{CMS:2018bvr,Treado:2748283} respectively.  This motivates developing consistent benchmark models that can serve as reference points for designing experimental analyses. In \cref{section:semivisible}, we show how glueball production can also realize both signatures and elucidate some properties of the resulting signals. In particular, 
the parameter $r_\text{inv}$ (which characterizes the collider-stable component of semivisible jets) can be predicted as a distribution, and we provide a number of examples in the results presented below.  

The rest of this paper is structured as follows. We introduce the hadronization model in \cref{section:hadronizationmodels}, first reviewing the Lund string model in \cref{section:lund} and then showing our modifications of this approach for pure-glue hadronization in \cref{section:algorithm} (further discussion is provided in \cref{appendix:algorithm}). We demonstrate that the expected thermal distribution of glueball species dynamically emerges in \cref{section:fits}.  We provide benchmark values of the hadronization model parameters in \cref{section:benchmarks}, which span a range of outputs representative of theoretical uncertainties from unknown hadronization details.
In \cref{section:glueballs}, we introduce glueball production and decay mechanisms through the Higgs portal that are relevant for collider signals.
We then apply our glueball production and decay simulations to two phenomenology studies: glueball LLP decays in the proposed MATHUSLA experiment in \cref{section:MATHUSLA} and semivisible/emerging jet signatures in \cref{section:semivisible}. We conclude in \cref{section:conclusion}. In \cref{appendix:benchmarks}, we elaborate on the glueball species and momentum variations from our suggested benchmark parameters. A few additional distributions of interest for the semivisible jet scenario are provided in \cref{appenidx:jetVars}.

\section{Glueball Hadronization}
\label{section:hadronizationmodels}
The non-perturbative nature of hadronization necessitates introducing phenomenological models to make predictions for the collider signatures of confining sectors. This approach has a long history going back to Field and Feynman's ``independent fragmentation" \cite{FIELD19781}, where the model is simply that individual quarks fragment into the mesons making up a jet. The modern Monte Carlo event generator \textsc{Pythia} uses the ``Lund string model'' \cite{Andersson:1983ia} (described below), while its contemporaries \textsc{Herwig} \cite{Bahr:2008pv} and \textsc{Sherpa} \cite{Gleisberg:2008ta} each use versions of ``cluster hadronization" \cite{Webber:1983if,Kupco:1998fx}, where color-singlet clusters of partons that are close together in phase space decay into hadrons. 
Hadronization models introduce phenomenological nuisance parameters.  For SM QCD, one can perform elaborate tunes to data to constrain these parameters.  Of course, we do not have the luxury of data to fit these parameters for a dark sector. Fortunately, as we show in \cref{section:results}, the observables of interest here are only moderately sensitive to the nuisance parameters. 
Developing theoretical predictions with error estimates and search strategies for dark sector jets that are less sensitive to nuisance parameters is an area of active interest, see e.g.~\cite{Cohen:2020afv, Cohen:2023mya}.

The rest of this section is devoted to a description of string hadronization. First, we review the Lund string model used in \textsc{Pythia 8}.  This provides the context for the description of our dark glueball hadronization model that follows. In all of our analyses, we consider the case where the number of colors $N_c$ is 3, but generalization to other values of $N_c$ is straightforward (given the glueball mass spectrum and its scaling with the confinement scale from lattice studies). 

\subsection{Lund String Model for Mesons and Baryons}
\label{section:lund}

We start by briefly summarizing the Lund string model of hadronization as implemented in \textsc{Pythia} 8. A more detailed explanation can be found in \cite{Bierlich:2022pfr} with additional context in \cite{Andersson:1983ia}. Before hadronization, partons undergo a perturbative shower, iteratively splitting until the characteristic energy scale (transverse momentum $p_T$ relative to the parent parton in \textsc{Pythia}'s implementation) reaches an IR cutoff $p_{T\text{min}}$, which parameterizes the scale where the shower approaches  strong coupling  and must be matched onto the hadronization model.  The shower is executed in the leading color ($N_c\rightarrow\infty$) approximation, such that each (anti-)quark has a unique (anti-)color label, and each gluon has unique color and anti-color labels. Therefore, for each color label, there is exactly one parton with the compensating anti-color label at each step of the shower.

Partons are grouped into color singlets by ``Lund strings.'' These are simple representations of flux tubes dictating the flow of color charge. Quarks and anti-quarks live at the ends of strings, while gluons are represented as kinks in the strings. Strings are therefore comprised of ``string pieces,'' the segments of the string that connect individual quarks and gluons. Each string piece has momentum 

\begin{equation}\label{equation:ppiece}
p^\mu_\text{piece} = a_1\, p^\mu_1 + a_2\, p^\mu_2 \,, \qquad\text{with}\quad a_i = 
\begin{cases}
    1\,,\, i^\text{th} \text{ parton is a (anti-)quark} \\
    \frac{1}{2}\,,\, i^\text{th} \text{ parton is a gluon}
\end{cases} ,
\end{equation}

\noindent where $p^\mu_1$ and $p^\mu_2$ are the momenta of the partons connected by the string piece. This momentum defines a string piece mass $m_{\text{piece}}$ via $p_{\text{piece}}^2 = m_{\text{piece}}^2$. Each (anti-)quark effectively donates all its momentum to the string piece that terminates on it, and each gluon donates half its momentum to each of the two string pieces that are connected to it. 

\begin{figure}[t!]
    \centering
    \begin{subfigure}[t]{0.3\textwidth}
       \includegraphics[width=\textwidth]{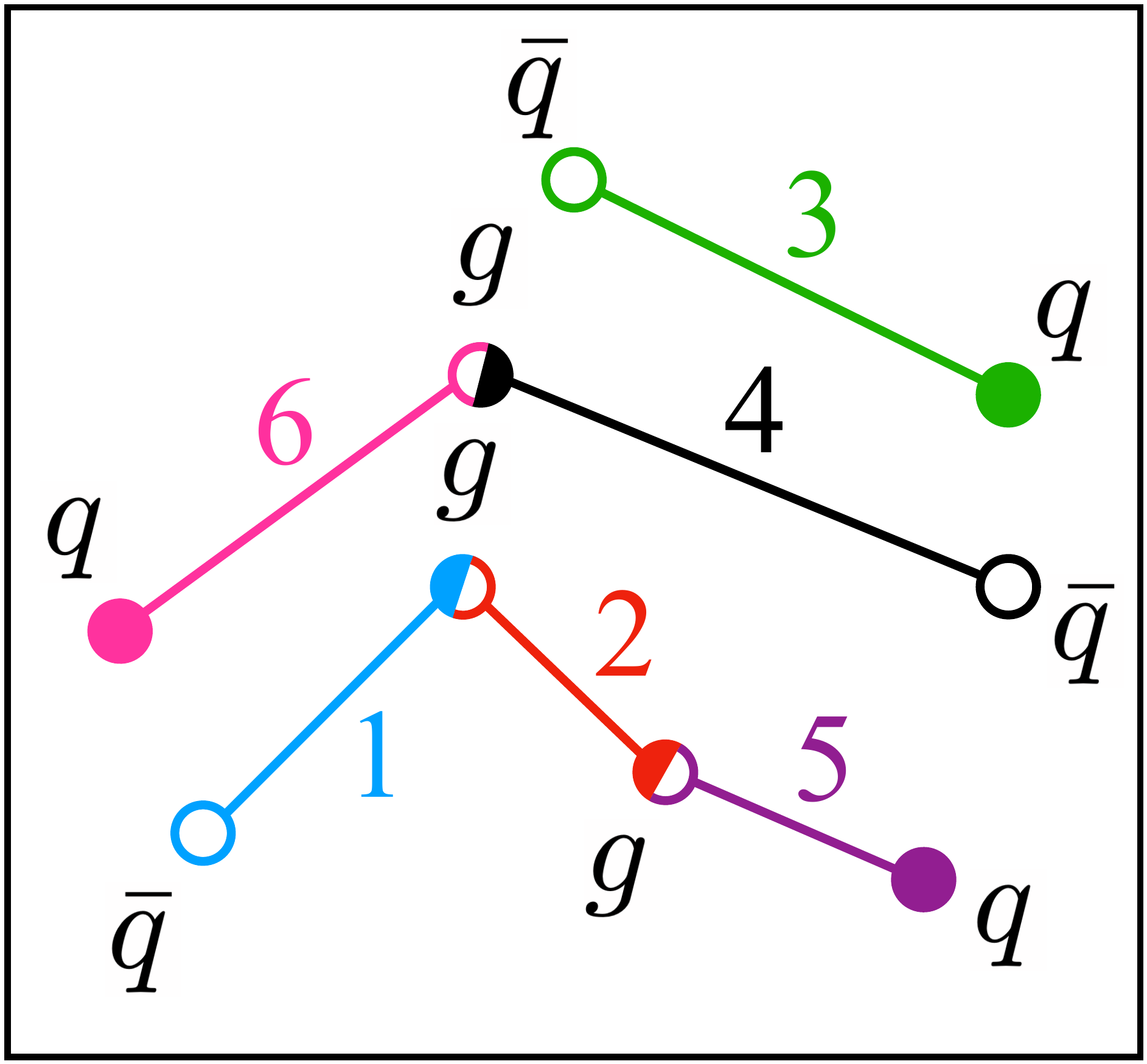} 
       \caption{Partons at the end of the shower with $N_c \to \infty$ have unique color and anti-color labels.\\}
       \label{figure:CR1}
    \end{subfigure}
    \hfill
    \begin{subfigure}[t]{0.3\textwidth}
       \includegraphics[width=\textwidth]{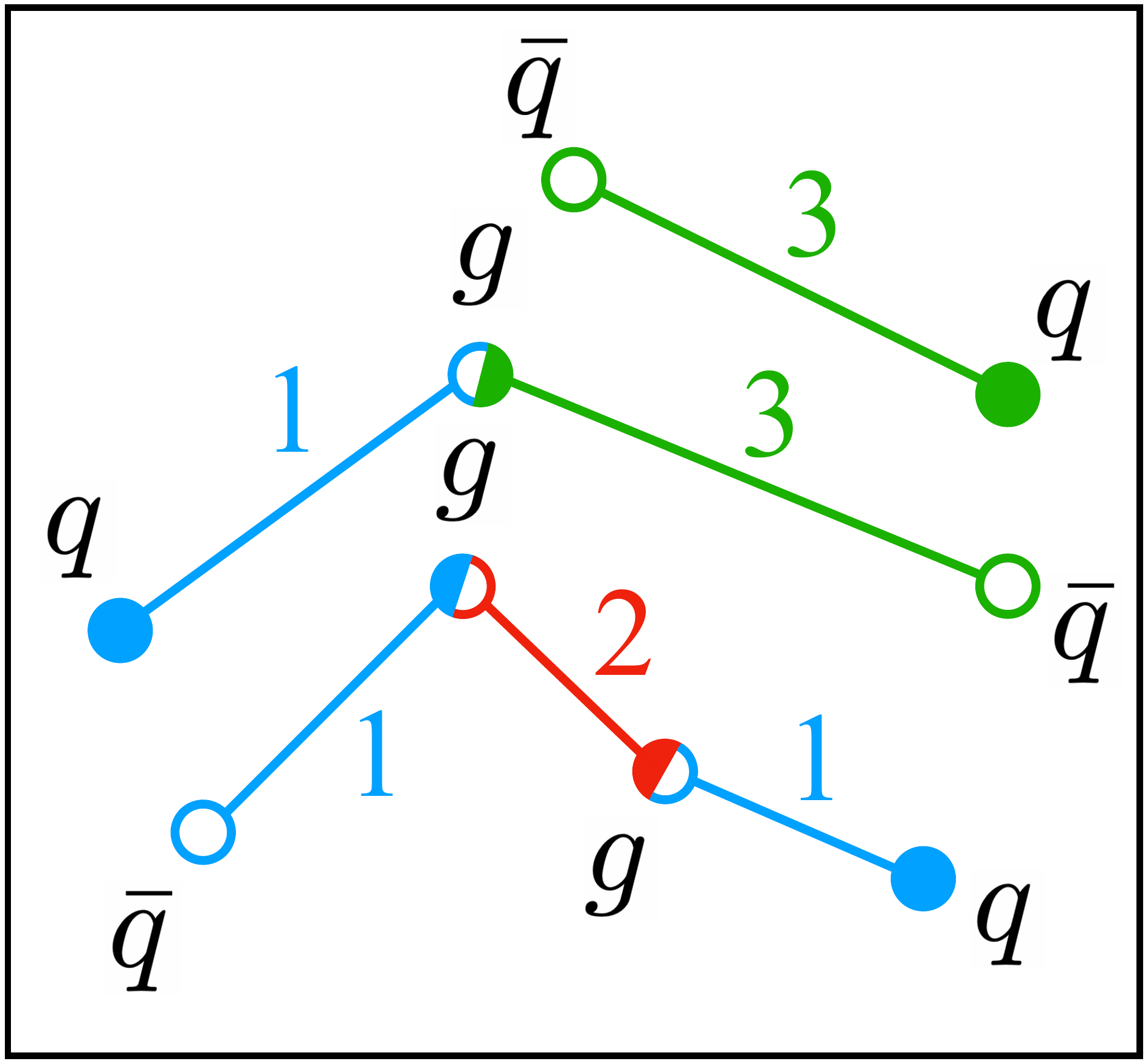} 
       \caption{String pieces are randomly reassigned a new color, now restricted to $N_c=3$ choices.}
       \label{figure:CR2}
    \end{subfigure}
    \hfill
    \begin{subfigure}[t]{0.3\textwidth}
       \includegraphics[width=\textwidth]{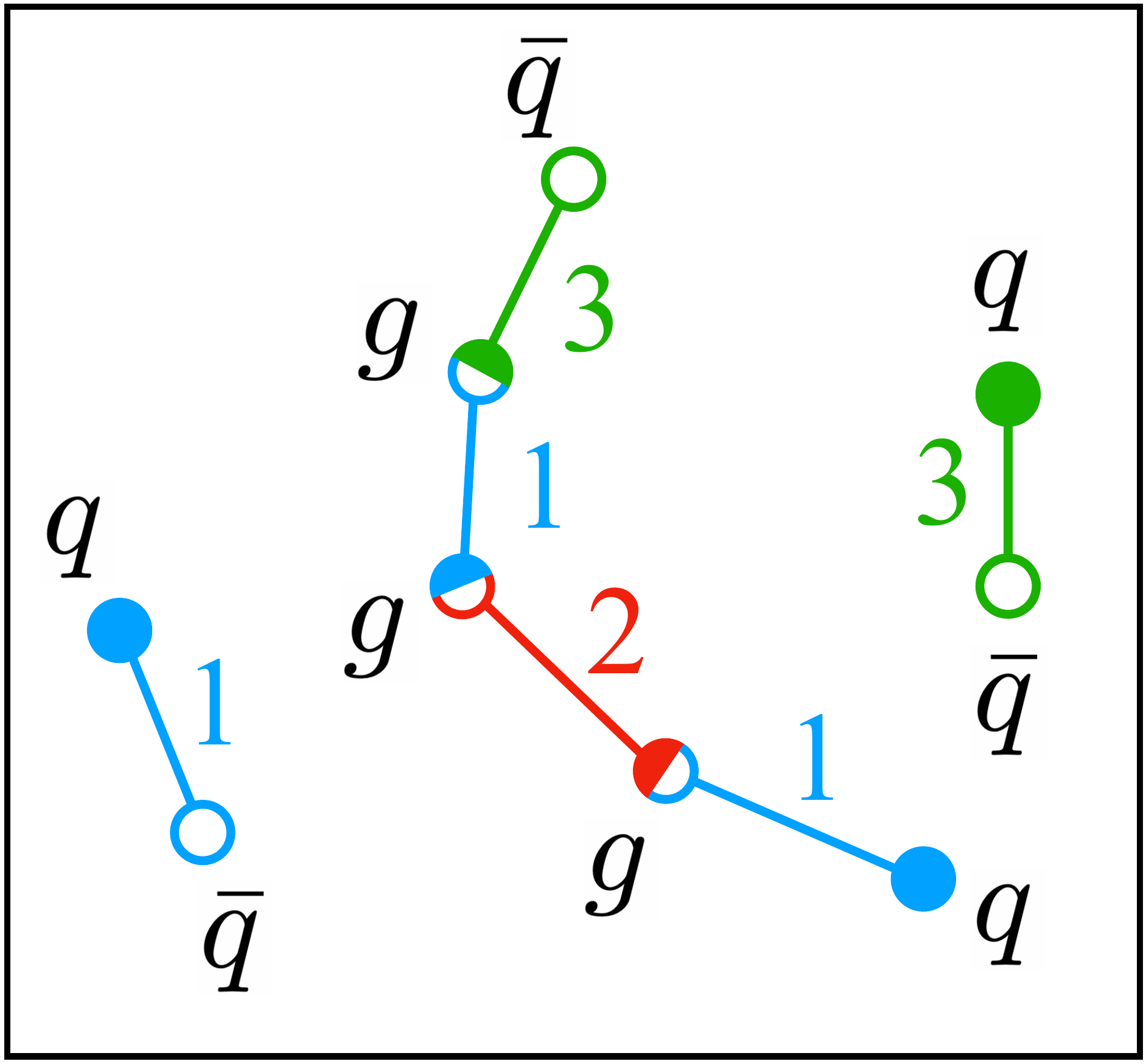} 
       \caption{Connections between color/anti-color pairs are swapped if this reduces the string-length $\lambda$.}
       \label{figure:CR3}
    \end{subfigure}    
    \caption{A sketch of QCD-based color reconnection. Color and anti-color labels are represented by filled and empty (semi-)circles, respectively. Quarks $q$ and anti-quarks $\bar{q}$ form the ends of strings, while gluons $g$ are kinks in the string. String pieces span between the color/anti-color pairs of individual partons and are labeled with displayed colors (colors have numerical labels so that gray-scale versions contain the same information). Each parton is displayed with a fixed location in an abstract color-connection space where the displayed lengths of the string pieces correspond to the string-length $\lambda$. Between \ref{figure:CR2} and \ref{figure:CR3}, the pairs of blue (1) and green (3) connections are swapped.}
    \label{figure:CR}
\end{figure}

In order to account for the fact that QCD has a finite number of colors, \textsc{Pythia} implements a procedure called ``color reconnection''
between the end of the perturbative shower and the formation of hadrons.  Out of the few different options \textsc{Pythia} 8 offers for color reconnection, the so-called ``QCD-based" version is the best-motivated for our regime without beam remnants or light quarks.\footnote{Part of the purpose of color reconnection is to treat states at the end of showers and beam remnants consistently, so \textsc{Pythia}'s default implementation was formulated with beam remnants specifically in mind.} Here, color/anti-color pairs are randomly reassigned a new label, which permits more than one possible grouping of partons into color singlets.\footnote{More precisely, there are nine possible reassignments (with the restriction that gluons cannot be reassigned to have the same color and anti-color) to reflect the $1/9$ probability of an $SU(3)$ fundamental being able to form a singlet with an $SU(3)$ anti-fundamental.  More detail can be found in \cite{Bierlich:2022pfr}.} Color reconnection seeks to minimize a Lorentz-invariant effective free energy $\lambda$ called the string-length

\begin{equation} \label{equation:stringlength}
\lambda = \sum_{\text{pieces}} \ln \left( 1 + \frac{m_{\text{piece}}^2}{m_{\text{ref}}^2} \right) \,,
\end{equation}

\noindent where $m_{\text{ref}}$ is the mass of some reference hadron. The minimization of $\lambda$ is performed by swapping which color end of a color/anti-color pair is connected to which anti-color end.  We provide a sketch of this procedure in \cref{figure:CR} for a simple system with a few partons. 

The approach taken in the Lund string hadronization model is to consider these color connections between partons as oscillating classical strings that break up into hadrons. A string with two ends connecting a $q\bar{q}$ pair is called a ``yo-yo mode" and is identified with a meson.\footnote{Strings can also connect three (anti-)quarks when they have three ends joined by a junction, and these are identified with (anti-)baryons.} When a hadron fragments off of a string, it takes away some random fraction $z$ of the string's light-cone momentum $p_\pm = E\pm p_z$ (with the $z$-axis being a preferred direction which \textsc{Pythia} takes to be the string axis), which has convenient Lorentz transformation properties. This $z$ is sampled from the Lund Symmetric Fragmentation Function (LSFF)\footnote{The use of the term ``fragmentation function" here should be distinguished from its use in perturbative QCD literature (see e.g.~\cite{Elder:2017bkd}), where this concept describes the density of hadrons resulting from an individual final state parton.}

\begin{equation}
\label{equation:LSFF} f_{\text{LSFF}}(z) \propto \frac{(1-z)^a}{z} e^{-b\s m_\perp^2/z} \,,
\end{equation}

\noindent where $a$ and $b$ are phenomenological parameters, and the hadron's transverse mass $m_\perp$ appears due to quark tunneling effects explained in \cite{Bierlich:2022pfr}. Given that Lund fragmentation assumes strings break into pieces ending on quarks, this approach must be modified to accommodate a pure-glue sector.

\subsection{String Model for Glueball Hadronization}
\label{section:algorithm}

In this section, we provide a qualitative discussion of our glueball hadronization algorithm for $N_c = 3$, which can be easily generalized to other numbers of colors.  Our focus is on explaining the role of the adjustable nuisance parameters that parameterize incalculable effects. A more detailed description with further justifications of the choices made here are is given in \cref{appendix:algorithm}. 

We modified \textsc{Pythia 8} to simulate 
the branching of a Lorentz-singlet dark gluon pair (produced e.g.~in the decay of a heavy scalar) via a leading color, $p_T$-ordered, pure-dark-glue parton shower.
The shower cutoff scale is parameterized as $p_{T\text{min}} = c \,\LD$, where $c$ is an $\mathcal{O}(1)$ nuisance parameter, and $\LD$ is the dark sector confinement scale.\footnote{Specifically, $\LD$ is the scale when $\aD^{-1} \to 0$ in the $\overline{\mathrm{MS}}$ scheme at three loops.} This cutoff scale parametrizes the onset of strongly-coupled dynamics, where the dark sector coupling $\aD$ becomes non-perturbatively large. In practice, the dimensionful scale that determines all of the scales in the pure-glue theory is the lightest glueball mass $m_0$, as further described below.

Lattice studies have provided us with many glueball properties in pure $SU(3)$, and in some cases for other values of $N_c$ \cite{Morningstar:1999rf,Chen:2005mg,Athenodorou:2021qvs}. There are twelve species that are stable in the absence of external couplings, each with their own set of $J^{PC}$ quantum numbers~\cite{Jaffe:1985qp}. The lightest state is the $0^{++}$ with mass $m_0$. Each heavy state's mass is a multiple of $m_0$ between $\sim$~1.4 and $\sim$~2.8, which we take from \cite{Chen:2005mg}. 
The masses and spins of this glueball spectrum provide the inputs we will need to select the species of glueballs that are emitted during fragmentation, which will be described below. 
Lattice results also allow us to specify the boundary condition in the dark strong coupling's renormalization group evolution given a choice of $m_0$, since $m_0 = 6.28\s\LD$ in pure $SU(3)$ \cite{Athenodorou:2021qvs}. In this way, we derive $\LD$ from the physical scale $m_0$. We set the default value of $c$ so that $\aD$ evaluated at the default shower cutoff scale is 1. Non-default values of $c$ change $p_{T\text{min}}$ without affecting the running of $\aD$.

\begin{figure}[t!]
    \centering
    \begin{subfigure}[t]{0.3\textwidth}
       \includegraphics[width=\textwidth]{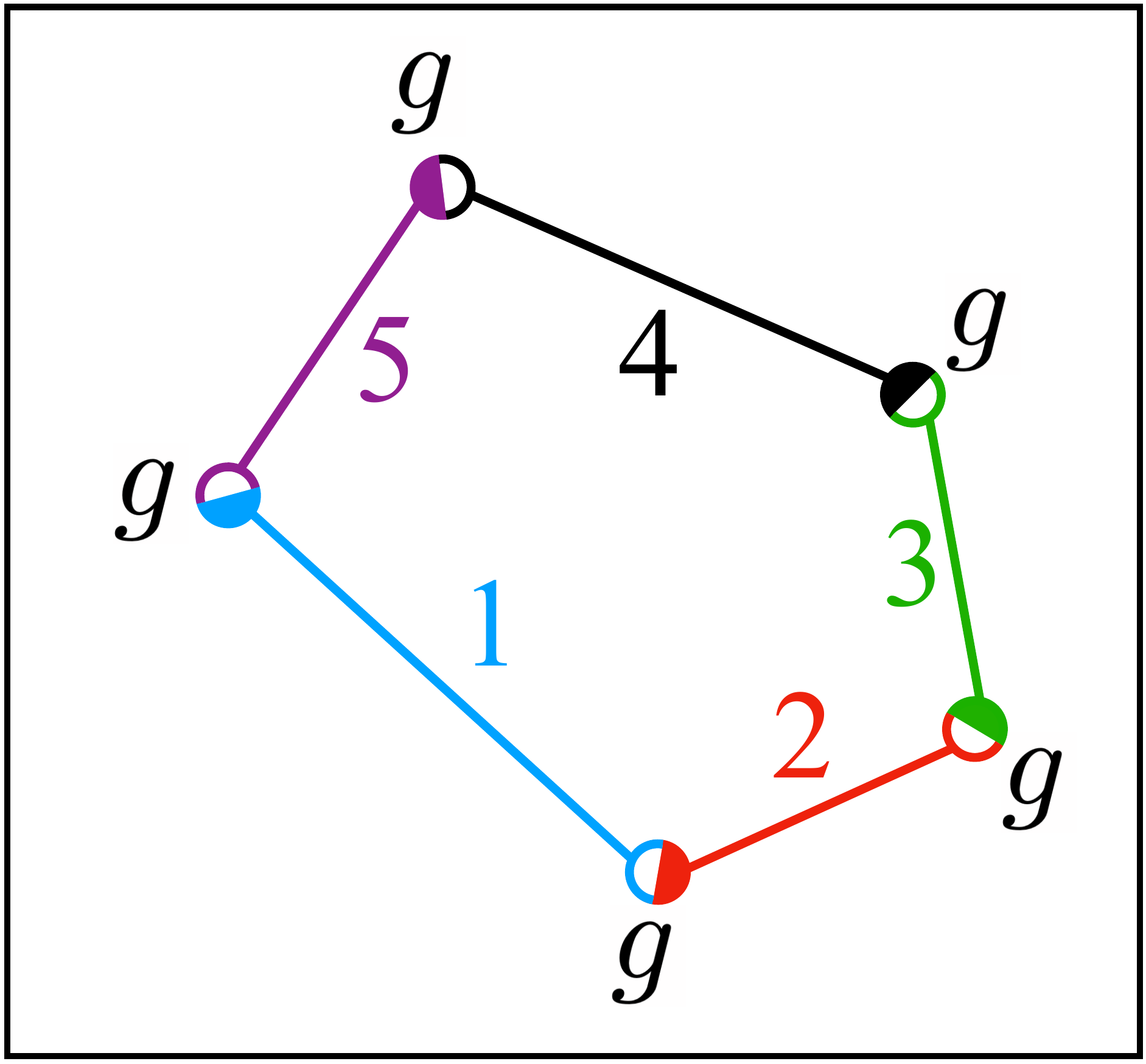} 
       \caption{Gluons at the end of the $N_c \to \infty$ shower.}
       \label{figure:CR1glue}
    \end{subfigure}
    \hfill
    \begin{subfigure}[t]{0.3\textwidth}
       \includegraphics[width=\textwidth]{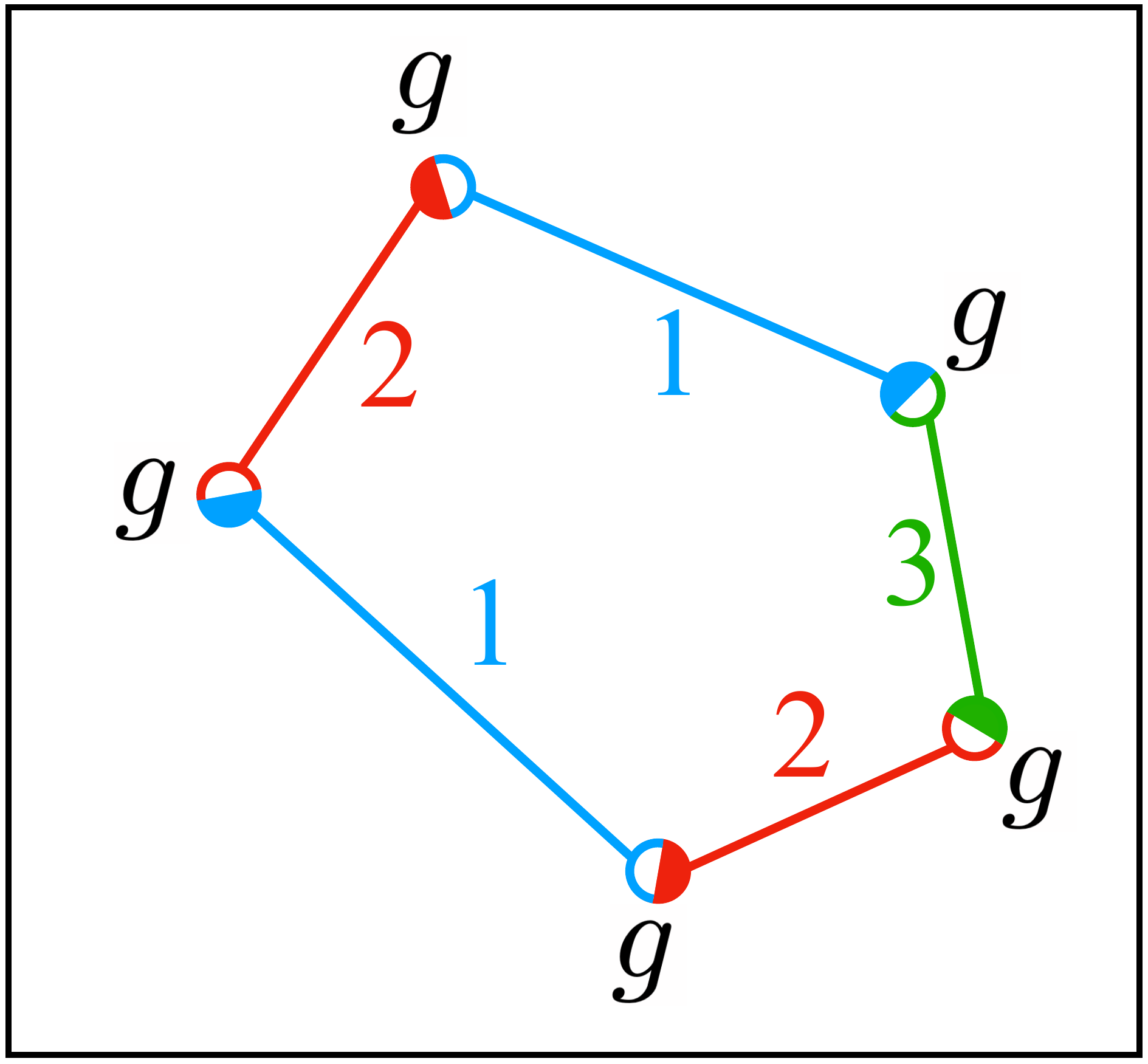} 
       \caption{Reassignment of string piece colors for $N_c = 3$.}
       \label{figure:CR2glue}
    \end{subfigure}
    \hfill
    \begin{subfigure}[t]{0.3\textwidth}
       \includegraphics[width=\textwidth]{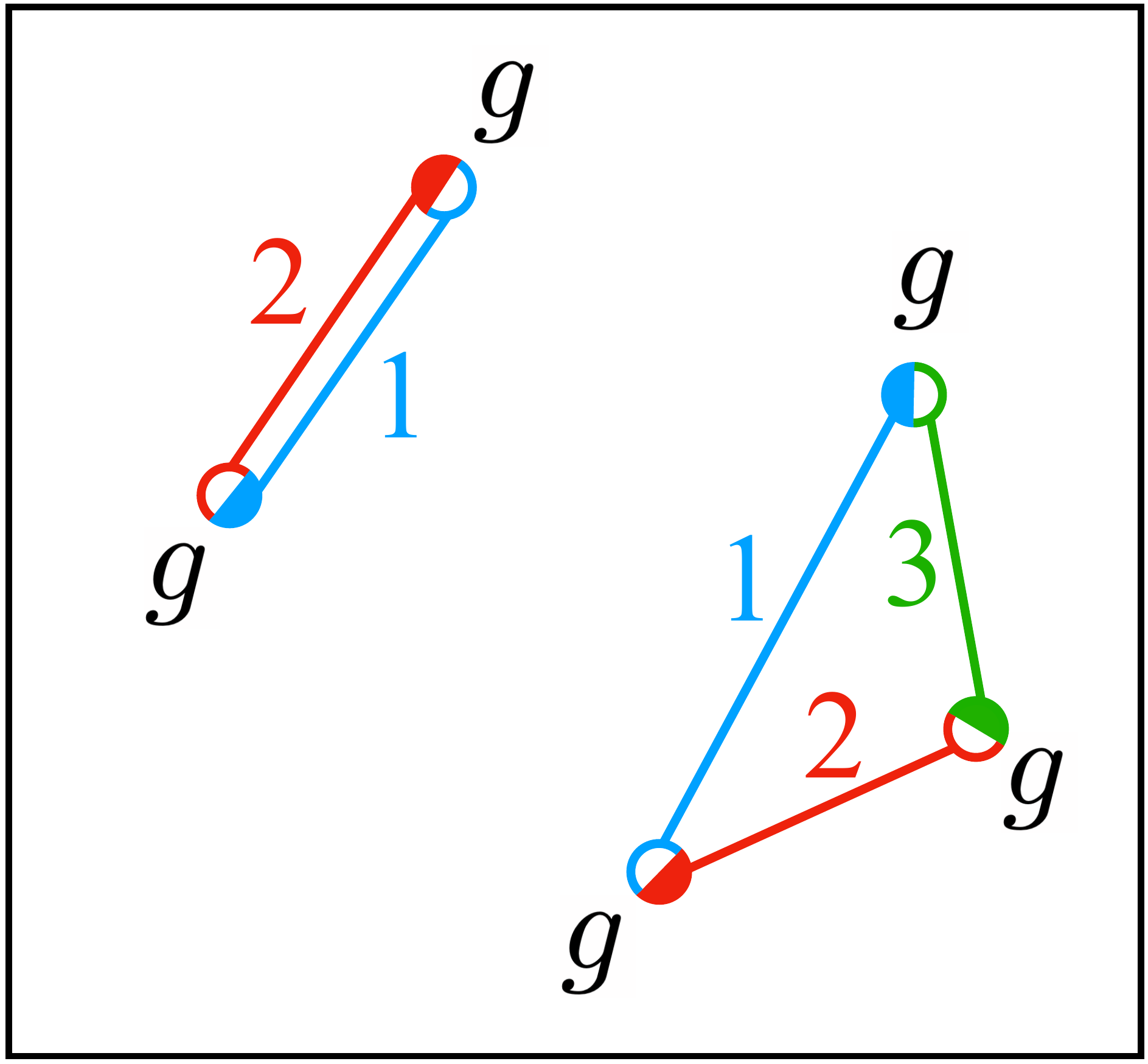} 
       \caption{Swapping of inter-parton connections.}
       \label{figure:CR3glue}
    \end{subfigure}    
    \caption{A sketch of QCD-based color reconnection for a system of only gluons, as in \cref{figure:CR}. Between \ref{figure:CR2glue} and \ref{figure:CR3glue}, the pair of blue (1) connections are swapped.}
    \label{figure:CRglue}
\end{figure}

Following the termination of the perturbative shower, we implement a version of QCD-based color reconnection. As visualized in \cref{figure:CRglue}, the only color-singlet Lund string topology in the pure-glue model is a closed ring. 
As in the SM, we swap color connections to minimize the string-length in \cref{equation:stringlength}, using the lightest glueball mass $m_0$ as $m_\text{ref}.$\footnote{The actual chosen value of $m_\text{ref}$ has negligible impact on all of our results.} This leaves us with 
color-singlet rings of string pieces that will fragment into glueballs. Both color reconnection and our hadronization algorithm are phenomenological models for flux tube rings twisting until they cross themselves and pinch into smaller rings. Without quarks, the strings are unable to break, so this pinching action is the only way the rings can divide into units with smaller invariant mass.

\begin{figure}[t!]
    \centering
    \begin{subfigure}[t]{0.3\textwidth}
       \includegraphics[width=\textwidth]{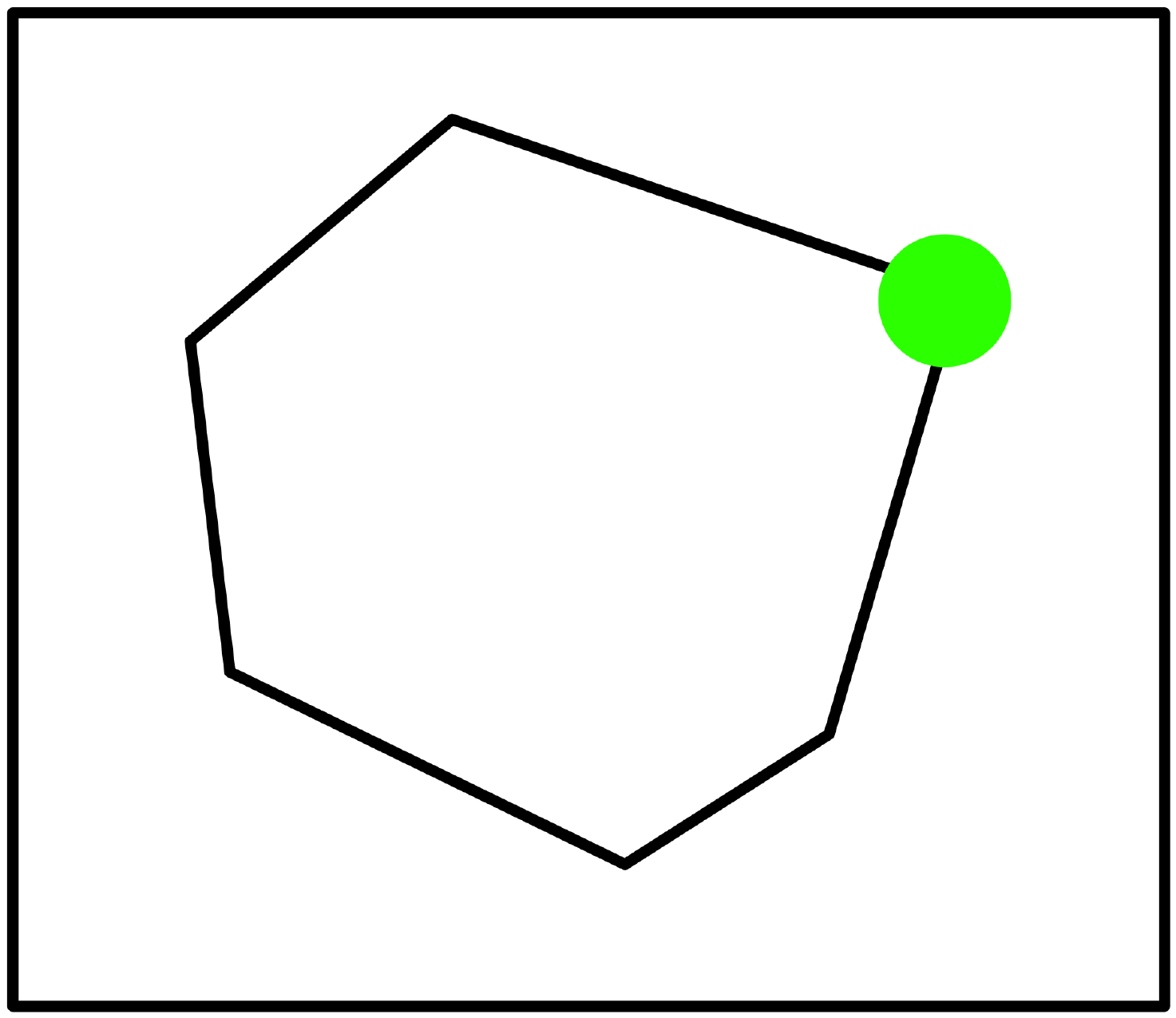} 
       \caption{The vertex (green) joining the string pieces with the largest string-length begins the fragmentation.\\ \\}
       \label{figure:had1}
    \end{subfigure}
    \hfill
    \begin{subfigure}[t]{0.3\textwidth}
       \includegraphics[width=\textwidth]{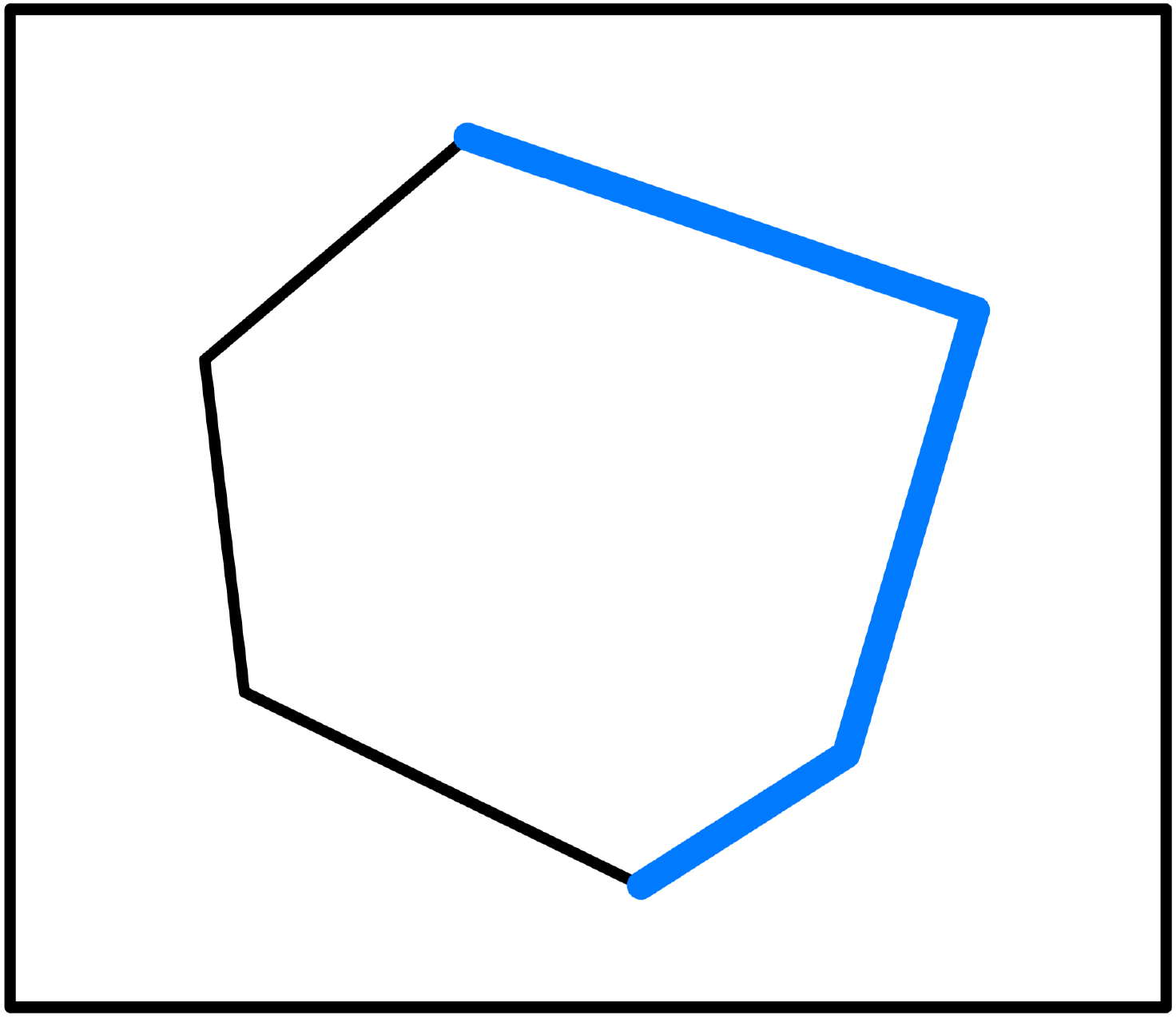} 
       \caption{A minimal set of string pieces  (blue)  with total mass $\geq m_0$ nearest to the vertex is selected to fragment into a glueball. \\}
       \label{figure:had2}
    \end{subfigure}
    \hfill
    \begin{subfigure}[t]{0.3\textwidth}
       \includegraphics[width=\textwidth]{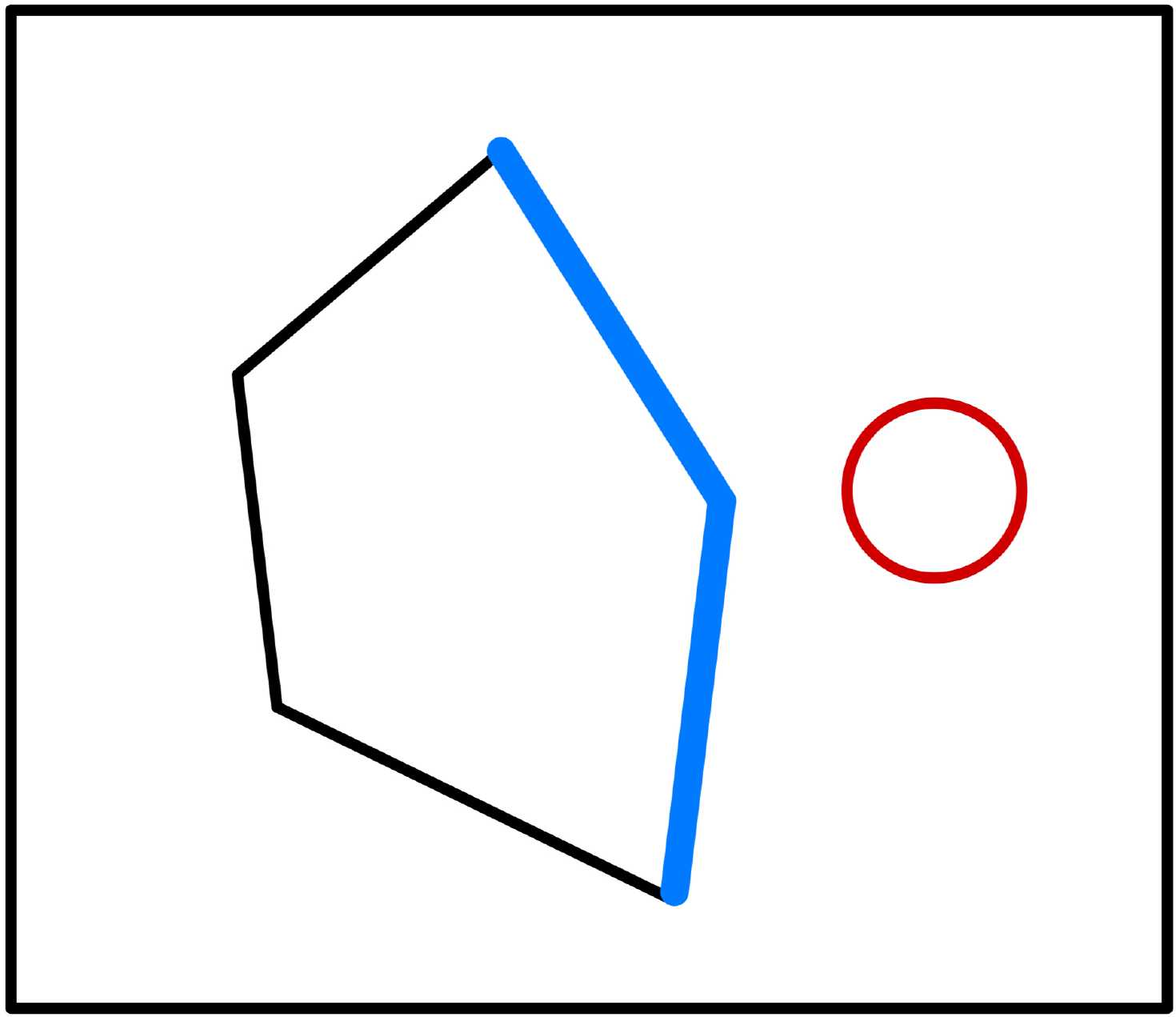} 
       \caption{The glueball (red circle) takes a fraction of the fragmenting pieces' momentum. The remaining momentum is distributed among two new string pieces (blue).}
       \label{figure:had3}
    \end{subfigure}   
    \caption{A visual depiction of our glueball hadronization mechanism. Color-singlet rings are shown as polygons whose edges are Lund string pieces. 
    The criterion for selecting the seed vertex (green) is chosen to allow for the most rapid reduction in overall string length with each step. 
    Glueball emission (depicted by the red circle in \ref{figure:had3}) is conceptualized as pinching off from the fragmenting string pieces (blue in \ref{figure:had2}).
    }
    \label{figure:had}
\end{figure}

The intuitive picture of glueball rings pinching off the color-singlet flux tube ring in order to most rapidly decrease the total string-length inspires the glueball hadronization  algorithm depicted in \cref{figure:had}.
First, we select string pieces with sufficient total invariant mass to be converted (or ``fragmented") into a glueball. 
We then determine the species of the glueball by randomly selecting  from among the species with mass less than that of the fragmenting string pieces, with probabilities weighted by the number of spin degrees of freedom. The emitted glueball's direction is along the total momentum of the fragmenting string pieces in the ring's rest frame, and its light-cone momentum is a random fraction $z$ of the light-cone momentum of the fragmenting strings. 
We study the effect of sampling $z$ from one of two fragmentation functions, the first being the LSFF in \cref{equation:LSFF} with $m_\perp$ replaced by the glueball mass $m_G$, and the second being a beta distribution

\begin{equation}
\label{equation:fbeta} f_\beta(z) \propto z^{\alpha-1}(1-z)^{k_\beta (m_0/m_G)^2} \,,
\end{equation}

\noindent where $\alpha$ and $k_\beta$ are nuisance parameters. Having specified the glueball's species, direction, and light-cone momentum, its four momentum is fully determined by the on-shell condition. Whatever momentum is left over from the fragmenting string pieces is distributed equally among two new string pieces, and the resulting new ring can emit the next glueball. If emitting a glueball with the selected momentum would result in a new ring that is kinematically forbidden from further fragmentation, the ring instead fragments into two glueballs with the second glueball's species randomly sampled as previously discussed. In summary, our model's nuisance parameters are the shower cutoff factor $c$, the fragmentation function, and the chosen function's two shape parameters. 

\subsection{Emergence of Thermally Distributed Production Rates}
\label{section:fits}

By analogy to the hadron spectra produced in SM QCD fragmentation~\cite{VanApeldoorn:1981gx}, a motivated expectation for the relative distribution of different glueball species is that they approximately follow a Maxwell-Boltzmann distribution~\cite{Falkowski:2009yz}:
\begin{equation}
\label{equation:thermal} P_J \propto (2J + 1) \left( \frac{m_J}{m_0} \right)^{3/2}e^{-(m_J-m_0)/T_{\text{had}}}\,,
\end{equation} 
\noindent where $P_J$ is the relative rate of producing the species with mass $m_J$ and spin $J$, and $T_{\text{had}}$ is some ``hadronization temperature.'' 
In~\cite{Falkowski:2009yz}, which analyzed glueball production in a dark matter indirect detection context without separating out the perturbative shower,  $T_\mathrm{had}$ was taken to be related to the center-of-mass energy of the initial hard process. 
For high enough initial energy, this would result in a power law distribution favoring heavier hadron masses, a behavior quite different from what we are used to in SM QCD\@. 
Instead, we would expect that the hadronization temperature is fairly unrelated to the high scale at which the original gluons are produced (provided this high scale is sufficiently greater than any kinematic thresholds), but rather is set by an intrinsic feature of the confining theory. In particular, for $N_f = 0$, $N_c = 3$, the critical or Hagedorn temperature of the QCD phase transition~\cite{Manes:2001cs,Blanchard:2004du,Noronha-Hostler:2010nut} is $T_{c} \simeq 1.2\, \LD$~\cite{Athenodorou:2021qvs,Lucini:2012wq}. 
This
motivated the approach taken by \texttt{GlueShower}~\cite{Curtin:2022tou}, which explicitly imposed this thermal distribution with $T_\mathrm{had} = d\s T_c$ for $d \sim \mathcal{O}(1)$. 
Remarkably, we will show that
our hadronization algorithm produces an approximately thermal multiplicity distribution of glueball species with very little dependence on the choice of fragmentation function or its parameters. Furthermore, the hadronization temperature is of the theoretically expected size,  $T_\mathrm{had} \simeq \LD$, increasing slowly with increasing shower cutoff scale for reasonable values of $c \sim \mathcal{O}(1)$. 

Our algorithm makes the minimal assumption that a local set of string pieces fragmenting into glueballs has no preference amongst the kinematically accessible glueball species, beyond the $2J+1$ spin multiplicity factor. The overall distribution of glueball species must therefore emerge from the kinematics of  color-singlet rings.
A color-singlet flux tube ring made of only soft string pieces that are close together in momentum space will predominantly produce only the lighter species, since each additional selected string piece will only modestly increase the total invariant mass available for fragmentation until the $m_0$ threshold is reached. On the other hand, fragmenting string pieces that are heavy compared to $m_0$, as well as combinations of color-connected string pieces that are far-separated in momentum space, will produce heavy and light glueballs without preference. The combination of these effects results in a net suppression of the heavy species. 

This idea of ``closeness" in momentum space (as determined by the invariant mass of sums of string piece momenta) elicits an intuitive geometric picture. If we imagine the color-singlet rings as polygons (as in \cref{figure:CR,figure:CRglue,figure:had}) whose side lengths are determined by the string-length in \cref{equation:stringlength}, but whose angles randomly fluctuate, then combinations of string pieces that are ``larger" in this sense have a greater propensity to cross each other and fragment off heavier glueballs. We would expect such a system to fragment in the order that most rapidly decreases its perimeter, which inspires our choice to begin fragmentation by selecting string pieces with the largest string-length.\footnote{In practice, we find that beginning selection of fragmenting string pieces with the highest string-length pieces or random pieces has minimal impact on the final species distribution.} This intuitive picture may serve as a good analogy for the dynamics of closed flux tube rings.

This illustrates qualitatively how our hadronization algorithm provides a plausible model of glueball fragmentation, but there is no \emph{a priori} reason to expect it to quantitatively produce an approximately thermal distribution. 
Furthermore, the above arguments suggest a significant dependence on the shower cutoff scale $p_{T\text{min}} = c \,\LD$, with higher values of $c$ producing fewer gluon splittings and therefore fewer string pieces that each have higher mass, resulting in overproduction of heavy glueball states.
Indeed, we observe a modest increase of the corresponding hadronization temperature with increasing $c$.

We now quantitatively demonstrate how this thermal species distribution emerges, and investigate the extent to which the glueball multiplicity distribution depends on the nuisance parameters of our hadronization model.  Recall that these parameters are the shower cutoff scale set by $c$, the choice of fragmentation function between \cref{equation:LSFF} or \cref{equation:fbeta}, and the two shape parameters for each function. 
Here, we focus on the fragmentation function and therefore set $c$ to its default  of 1.8.  The discussion will not change for other $\mathcal{O}(1)$ values, and when defining hadronization benchmarks in the next section, we will include different choices for $c$. 

The physical interpretation of the numerical values of the fragmentation function parameters is obscure, and there is no obvious correspondence between the parameters of the LSFF and that of the beta distribution. Therefore, we specify the mean $\mu_z$ and standard deviation $\sigma_z$ of the probability distributions for the $0^{++}$ species, which fix  values for the fragmentation function parameters and are easy to interpret. Another advantage of specifying the mean and standard deviation is that for each possible mean of a finitely-supported probability distribution, there is a maximum possible standard deviation. Thus, the space of all possible fragmentation function parameters is bounded when expressed this way.

\begin{figure}[t!]
\centering
\includegraphics[width=.46\textwidth]{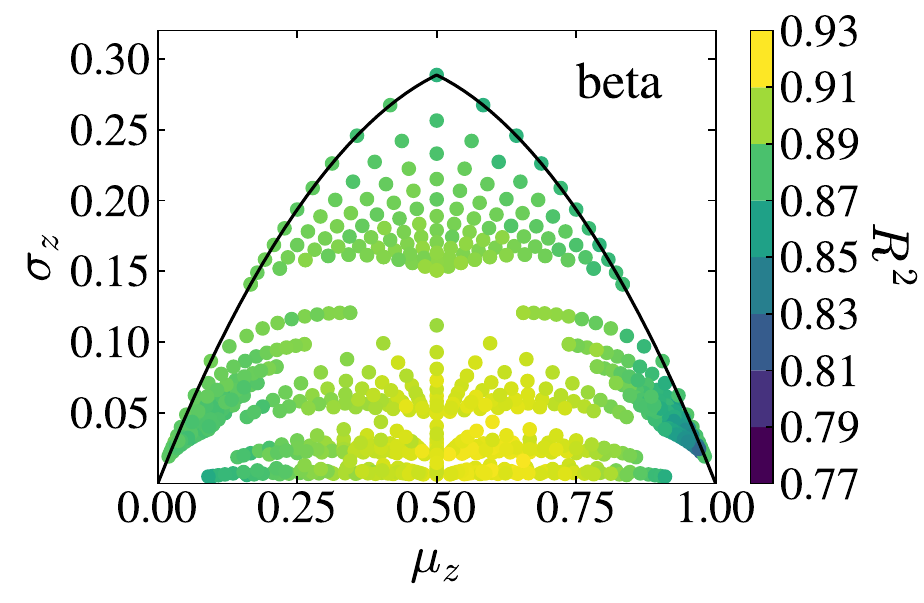}
\qquad
\includegraphics[width=.46\textwidth]{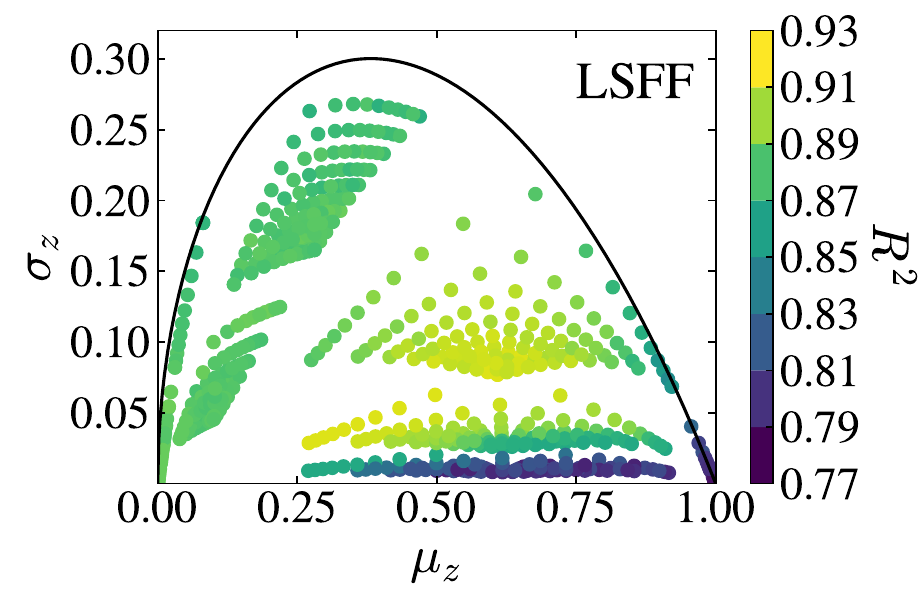}
\includegraphics[width=.46\textwidth]{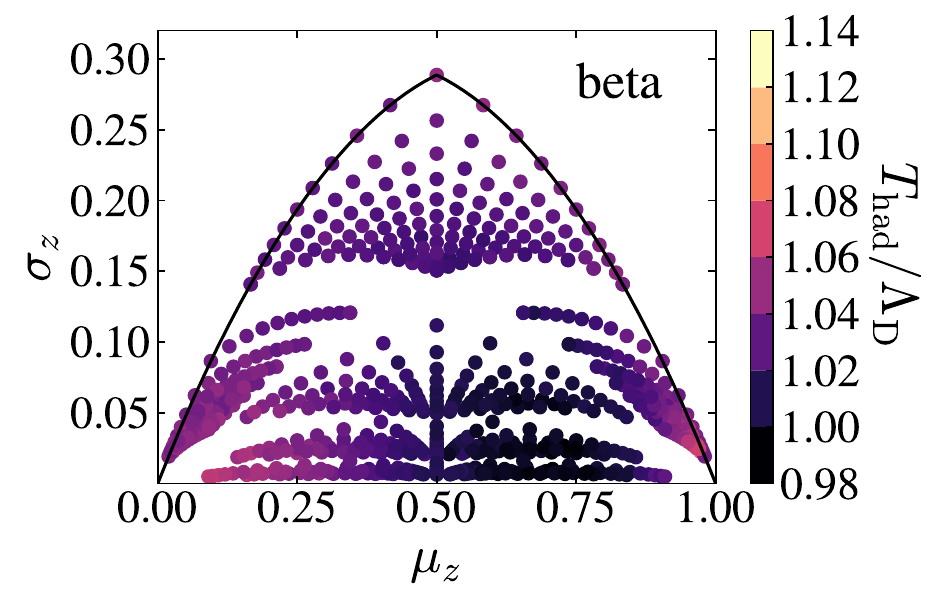}
\qquad
\includegraphics[width=.46\textwidth]{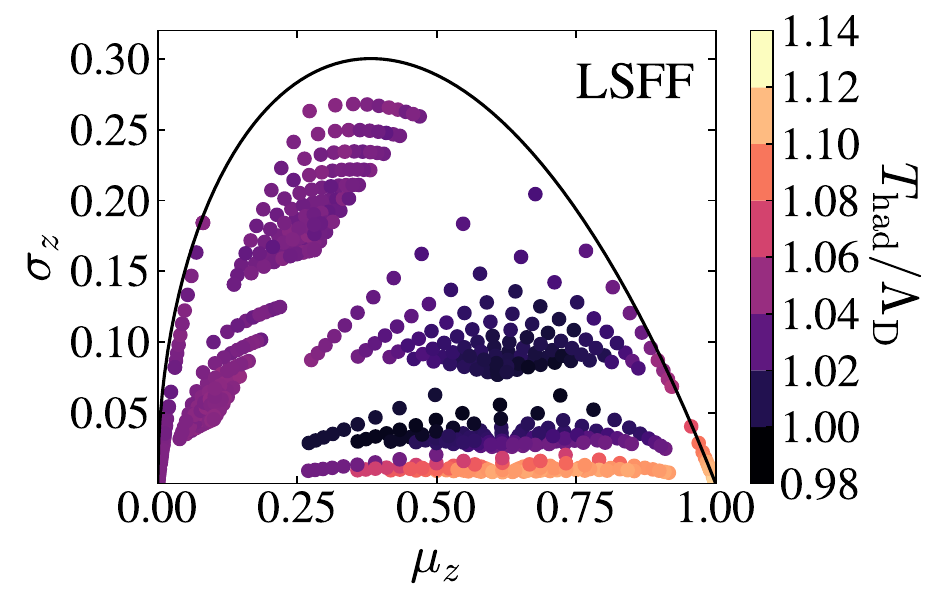}
\caption{Thermal model fit quality (top) and corresponding best fit hadronization temperature in units of $\LD$ (bottom) using the beta distribution (left) and the LSFF (right) for points in the plane of the mean $\mu_z$ and standard deviation $\sigma_z$ of the $0^{++}$ fragmentation function.
The shower center of mass energy is $125\,\text{GeV}$, the lightest glueball mass is $m_0 = 10\,\text{GeV}$, and the shower cutoff parameter is set to the default $c = 1.8$.
The black contours indicate an upper bound on $\sigma_z$ for a given $\mu_z$. For the beta distribution, this is the least upper bound for parameters where $f_\beta (0)$ and $f_\beta(1)$ are finite. For the LSFF, $\sigma_z$ does not saturate the bound. \label{figure:fits}}
\end{figure}

In \cref{figure:fits}, we present the results of fitting the distributions of species to \cref{equation:thermal} for many points in the $\mu_z$-$\sigma_z$ plane for the fragmentation function of the $0^{++}$ species. We find the best fit $T_{\text{had}}$ by maximizing

\begin{equation}
\label{equation:Rsquared} R^2 = 1 - \frac{\sum_i (y_i - P_J(x_i))^2}{\sum_i(y_i-\langle y \rangle)^2} \,,
\end{equation}

\noindent where $y_i$ is the total fraction of the $i^\text{th}$ species produced by the Monte Carlo, $P_J(x_i)$ is the thermal distribution in \cref{equation:thermal} evaluated at the mass $x_i$ of the $i^\text{th}$ species, and $\langle y\rangle$ is mean of the $y_i$'s. 
We chose $R^2$ to quantify the quality of the fit, rather than $\chi^2$, because we are interested in the infinite Monte Carlo statistics limit. Using $\chi^2$ would therefore artificially assign greater weight to the smallest species fractions. 
The point with the best fit was in the plane of the beta distribution at $\mu_z = 0.6$, $\sigma_z = 0.008$ with $R^2 = 0.93$ and $T_\text{had} = 0.99\s\LD$. 
We show results for shower center-of-mass energy $125\,\text{GeV}$ and glueball mass $m_0 = 10\,\text{GeV}$, but the results are very similar for center of mass energy of $1\,\text{TeV}$.
It is encouraging 
that the best-fit hadronization temperature lies close to the theoretical expectation, and that both $T_\mathrm{had}$ and the high quality of fit depend only very little on the choice of fragmentation function, its parameters, or the center of mass energy. This suggests that our hadronization model may represent a good analogy for the true non-perturbative fragmentation dynamics of crossing color strings.

\begin{figure}[t!]
\centering
\includegraphics[width=.8\textwidth]{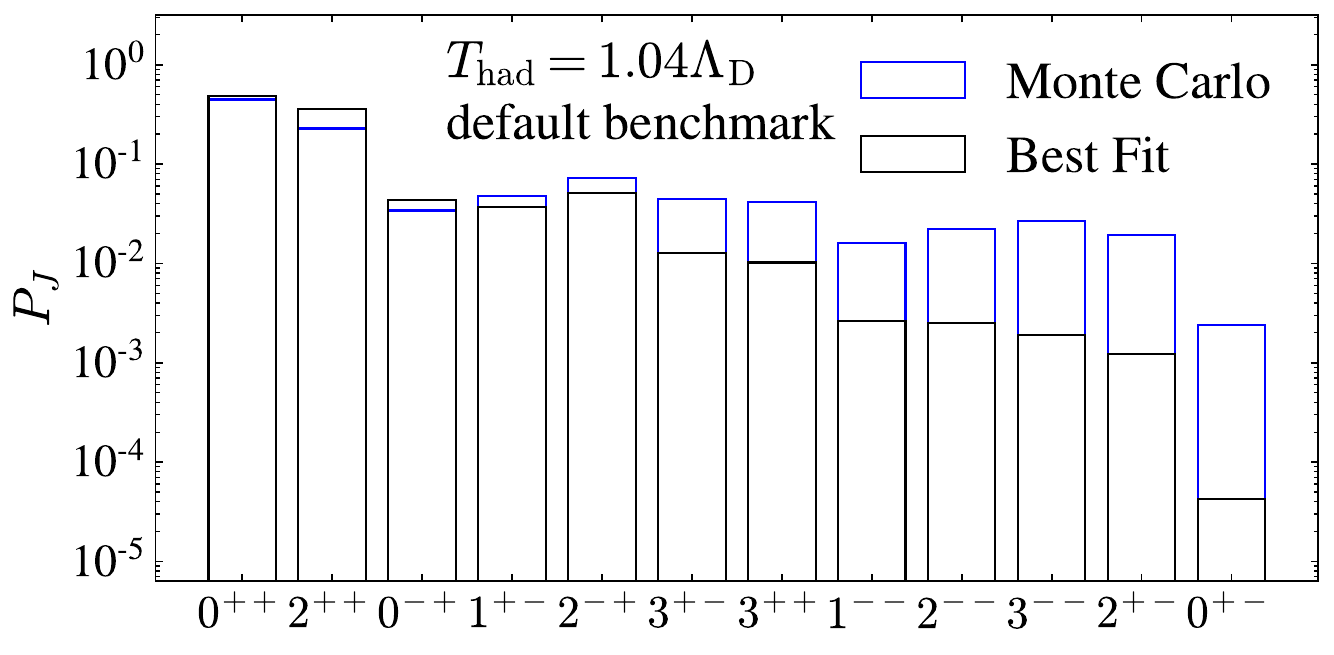}
\includegraphics[width=.8\textwidth]{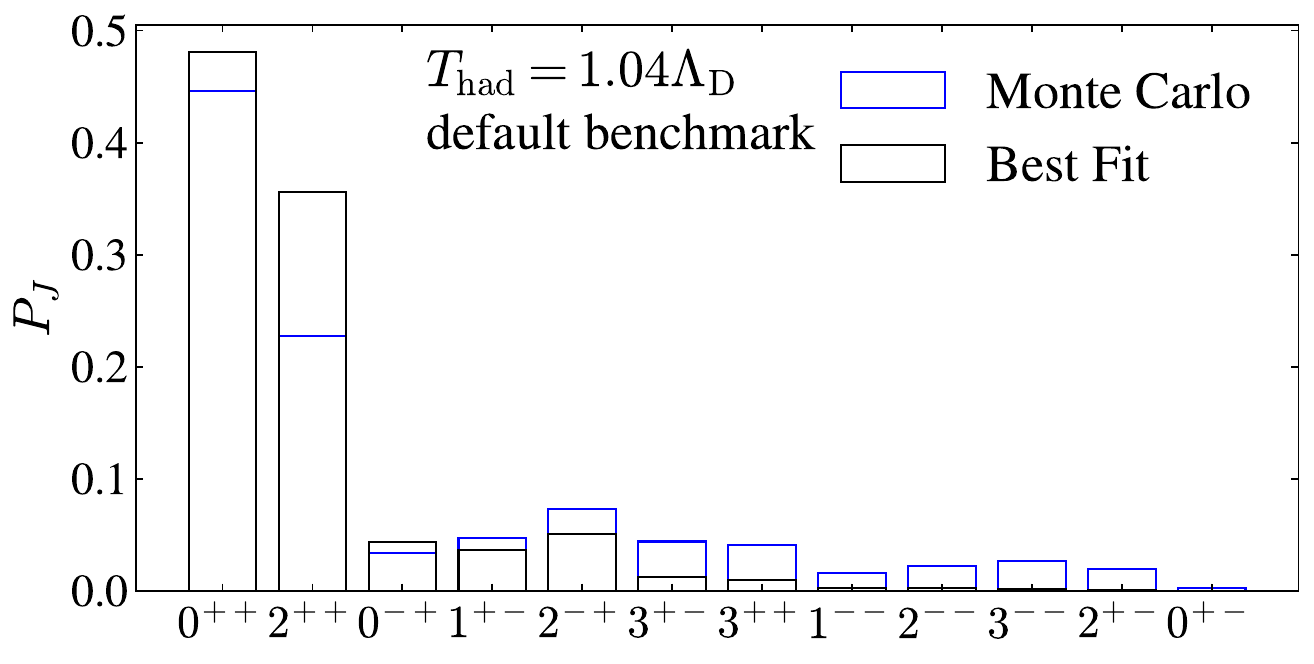}
\caption{Production rates $P_J$ of glueball species in ascending order of mass produced using the default parameters described in \cref{section:benchmarks} and the corresponding best fit to the thermal model in \cref{equation:thermal}. The upper and lower plots have identical information, with the upper plot on a logarithmic scale and the lower plot on a linear scale.
\label{figure:defaultfit}}
\end{figure}

\cref{figure:defaultfit} shows a representative example of these production rates as a function of  species, with the corresponding best fit using the default nuisance parameters described in \cref{section:benchmarks}.
Overall, the agreement between the thermal expectation and our model is fairly good. There is, however, a noticeable and consistent overproduction of very heavy glueball states compared to the Boltzmann expectation.

\begin{figure}[t]
\centering
\includegraphics[width=0.9\textwidth]{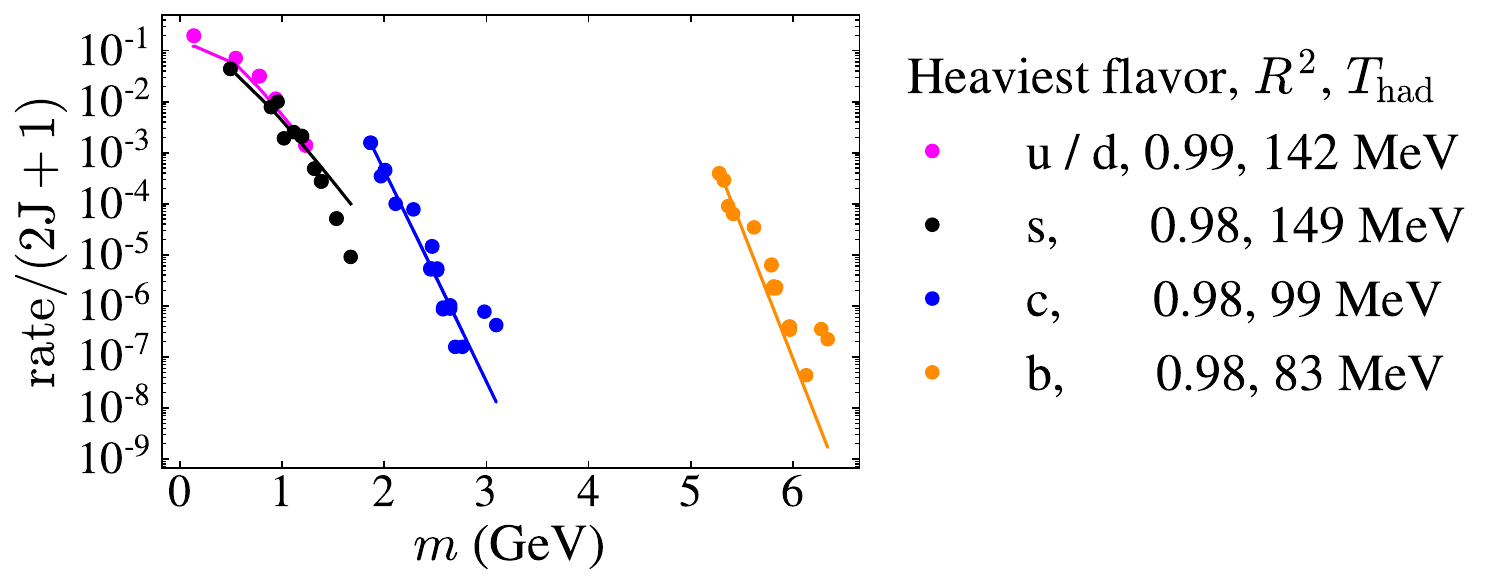}
\caption{Relative production rates of primary SM hadron species, normalized to number of spin degrees of freedom, following decay of a $1\,\text{TeV}$ scalar to two gluons in \textsc{Pythia} with electroweak interactions turned off. 
The grouping is determined by identifying the heaviest valence quark flavor within the hadron,
and each group is fit to its own Boltzmann distribution (solid lines) by maximizing the coefficient of determination $R^2$. 
The contribution to the fit from hadron/anti-hadron species pairs are averaged to avoid fitting the same masses twice. The Monte Carlo statistics are sufficiently high that the excesses of heavy hadrons compared to the fits are not due to random fluctuations. 
\label{figure:SMRates}}
\end{figure}

As an instructive comparison, \cref{figure:SMRates} shows production rates of different SM hadron species produced by \textsc{Pythia}'s default hadronization tune. The distribution organizes itself by the heaviest quark flavor in each species, and each of these groups appears to follow its own scaling relation as a function of mass. We fitted each group to its own Boltzmann distribution, generally finding good agreement between the fits and Monte Carlo, up to a few outliers. In fact, the heaviest SM hadrons appear to exhibit a mild enhancement compared to the thermal fit, which is consistent with the output of our glueball hadronization algorithm.
These fits demonstrate that glueball species production from our algorithm should be approximately Boltzmann-like, but some deviations are to be expected, and the parameters should not necessarily be tuned solely to achieve an optimal Boltzmann fit. Rather, we are encouraged that our model's parameters only mildly impact the fit quality, and we set benchmarks in \cref{section:benchmarks} to capture the most extreme possible variations of the model's output.

\subsection{Benchmark Parameters}
\label{section:benchmarks}

In this section, we suggest three benchmarks for setting the shower cutoff and fragmentation function parameters for collider studies. Our analysis of thermal production rates in \cref{section:fits} does not strongly favor a particular region in fragmentation function parameter space. Therefore, in addition to a well-motivated default choice, we define two bracketing variations that capture the different plausible outcomes of glueball hadronization. These are ``soft" and ``hard" scenarios, where the glueballs tend to have smaller and larger momentum in the dark shower rest frame, respectively.

First, we investigate how varying the hadronization scale parameter $c$ changes the glueball hardness. As the hadronization scale decreases, the gluons in the perturbative shower branch more often, and the Lund strings are softer, so the glueballs emitted by combining those strings are also softer. Thus, benchmarks for softer scenarios should correspond to smaller values of $c$. As for the fragmentation function parameters, it is simplest to interpret their effect in the $\mu_z$-$\sigma_z$ plane as in \cref{section:fits}, since parameters corresponding to a larger $\mu_z$ tend to take more momentum from the fragmenting string pieces, resulting in harder glueballs. It is not obvious \textit{a priori} which of the two fragmentation functions \cref{equation:LSFF} and \cref{equation:fbeta} would result in harder glueballs, but simulations show that the LSFF leads to harder kinematics by a small margin.
See further discussion of glueball hardness in \cref{appendix:benchmarks}.

\begin{table}[t!]
 \renewcommand{\arraystretch}{1.6}
  \setlength{\arrayrulewidth}{.3mm}
  \setlength{\tabcolsep}{0.35 em}
\centering
\begin{tabular}{c|c|c|c|c|c|c|c|c}
\hline
& $c$ & function & \multicolumn{2}{c|}{shape parameters} & $\aD(p_{T\text{min}})$ & $\mu_z$ & $\sigma_z$ & $T_\text{had}/\LD$ \\
\hline
default & 1.8 & LSFF & $a = 1.9\times10^{-4}$ & $b\s m_0^2 = 0.26$ & 1.0 & 0.5 & 0.3 & 1.04 \\
soft & 1.4 & beta & $\alpha = 90.$ & $k_\beta = 810$ & 1.6 & 0.1 & 0.01 & 0.911 \\
hard & 2.1 & LSFF & $a = 82$ & $b\s m_0^2 = 660$& 0.76 & 0.9 & 0.01 & 1.38 \\
\hline
\end{tabular}
\caption{\label{table:benchmarks}Suggested benchmarks to set the hadronization scale and fragmentation function nuisance parameters. Also shown are the corresponding values of the dark sector coupling $\aD$ evaluated at the shower cutoff scale $p_{T\text{min}}=c\,\LD$, the mean $\mu_z$ and standard deviation $\sigma_z$ of the $0^{++}$ fragmentation function, and the best-fit $T_\text{had}$ in \cref{equation:thermal} for the relative species production rates.}
\label{tab:hadbenchmarks}
\end{table} 

With these qualitative effects in mind, we make some concrete suggestions for benchmarks. The default value of $c$ is taken to be $1.8$, since this sets our $p_{T\text{min}}$ at the scale where $\aD = 1$. \textsc{Pythia}'s default settings provide useful guidance in the variation of $c$ because the SM $\alpha_S$ evaluated at the default $p_{T\text{min}}$ for final state radiation is $\simeq 1.6$. The value of $c$ for which our $\aD$ satisfies the same condition is 1.4, which we therefore adopt as our soft benchmark. We choose our hard benchmark $c$ to be 2.1 so that the default is equidistant from the two variations. For the default fragmentation function, we choose the LSFF because this is what is used in the Lund string model. We choose the default function parameters to satisfy $\mu_z = 0.5$ and $\sigma_z = 0.3$ since this is about as close to a uniform distribution as possible with this fragmentation function, and we do not have any reason to favor large or small $z$. For the soft and hard scenarios, we want to capture the extremes of the possible variations, so we choose $\mu_z = 0.1$ and 0.9, respectively, with $\sigma_z = 0.01$. These numerical values, and their corresponding fragmentation function parameters, are summarized in \cref{table:benchmarks}.

\subsection{Total Multiplicity Distributions}
\label{section:totalN}

\begin{figure}[t!]
    \centering
    \includegraphics[width=0.55\textwidth]{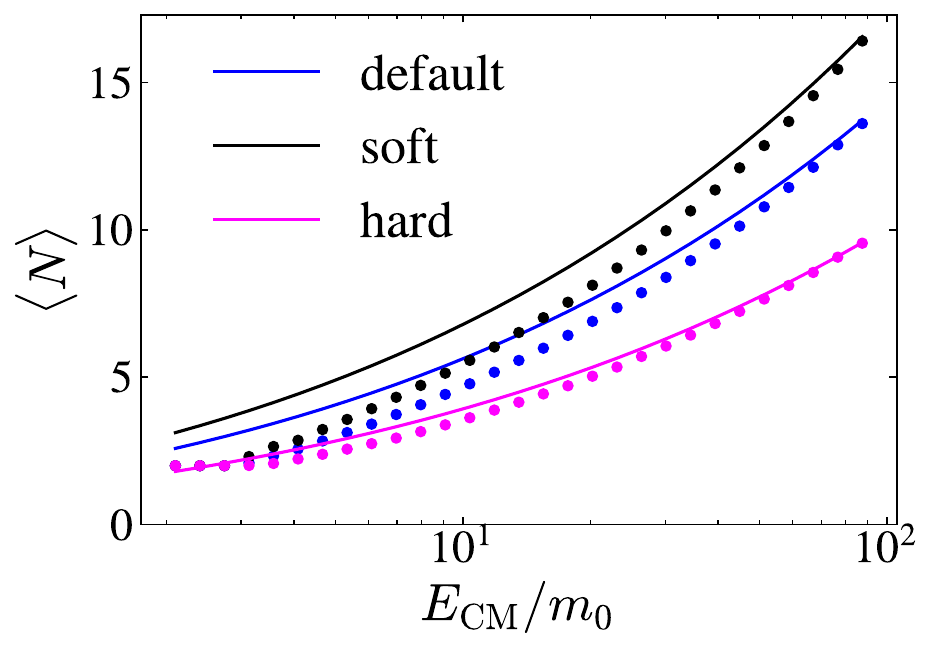}
    \caption{Average glueball multiplicity as a function of $E_\text{CM}$. The points show the output of our algorithm, and the solid lines show the QCD prediction in \cref{equation:Nscaling} normalized to match the Monte Carlo at the largest $E_\text{CM}$.
    \label{figure:multscaling}}
\end{figure}

In addition to the relative production rates of different glueball species in \cref{section:fits}, we can compare distributions of the total number of glueballs per event $N$ from our algorithm to analytical predictions for QCD with zero flavors. In the pure-glue limit with small hadron masses, the average number of hadrons is expected to scale with center of mass energy $E_\text{CM}$ as \cite{Ellis:1991qj}

\begin{equation} \label{equation:Nscaling}
\langle N (E_\text{CM})\rangle \propto \exp \left[ \frac{12\s \pi}{11\s C_A}\sqrt{\frac{2\s C_A}{\pi \aD(E_\text{CM})}} + \frac{1}{4} \ln \left(\aD(E_\text{CM}) \right) \right]\,,
\end{equation}

\noindent where $C_A$ is the quadratic Casimir factor for the adjoint representation, which is $N_c$ for $SU(N_c)$. \cref{figure:multscaling} shows a comparison of $\langle N \rangle$ at various values of $E_\text{CM}$ between our algorithm with each set of benchmark parameters and the QCD prediction, with the normalization of \cref{equation:Nscaling} fixed by matching to the Monte Carlo at the largest $E_\text{CM}$. 
The analytic prediction is slightly larger than our model at low energies, except where $E_\text{CM}$ is near $2m_0$ and the prediction falls below the kinematic threshold of two hadrons.
This is an expected consequence of the finite hadron masses, since for smaller $E_\text{CM}$ and parameters that tend to produce more glueballs, a larger portion of the energy is taken up by glueball masses.
The reproduction of this standard result is a useful check on the validity of our algorithm.
Notably, our algorithm tends to produce a greater number of glueballs than \texttt{GlueShower}, which generated $\langle N\rangle\sim 7$ at $E_\text{CM} = 100\,m_0$ \cite{Curtin:2022tou}. Therefore, our more physically-motivated approach predicts greater discovery potential.

\section{Dark Glueball Decay via Higgs Portal}
\label{section:glueballs}

We now have a concrete numerical method to simulate the production of dark glueballs.  In order to connect with phenomenology, we need to specify the portal between the dark sector and the SM that will determine how the dark gluons are produced in a hard interaction and how the dark glueballs decay.
We consider a pure-glue dark sector that couples to the SM via the Higgs portal, since this is the lowest-dimension portal that can connect the pure-glue sector to the SM. We compute the resulting glueball lifetimes
and show them in the parameter space of two neutral naturalness scenarios. 

\subsection{Dark Glueball Decay Widths}
\label{s.higgsportal}

In this section, we briefly summarize the pertinent results of \cite{Juknevich:2009gg} used in our estimation of dark sector glueball lifetimes and decay branching ratios. Strongly-coupled dark sectors that include heavy fermions coupling to the Higgs give rise to the effective dimension-6 Higgs portal operator\footnote{Typically, one would define an effective scale $\Lambda =M/y$ when writing a higher-dimensional portal like this.  We leave the $M/y$ explicit here to avoid confusion with $\LD$.}

\begin{equation} \label{equation:dim6}
\delta\mathcal{L}^{(6)} = \frac{\aD}{3\pi} \: \frac{y^2}{M^2} \: H^\dagger H \: \text{tr}\left(G^{\mu\nu}_{(\text{D})}G_{(\text{D})\mu\nu}^{\vphantom{\mu}}\right),
\end{equation}

\noindent where $H$ is the SM Higgs doublet, $G^{\mu\nu}_{(\text{D})}$ is the dark gluon field strength, $\aD$ is the dark sector strong coupling, $M$ is the mass scale of the dark sector fermions, and $y$ is an effective coupling that is determined by a model-dependent combination of the dark sector fermion Yukawa couplings with the Higgs (see~\cite{Juknevich:2009gg} for explicit expressions). This operator can mediate both dark gluon production at the LHC and subsequent glueball decay to the SM. 

\begin{table}[t!]
\centering
 \renewcommand{\arraystretch}{1.6}
  \setlength{\arrayrulewidth}{.3mm}
  \centering
  \small
  \setlength{\tabcolsep}{0.35 em}

\begin{tabular}{c|c|c}
\hline
 Glueball & Mass ($m_0$) & Higgs Portal \\ \hline
 $0^{++}$ & 1.00 & $h^*\rightarrow \rm{SM,SM}$ \\ \hline
 $2^{++}$ & 1.40 &$0^{++} + h^*$ \\ \hline
 $0^{-+}$ & 1.50 & -  \\ \hline
 $1^{+-}$ & 1.75 & -  \\ \hline
 $2^{-+}$ & 1.78 & $0^{-+} + h^*$  \\ \hline
 $3^{+-}$ & 2.11 & $1^{+-} + h^*$  \\ \hline
 $3^{++}$ & 2.15 & $\{2^{++},0^{-+},2^{-+}\} + h^*$ \\ \hline
 $1^{--}$ & 2.25 & $1^{+-} + h^*$  \\ \hline
 $2^{--}$ & 2.35 & $\{1^{+-},3^{+-},1^{--}\}+h^*$  \\ \hline
 $3^{--}$ & 2.46 & $\{1^{+-},3^{+-},1^{--},2^{--}\} + h^*$  \\ \hline
 $2^{+-}$ & 2.48 & $\{1^{+-},3^{+-},1^{--},2^{--},3^{--}\} + h^*$ \\ \hline
 $0^{+-}$ & 2.80 & $\{1^{--},3^{--},2^{+-}\} + h^*$ \\ \hline
\end{tabular}

\caption{Table of masses and decay channels for each glueball; $h^*$ indicates an off-shell Higgs.}
\label{table:decay_table}
\end{table}

The decay channels for each of the twelve glueballs are summarized in \cref{table:decay_table}. The $0^{++}$ species decays into SM states $\xi$ by mixing with the Higgs boson, $0^{++}\rightarrow h^* \rightarrow \xi\xi$, with the decay width

\begin{equation}
    \Gamma_{0^{++}\rightarrow\xi\xi}=\frac{y^4}{M^4}\bigg(\frac{ v \aD \mathbf{F_{0^{++}}}}{3\pi(m_h^2-m_0^2)}\bigg)^2\:\Gamma^{\text{SM}}_{h\rightarrow\xi\xi}(m_0),
\end{equation}
where $m_h$ is the Higgs mass, $\Gamma^{\text{SM}}_{h\rightarrow\xi\xi}(m_0)$ is the decay width for a Higgs-like scalar of mass $m_0$, which we calculate using HDECAY \cite{Djouadi:2018xqq}, and $\mathbf{F_{0^{++}}}$ is the non-perturbative decay constant with mass dimension 3. The heavier species (with the exceptions of the stable $0^{-+}$ and $1^{+-}$) decay into lighter glueballs via emission of an off-shell Higgs, $J\rightarrow J'+h^*(\rightarrow \xi\xi)$. The decay width for a glueball with spin $J$ to a lighter glueball with spin $J'$ and the SM is given by

\begin{multline}
    \Gamma_{J\rightarrow J'\xi\xi}=\frac{1}{16\pi\s m_J (2J+1)}\frac{y^4}{M^4}\bigg(\frac{v \aD|\mathbf{M}_{J,J'}| }{3\pi}\bigg)^2 \times \\[3pt]
    \int \text{d}m^2_{12} \frac{m_{12}}{\pi} \frac{\Gamma_{J,J'}^{(i)}[g(m_{J'}^2,m_{12}^2;m_J^2)]^{1/2}}{(m_{12}^2-m_h^2)^2+m_h^2(\Gamma_h^{SM})^2}\Gamma^{\text{SM}}_{h\rightarrow\xi\xi}(m_{12})\,,
\end{multline}
where $g(x,y;z)=(1-x/z-y/z)^2-4\s x\s y/z^2$, $|\mathbf{M}_{J,J'}|$ is the averaged non-perturbative transition matrix element, and $\Gamma_{J,J'}^{(i)}$ are dimensionless functions of the glueball masses that depend on the angular momentum transfer associated with each transition, which can be found in \cite{Juknevich:2009gg}.
The mass splitting of the glueball states can be a few GeV at small $m_0$, where perturbative SM QCD breaks down, so we use $\Gamma^{\text{SM}}_{h\rightarrow\xi\xi}(m_{12})$ values calculated from chiral perturbation theory \cite{Winkler:2018qyg} for this range. 

The glueball decay widths depend on the theory parameters $m_0$ and $M/y$, as well as the non-perturbative decay constants and transition matrix elements. The matrix elements corresponding to the decays of the lightest glueballs have been calculated on the lattice \cite{Chen:2005mg}, e.g.\ $4\pi\s \alpha_{\rm{D}}\mathbf{F_{0^{++}}}=2.3\s m_0^3$, which we use in this work. However, the matrix elements for the heavier states have not been computed.  We use dimensional analysis to approximate the remaining heavy species' transition elements up to dimensionless prefactors, thus obtaining the correct scaling with $m_0$.  We set  $|\mathbf{M}_{J,J'}|= m_0^3$ for our decay widths, vary each matrix element independently by a factor of two, and marginalize over the variation as part of our hadronization uncertainty. One could in principle also incorporate the dimension-8 operators listed in \cite{Juknevich:2009gg}, which render the $0^{-+}$ and $1^{+-}$ unstable.  However, corrections due to these operators are suppressed by multiple orders of magnitude in the parameter space of interest, and these species only make up a few percent of produced glueballs, so we do not include these decays in our study.


\begin{figure}[t!]
\centering
\includegraphics[width=0.49\textwidth]{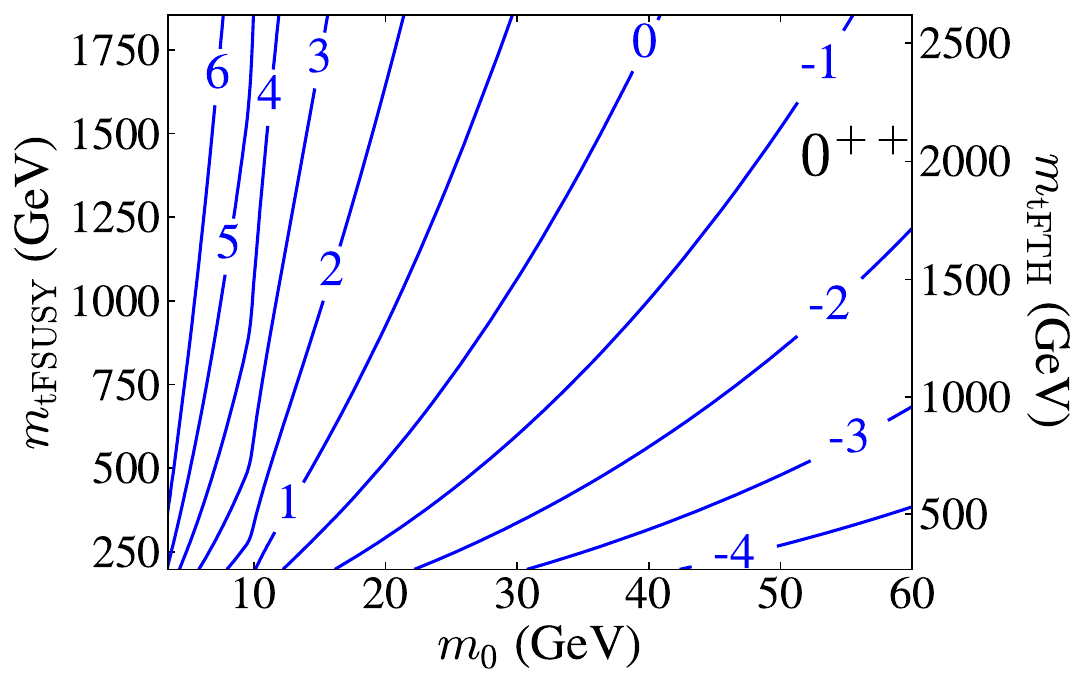}
\includegraphics[width=0.49\textwidth]{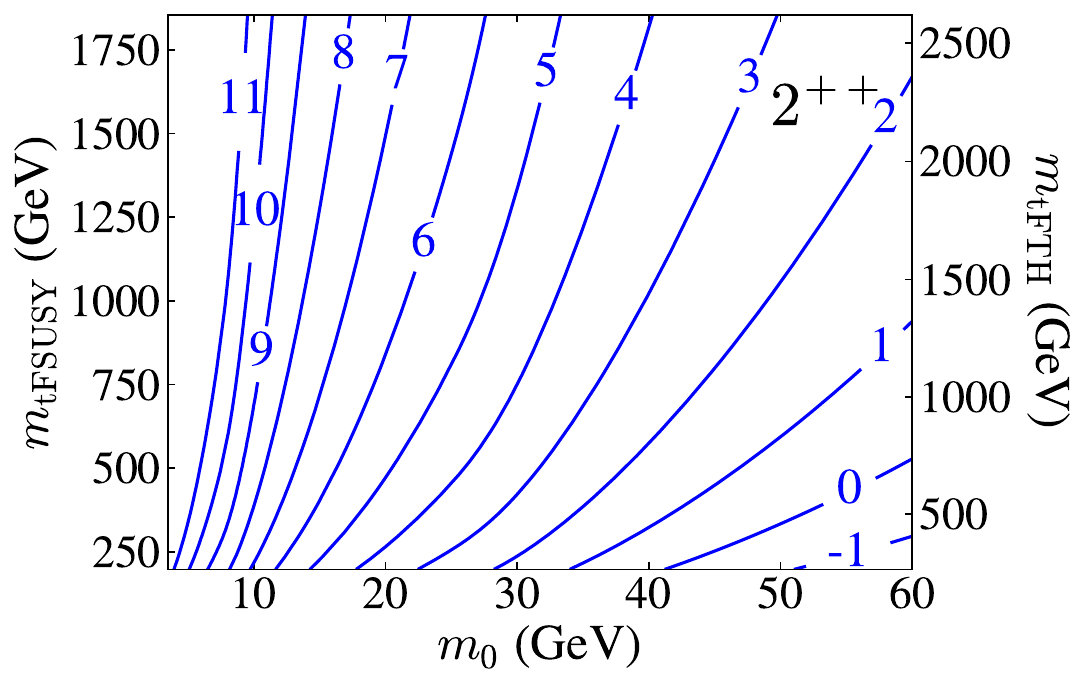}
\includegraphics[width=0.49\textwidth]{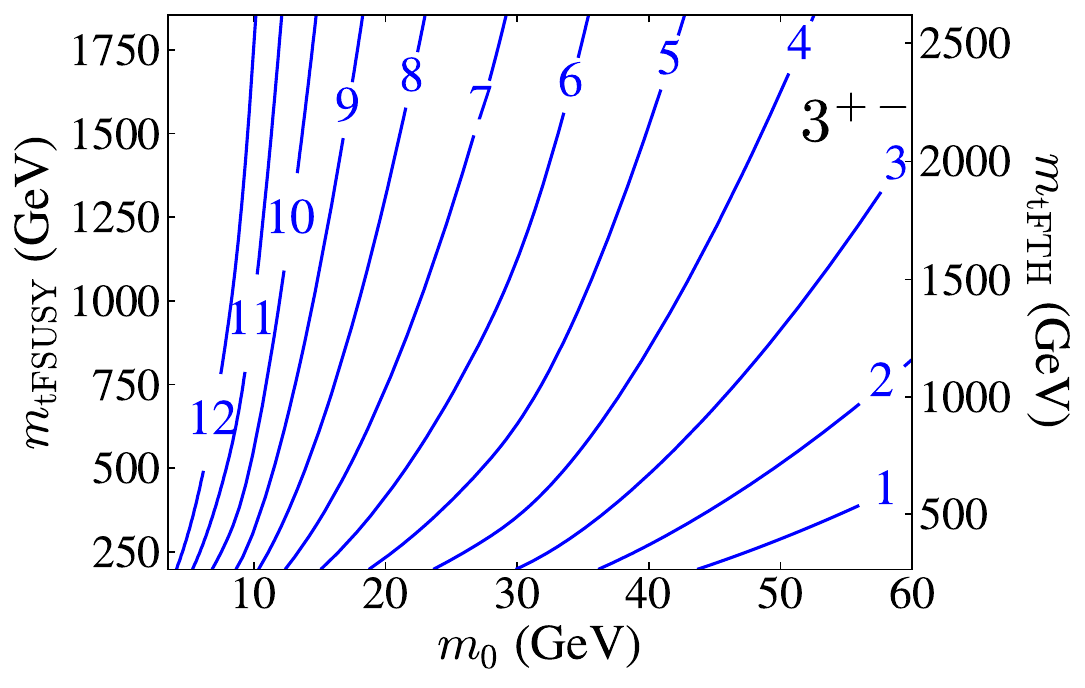}
\includegraphics[width=0.49\textwidth]{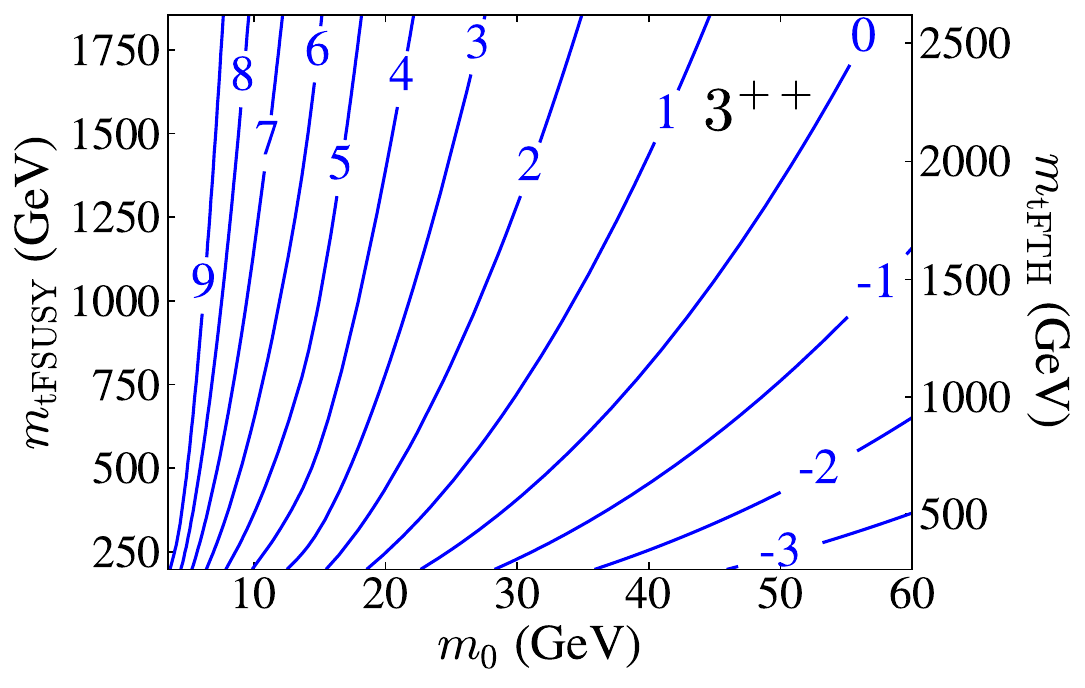}
\caption{
Contours show $\log_{10}(c\tau/\text{m})$, where c$\tau$ is the mean decay length of the glueball in the space of the lightest glueball mass $m_0$ and top partner masses assuming folded supersymmetry $m_{\text{tFSUSY}}$ or fraternal twin Higgs $m_{\text{tFTH}}$. The top plots show the two lightest species, and the bottom plots show representative examples of heavier species.\label{figure:lifetimes}}
\end{figure}

\subsection{Neutral Naturalness Models}
\label{section:maps}

The results of \cref{s.higgsportal} can be readily mapped onto parameters in neutral naturalness models because they generate the Higgs portal operator in \cref{equation:dim6}~\cite{Curtin:2015fna}.
For the Fraternal Twin Higgs (FTH)~\cite{Craig:2015pha}, 

\begin{equation}
\frac{M}{y} = \frac{2v \: m_\text{tFTH}}{m_t \sqrt{\cos{\theta}}}\,,
\end{equation}

\noindent where $m_\text{tFTH}$ is the twin top  mass, and $\theta=\tan^{-1}(m_t/m_\text{tFTH})$. For Folded Supersymmetry (FSUSY)~\cite{Burdman:2006tz},

\begin{equation}
\frac{M}{y} =  \frac{\:v\sqrt{8}\:m_{\text{tFSUSY}}}{m_{t}} \,,
\end{equation}

\noindent where $m_t$ is the SM top quark mass, $m_{\text{tFSUSY}}$ is the folded stop mass, and $v$ is the SM Higgs vacuum expectation value. 
This mapping allows us to predict lifetimes and branching ratios for each glueball species given a choice of $m_0$ and the mass of the FSUSY or FTH top quark partner. \cref{figure:lifetimes} shows some representative examples. The plots show that $0^{++}$ state is the shortest lived state for each point in parameter space because it mixes directly with the Higgs. The next few heavy species that can decay via the dimension-6 operator ($2^{++},2^{-+},3^{+-}$) have much longer lifetimes due to only being able to radiate an off-shell Higgs. The remaining heaviest states ($3^{++}$ and above) have slightly shorter lifetimes due to having more available decay channels with larger mass splittings for the off-shell Higgs.


\section{Results}
\label{section:results}

One of the most important characteristics of the LHC signatures of dark glue showers is the distribution of glueball lifetimes.  Depending on the fundamental parameters of the model, the predictions range from semivisible jets (all the decays that are visible to ATLAS/CMS are prompt) to emerging jets (the glueball decays occur within the ATLAS/CMS detector) to the long lifetime regime (all glueballs escape the main detectors). 
For small $m_0$ and large $M/y$, all species are sufficiently long-lived that a dedicated long-lived particle experiment such as MATHUSLA~\cite{Chou:2016lxi,MATHUSLA:2018bqv, MATHUSLA:2020uve} can significantly extend sensitivity beyond main detector searches due to its large volume. There are also regions where one kind of jet signature dominates over another, or a mixture of both strategies is potentially viable. Given that a generic strongly-coupled sector can have a spectrum of many hadrons with a broad hierarchy of lifetimes (as in the SM itself), an optimal search could incorporate methods from the semivisible jet, emerging jet, and external LLP detector strategies simultaneously.

In the remainder of this section, we discuss two production mechanisms for glueballs at the LHC: through the Higgs and through a new heavy $Z'$. For each mechanism, we show the parameter space relevant for semivisible or emerging jet searches, and we discuss predictions for two different classes of glueball collider phenomenology. For Higgs portal production, we make predictions for glueball decays that could be observed in the proposed MATHUSLA experiment, considering glueball production and decays through a Higgs portal within FSUSY and FTH as discussed in \cref{section:maps}. This will supersede the rudimentary MATHUSLA sensitivity estimates for neutral naturalness presented in~\cite{Curtin:2018mvb}.

For the $Z^\prime$ production, we show that this model could yield a good benchmark for semivisible jet and emerging jet searches. One phenomenological parameter used in the studies of semivisible jets is $r_{\text{inv}}$, the average ratio of the number of dark hadrons that are stable on collider scales compared to the total number of total dark hadrons produced.  In current searches, one takes $r_{\text{inv}}$ as a simplified model-like input parameter and models the distribution of the invisible fraction of hadrons as Poissonian. Our ability to model hadronization and decay of dark glueballs allows us map parameters in the fundamental description onto a prediction for $r_{\text{inv}}$ event by event. Thus, we provide a theoretically motivated range of $r_{\rm{inv}}$ distributions to consider for future semivisible jet searches.

\subsection{Dark Glueballs via Higgs Production}
\label{section:MATHUSLA}

\begin{figure}[t]
\centering
\includegraphics[width=0.49\textwidth]{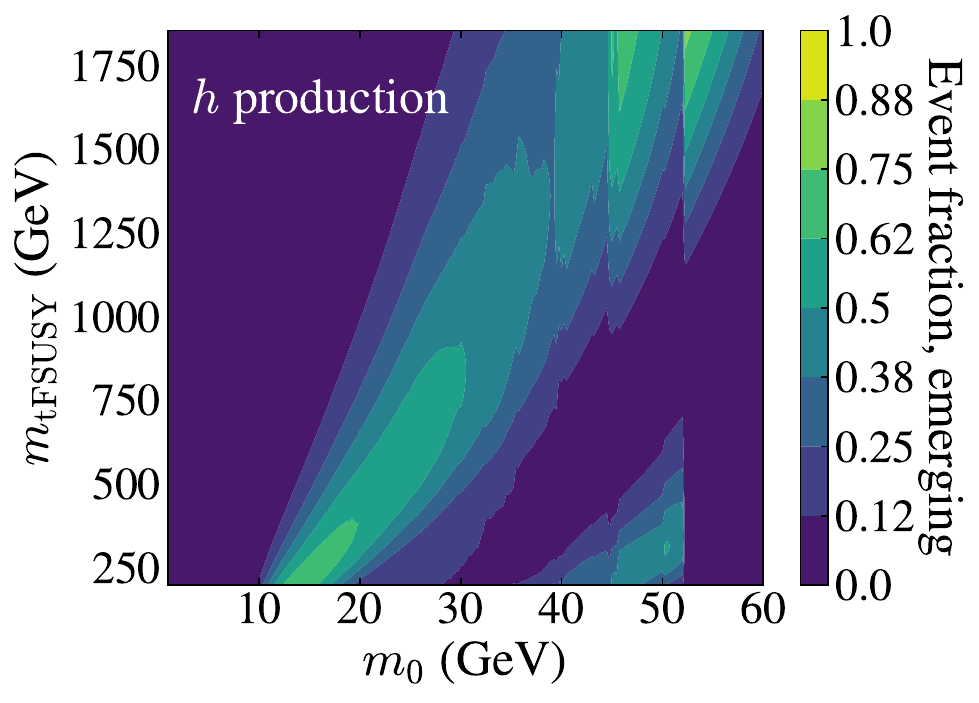}
\includegraphics[width=0.49\textwidth]{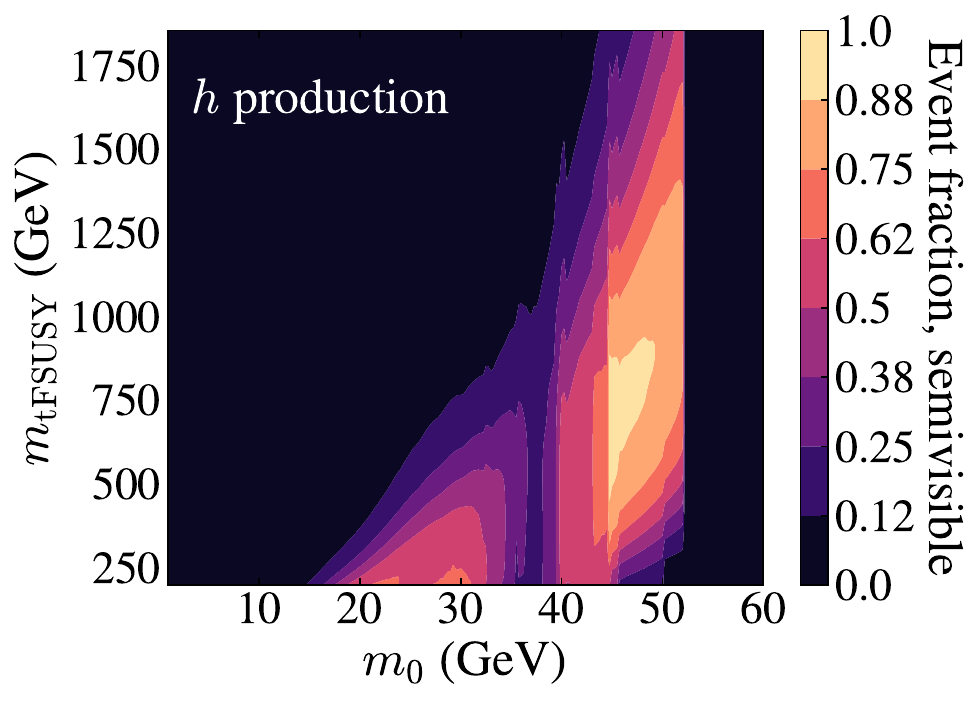}
\caption{Fractions of dark glueball events for the Higgs production scenario satisfying necessary but not sufficient conditions to produce emerging jet signals (left) or semivisible but not emerging jet signals (right). For the emerging jet fractions, events were required to have at least one glueball decay within the CMS tracker with transverse displacement of at least $50\,\text{mm}$ \cite{Schwaller:2015gea}. For the semivisible jet fraction, events were required to have at least one glueball escape the tracker, at least one prompt glueball decay within the tracker, and no glueball decays within the tracker with transverse displacement $>50\,\text{mm}$.
\label{figure:eventFracsh}}
\end{figure}

In this section, we discuss the signatures of dark glueball showers produced via the Higgs portal, which we outline in \cref{sec:higgs}. \cref{figure:eventFracsh} shows fractions of dark glueball events that could possibly give rise to an emerging jet signature, as well as fractions of events that could possibly have a semivisible jets signature with no displaced decays. These plots reveal how different regions of parameter space motivate different combinations of main detector search strategies depending on the glueball lifetime hierarchy. Therefore, the below sensitivity analysis for MATHUSLA will demonstrate where in parameter space the dedicated LLP strategy has reach beyond the main detectors and where these strategies have potential overlap.

\subsubsection{Higgs Production} 
\label{sec:higgs}
To model Higgs production of dark glueballs we simulate gluon-gluon fusion and VBF in \textsc{MadGraph5}\_a\textsc{mc@nlo} \cite{Alwall:2014hca} + \textsc{Pythia} 8~\cite{Bierlich:2022pfr}. 
Gluon fusion is implemented via the effective $ggh$ operator, with jet matching for up to one extra hard jet and slight event reweighting to  reproduce the NLO+NNLL Higgs $p_{T}$ spectrum computed by HqT 2.0~\cite{Bozzi:2005wk,deFlorian:2011xf}. 
As discussed in \cite{Curtin:2015fna}, the Higgs-to-dark gluon branching ratio can be found by a rescaling of the SM Higgs-to-gluon branching ratio of $8.5\%$~\cite{ATLAS:2015egz,Workman:2022ynf}.


\subsubsection{Dark Glueballs at MATHUSLA}

\begin{figure}[t!]
\centering
\includegraphics[width=0.49\textwidth]{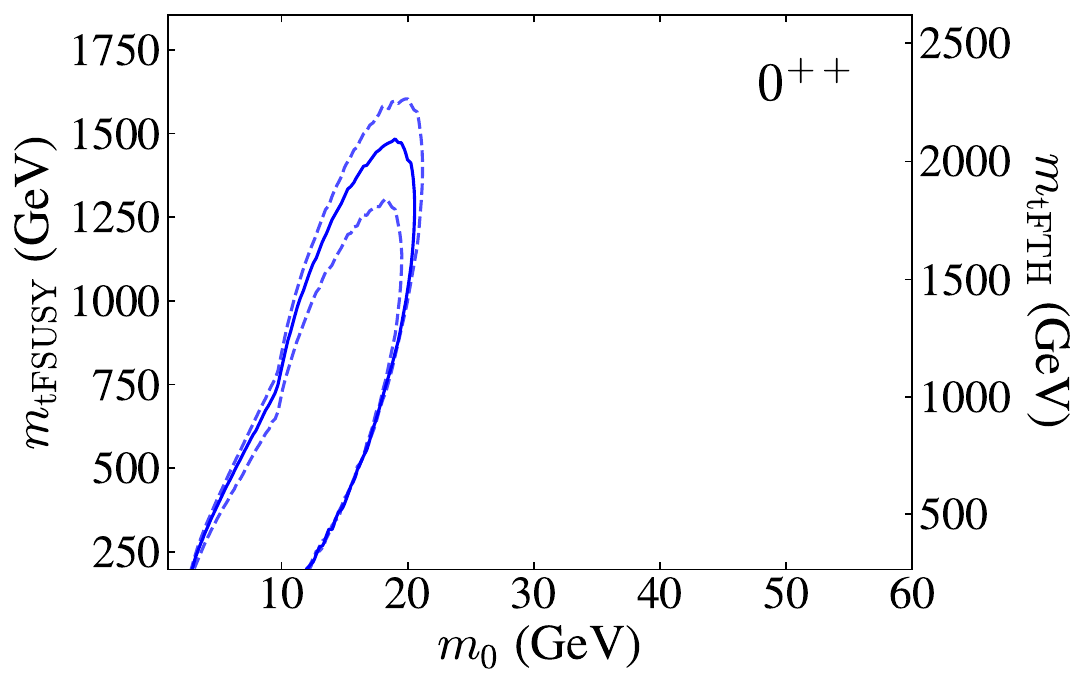}
\hspace{-0.3cm}
\includegraphics[width=.49\textwidth]{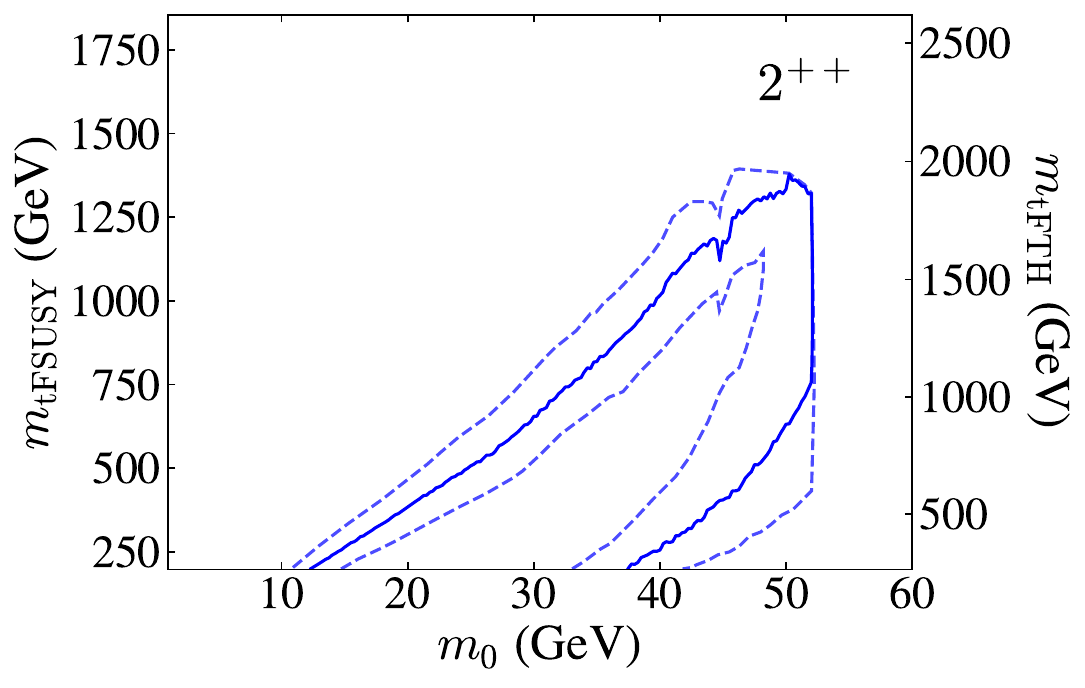}
\includegraphics[width=.49\textwidth]{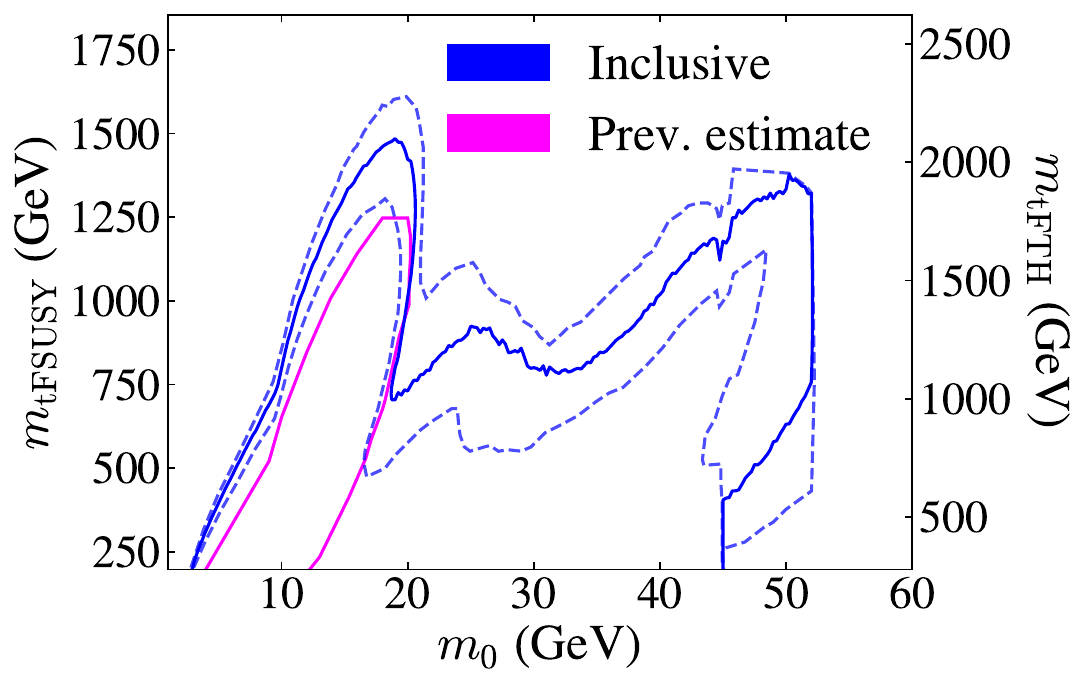}
\caption{Sensitivity curves for glueball decays in MATHUSLA in the space of the lightest glueball mass $m_0$ and top partner masses in the fraternal twin Higgs model $m_{\text{tFTH}}$ or folded supersymmetry model $m_{\text{tFSUSY}}$.  We take 4 events within the decay volume as the exclusion limit. The top plots show exclusive decays of the two lightest glueball species, and the bottom is inclusive of all species. The dashed contours reflect uncertainties due to variation of both the hadronization benchmark and the decay matrix elements. The inclusive plot also shows the previous estimate from \cite{Curtin:2018mvb} based on the simplifying conservative assumption of two-body exotic Higgs decays $h\rightarrow 0^{++}0^{++}$ only.\label{figure:MATH}}
\end{figure}

In \cref{figure:MATH}, we show sensitivity curves for decays within the $100\,\text{m} \times 100\,\text{m} \times 25\,\text{m}$ MATHUSLA decay volume as specified in~\cite{MATHUSLA:2020uve}, assuming an integrated luminosity of $3\,\text{ab}^{-1}$ at $\sqrt{s} = 14\,\text{TeV}$. The contours we show correspond to 4 decays in MATHUSLA's decay volume, illustrating the exclusion reach in the absence of backgrounds, which is expected for LLP decays to high multiplicities of SM hadrons.
The experimental bound on the Higgs-to-invisible branching ratio  of $18\%$ \cite{CMS:2022qva} excludes the parameter space of top partner masses below the range shown in the plots. We account for uncertainty in the various heavy glueball lifetimes by varying the corresponding decay constants independently by a factor of 2 in each direction. Our uncertainty bands also include the variation obtained by running simulations with the different hadronization benchmarks introduced above in \cref{tab:hadbenchmarks}. 

A striking feature of these results is the importance of including the heavier glueball species, and the resulting dramatic increase to MATHUSLA's estimated sensitivity in neutral naturalness parameter space.
Since the heavier glueballs have longer lifetimes than the $0^{++}$, MATHUSLA is able to probe an entirely complementary mass regime with heavier glueball decays, extending its reach up to $m_0\sim 50\,\text{GeV}$ compared to the $\sim20\,\text{GeV}$ maximum probed by the $0^{++}$ alone.

Previous studies of dark glueball phenomenology \cite{Curtin:2015fna,Curtin:2018mvb} made the conservative simplifying assumption that only two glueballs were produced in exotic Higgs decays, some fraction of which was $h\rightarrow 0^{++}0^{++}$, the only channel assumed to be observable. 
This was necessitated by the absence of a realistic simulation framework for glueball production.
Our work significantly improves on these previous sensitivity estimates by including the dark gluon shower and Lund string-inspired glueball hadronization, allowing for both the higher glueball multiplicity in each Higgs decay and the contribution of heavier, more long-lived glueball states to be systematically taken into account. 
This leads to the improved projections for the inclusive total reach of MATHUSLA to all glueball decays, shown in the bottom panel of \cref{figure:MATH}. 
It is interesting to note that this updated inclusive MATHUSLA reach therefore not only includes the long $0^{++}$ lifetime regime below $20\,\text{GeV}$, but also  exceeds, or is at least comparable to, the total projected coverage of main detector searches relying on LLP decays in the tracker or muon system for $m_0 \lesssim 50\,\text{GeV}$, computed with the above simplifying two-body-decay assumption~\cite{Curtin:2015fna}. While the main detector search sensitivities would be expanded due to increased glueball multiplicity in our updated simulations, the inclusion of heavier glueballs would have a much smaller effect than it did for MATHUSLA, since the main detector is most sensitive to short lifetimes.  While the detailed study of main detector sensitivities to dark glueballs is an important subject of future study with our updated simulation framework, this nonetheless already suggests that MATHUSLA's LLP sensitivity may dramatically enhance new physics coverage in a large region of dark glueball parameter space.

Our results also motivate further study into the properties of the heavier species. In particular, lattice computations to determine the decay matrix elements would reduce uncertainty in the glueball lifetimes. In the regions of parameter space where the heavier species dominate decays in MATHUSLA, the uncertainty due to lifetime variation is larger than that due to the hadronization benchmark variation.

Note that the final states of  $2^{++}$  decay always include a $0^{++}$, and the region of parameter space where the $2^{++}$ dominates decays in MATHUSLA is also where the $0^{++}$ has short $\mathcal{O}(\mathrm{cm})$ lifetimes, see \cref{figure:lifetimes}. 
Therefore, given the cm-scale tracking resolution of the MATHUSLA experiment, the $2^{++}$ decay can be treated as a single vertex.
This region of parameter space is also interesting because any $0^{++}$ produced would decay within CMS. These can be searched for with dedicated searches using CMS detector information alone~\cite{Curtin:2015fna,CMS:2020iwv} (though with significant signal penalty due to trigger limitations) or a combined MATHUSLA-CMS search if MATHUSLA provides a trigger signal to CMS~\cite{Barron:2020kfo}. In the latter case, simultaneous reconstruction of the  $0^{++}$ and $2^{++}$ decay would allow a detailed characterization of the dark sector and provide strong evidence that the newly discovered LLP states are in fact dark glueballs. 

\subsection{Dark Glueballs via \texorpdfstring{$Z'$}{Z'} Production}
\label{section:semivisible}

\begin{figure}[t!]
\centering
\includegraphics[width=0.49\textwidth]{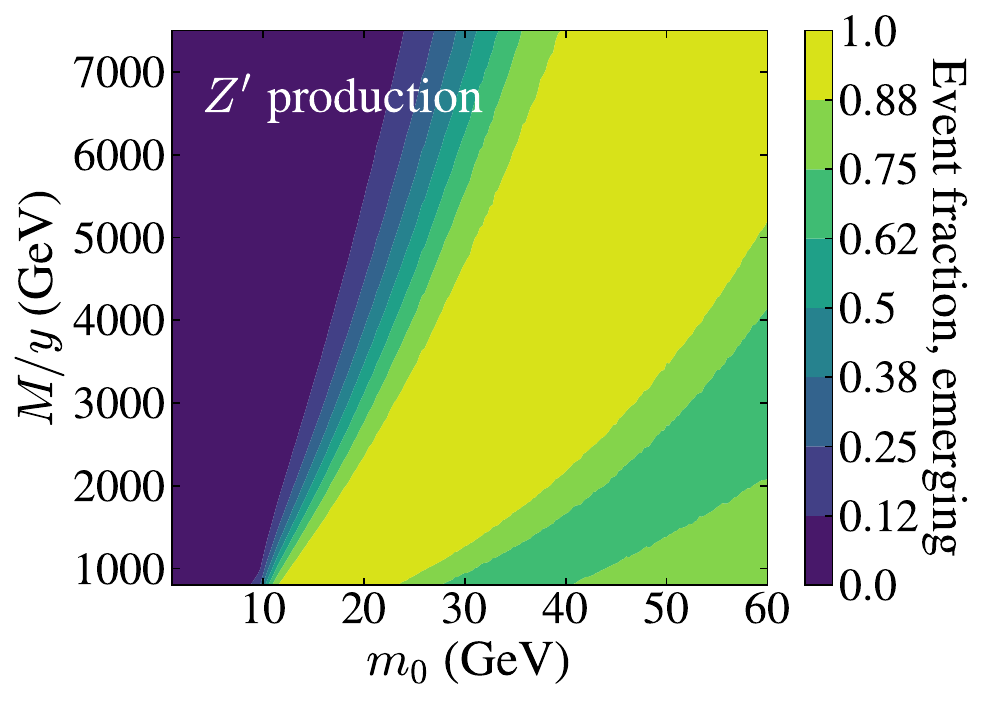}
\includegraphics[width=0.49\textwidth]{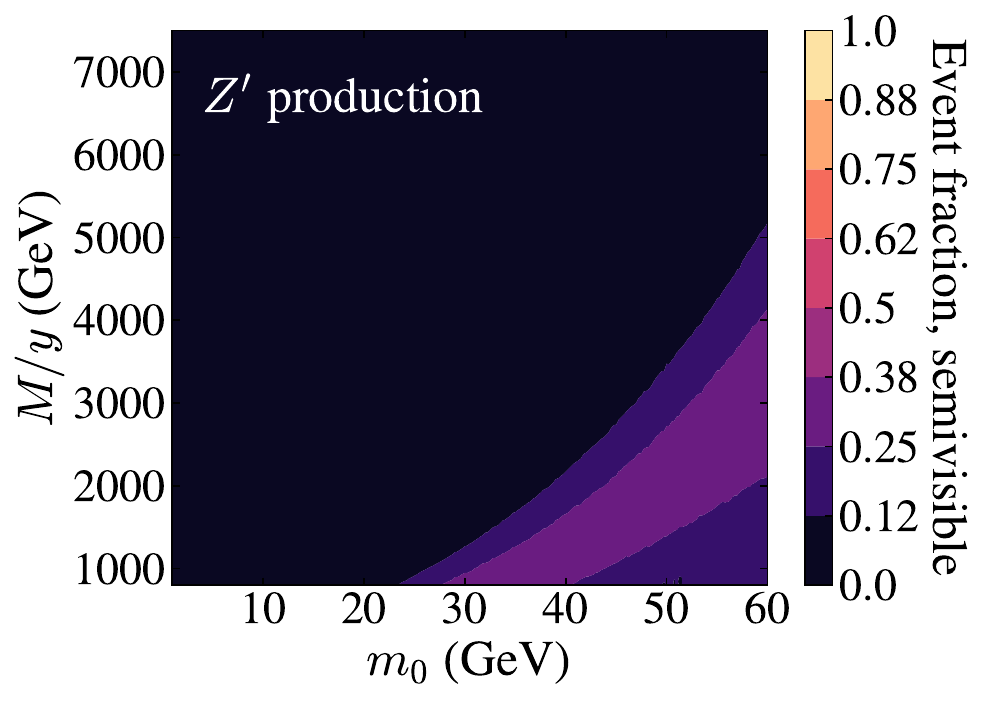}
\caption{Fractions of dark glueball events in the $Z'$ production scenario with $m_{Z^\prime} = 3\,\text{TeV}$ satisfying necessary but not sufficient conditions to produce emerging jet signals (left) or semivisible but not emerging jet signals (right). For the emerging jet fractions, events were required to have at least one glueball decay within the CMS tracker with transverse displacement of at least $50\,\text{mm}$ \cite{Schwaller:2015gea}. For the semivisible jet fraction, events were required to have at least one glueball escape the tracker, at least one prompt glueball decay within the tracker, and no glueball decays within the tracker with transverse displacement $>50\,\text{mm}$.\label{figure:eventFracsZ}}
\end{figure}

In some of our parameter space, the lightest dark glueballs can decay promptly while the rest are either stable or very long lived.  This would lead to LHC events where the visible jet transverse momentum $\vec{p}_T^{\,J}$ and missing transverse momentum $\vec{p}_T^{\,\,\text{miss}}$ are aligned, which is the characteristic property of so-called semivisible jets \cite{Cohen:2015toa}.  
To define a benchmark for future semivisible jet searches, we consider the simplified signal model used in the recent CMS semivisible jet search~\cite{CMS:2021dzg}. 
This search assumed the resonant production of a $Z^\prime$ mediator that decayed to dark sector quarks, which showered and formed dark hadrons that decayed to SM quarks. We retain the $Z^\prime$ mediator, but we work in the region of parameter space that produces dark glueballs, and we introduce the Higgs portal to facilitate the glueball decays. Further details of the production mechanism via a heavy $Z'$ are outlined below in \cref{section:Zprime}. In \cref{figure:eventFracsZ}, we show the fractions of dark glueball events that could give rise to a emerging jet and/or a semivisible jet signature through $Z'$ production. Semivisible jet searches may be able to probe the large $m_0$ regime, while emerging jet search strategies may be able to probe parameter space that includes lower $m_0$ values. The actual ATLAS or CMS sensitivity to these search strategies requires detailed modeling of the emerging or semivisible jets including SM backgrounds, which is beyond the scope of this paper.

\subsubsection{$Z'$ Production}
\label{section:Zprime}

The signal model features a $Z^\prime$ that couples to both SM quarks and dark sector quarks $Q_{\rm{D}}$ charged under the dark QCD with confinement scale $\LD$.
This allows for dark quark pair production in LHC collisions $p\s p \rightarrow Z' \rightarrow Q_{\rm{D}} \bar{Q}_{\rm{D}}$.
In the parameter regime analyzed by the CMS search, the dark sector quarks have mass $M_Q < \LD \ll M_{Z'}$, which hadronize into jets of dark mesons (bound states of quark--anti-quark pairs). 
For the semivisible jet benchmark introduced here, we instead consider the quirk-like regime~\cite{Kang:2008ea}, with  $\LD \ll M_Q \sim M_{Z'}/2$. This implies that $Q_{\rm{D}} \bar Q_{\rm{D}}$ pair production via the $Z'$ resonance results in a quirk bound state. The $Q_{\rm{D}} \bar Q_{\rm{D}}$ pair are connected by an oscillating flux tube, which de-excites by radiating glueballs (and angular momentum) before the dark quark pair annihilates into dark gluons.  Since the de-excitation sheds angular momentum, the final annihilation is anticipated to be dominated by the $s$-wave.
We also assume that the dark quarks couple to the SM Higgs with a Yukawa coupling $y$.  Integrating out the dark quarks generates the Higgs portal operator, which we assume provides the dominant channel for the dark glueball decays.

The dynamics of quirk de-excitation via glueball emission are not well understood, but as a na\"{i}ve first guess, we assume the glueball radiation from de-excitation is highly subdominant compared to the glueballs produced in the ultimate $s$-channel annihilation of the $Q_{\rm{D}} \bar Q_{\rm{D}}$ pair. This can be guaranteed by setting $M_Q$ just below $M_{Z'}/2$, where a tiny mass difference is required to allow emission of a single glueball to shed the quirk's orbital angular momentum. 

This model technically contains both Higgs and vector portals to the dark sector. In practice however, the vector portal dominates dark quark pair production for $M_{Z^\prime}$ up to a few TeV, while the Higgs portal dominates glueball decays with lifetimes shown in \cref{section:maps}. The vector portal glueball decays are phase space suppressed, since they induce four body decays compared to the two body decays that are induced by the Higgs portal.  Additionally, the vector portal decays have lower rates due to the higher dimension of the corresponding effective operator.

The existence of the Higgs portal accommodates the heavy quarks being vector-like doublets under SM $SU(2)_L\times U(1)_Y$, but one can also consider the same effective operator in an FTH-like scenario where the dark quarks are SM singlets that couple to a scalar that mixes with the SM Higgs boson, resulting in an analogous $M_Q/y$ that sets the glueball lifetimes in combination with $m_0$. Whether the UV model has $SU(2)_L$ doublet quarks that do not get all of their mass from SM electroweak symmetry breaking, or FTH-like quarks whose $M_Q/y$ is not fixed by neutral naturalness considerations, we will simply vary $M_Q \simeq M_{Z'}/2$ and $y$ independently for the purpose of studying the semivisible jet signal at the LHC. 

\subsubsection{Dark Glueball Semivisible/Emerging Jets}

\begin{figure}[t!]
\centering

\includegraphics[width=\textwidth]{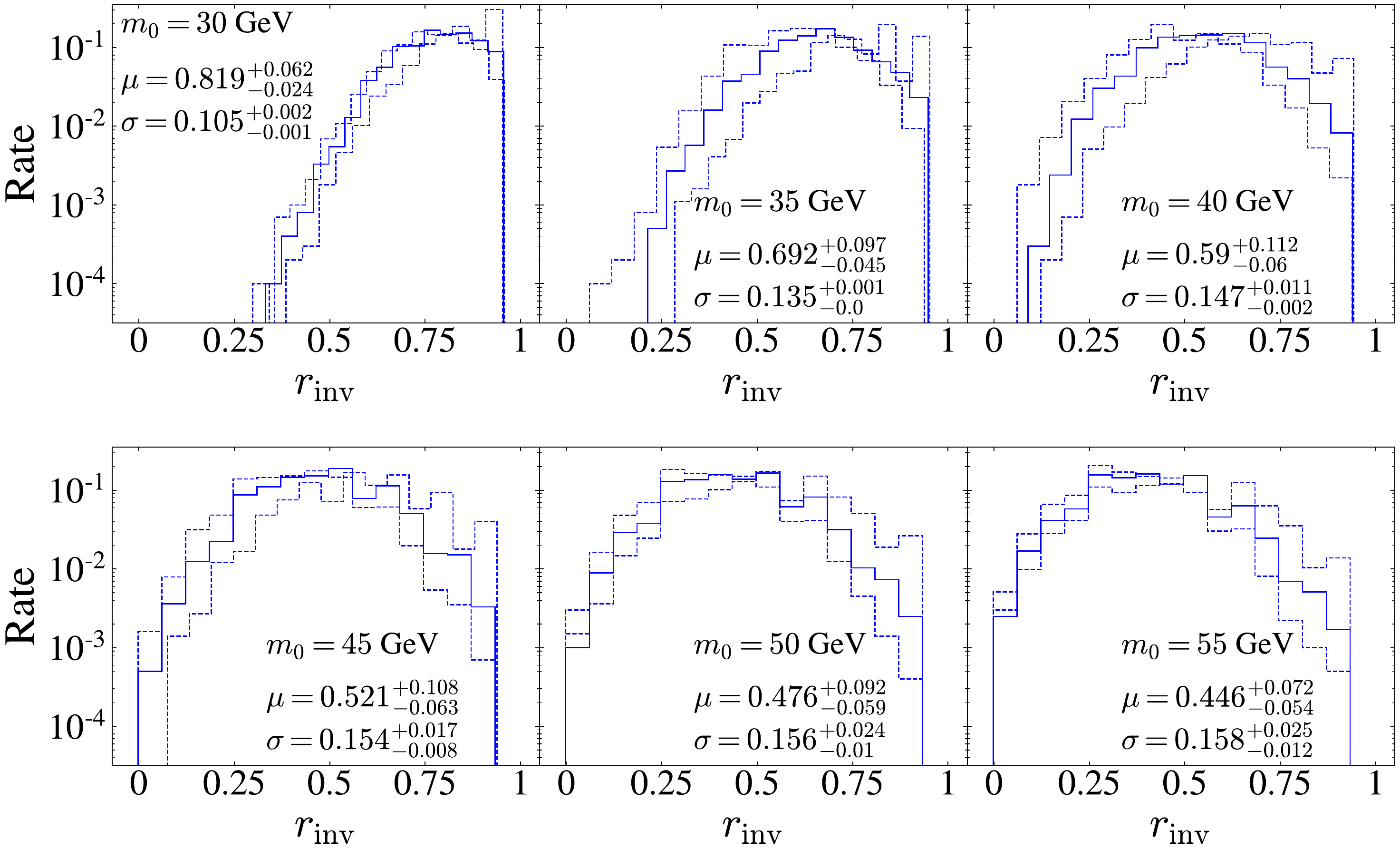}

\caption{Distributions of $r_{\text{inv}}$ for various values of the lightest glueball mass $m_0$ in the $Z'$ production model with $m_{Z^\prime} = 3$~TeV and $M_Q \sim M_{Z^\prime}/2$, where $r_{\text{inv}}$  is the fraction of dark hadrons that are invisible to the semivisible jet reconstruction. Solid histograms come from using the default hadronization benchmark, and dashed histograms come from the soft and hard variations. Means $\mu$ and standard deviations $\sigma$ are displayed, with uncertainties corresponding to hadronization variations.\label{figure:rinv}}
\end{figure}

For this study, we assume the model described in \cref{section:Zprime}.  We take a benchmark where the $Z^\prime$ has mass $M_{Z^\prime} = 3\,\text{TeV}$, and the dark quarks have $M = M_Q \sim M_{Z^\prime}/2$. We choose $M / y = 4.5\,\text{TeV}$, which fixes the dimension-6 glueball lifetimes for any choice of $m_0$. To generate events, we use \textsc{MadGraph5}\_a\textsc{mc@nlo} \cite{Alwall:2014hca} with a $Z^\prime$ model \cite{Altarelli:1989ff,Fuks:2017vtl} to simulate $pp\rightarrow Z^\prime$ production at a $14\,\text{TeV}$ proton collider. 
We run our dark shower and glueball hadronization algorithm as though the $Z^\prime$ were a heavy scalar decaying to two dark gluons, which models $s$-wave quirk annihilation as discussed in \cref{section:Zprime}.\footnote{Similar signals could result if $M_Q > m_{Z^\prime}/2$ and the $Z'$ instead decays via a loop to three gluons. } 
We take events with $Z^\prime$ production and decay to glueballs as the hard process and pass them to \textsc{Pythia} version 8.307~\cite{Bierlich:2022pfr} to handle SM QCD initial- and final-state radiation and jet clustering with the FastJet version 3.4.0 plugin \cite{Cacciari:2011ma}.
Following the procedure from the CMS search in \cite{CMS:2021dzg}, we use the anti-$k_T$ jet clustering algorithm \cite{Cacciari:2008gp} with jet radius $R=0.8$.  SM final states in each event are characterized as ``invisible" if they are neutrinos or they have a glueball ancestor that decayed outside of the cylinder spanned by the CMS tracker~\cite{CMS:1997ema}. 
We consider all other SM final states as ``visible" and cluster them into jets. This is a highly simplified picture of the detector and semivisible jet reconstruction, but we performed the same analysis using the central hadronic calorimeter as the border between visible and invisible states and found qualitatively similar results. The invisible SM final states, as well as any stable glueballs, contribute to the missing transverse momentum $\vec{p}_T^{\,\,\text{miss}}$.

\begin{figure}[t!]
\centering

\includegraphics[width=\textwidth]{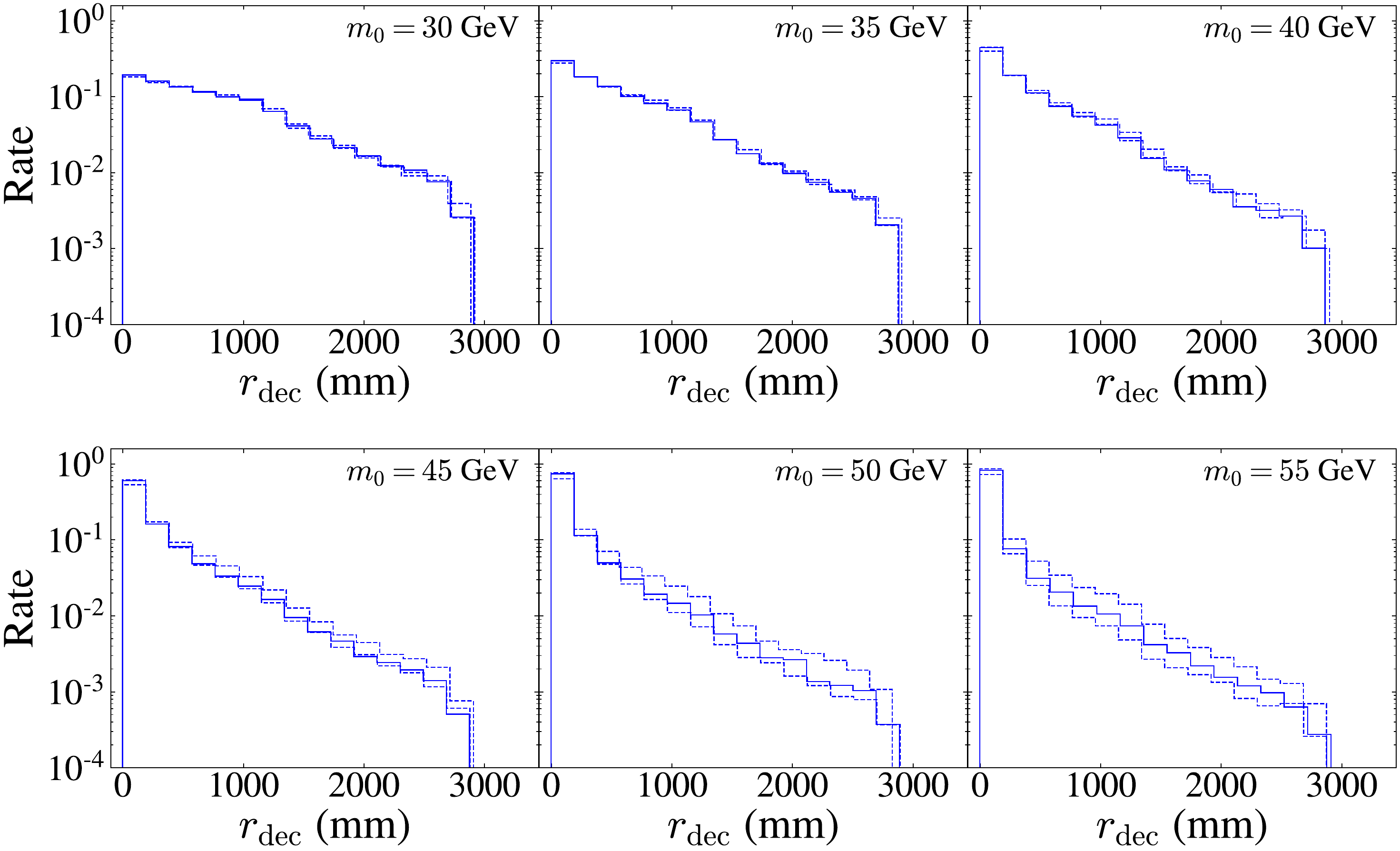}

\caption{Distributions of $r_\text{dec}$ for various values of the lightest glueball mass $m_0$ in the $Z'$ production model with $m_{Z^\prime} = 3\,\text{TeV}$ and $M_Q \sim M_{Z^\prime}/2$, where $r_\text{dec}$  is the distance of glueball decay vertices within the CMS tracker to the IP. Solid histograms come from using the default hadronization benchmark, and dashed histograms come from the soft and hard variations. \label{figure:rdec}}
\end{figure}

Most of the glueball species decay by emitting a lighter glueball, resulting in cascade decays. However, only the primary glueballs (i.e.~those produced from hadronization) contribute to $r_\text{inv}$. 
In order to understand the role of displaced vertices, we tracked the distances $r_\text{dec}$ between the interaction point and the decay vertices of glueballs that decayed within the tracker. We also computed a few useful observables for semivisible jet searches and describe them further \cref{appenidx:jetVars}. The distributions of these other observables have the expected qualitative form, so here we focus on the novel results of $r_{\text{inv}}$ and $r_\text{dec}$ shown in \cref{figure:rinv} and \cref{figure:rdec}, respectively.  

We see that the average value of $r_{\text{inv}}$ ranges from $\sim 0.45$ to $\sim 0.82$. This spread of average $r_{\text{inv}}$ demonstrates that a constraint on $r_\text{inv}$ from a semivisible jet search could potentially be recast as a constraint on our model's microscopic parameters.  It would also be interesting to investigate the extent to which the shape of the $r_{\text{inv}}$ distribution predicted here  would impact limits set by existing analyses. As $m_0$ increases, the mean $r_{\text{inv}}$ approaches $\sim 0.45$, and we expect from \cref{figure:defaultfit} that about half of the glueballs produced are $0^{++}$. Therefore, this behavior of $r_{\text{inv}}$ shows that as $m_0$ increases through this range, all the $0^{++}$ glueballs decay within the tracker while all heavier species tend to escape. The $r_\text{dec}$ plots emphasize the importance of displaced vertices in this analysis. There are two ways for a dark sector to generate missing momentum aligned with a jet: the jet contains states with long lifetimes compared to the detector scale as well as states that decay promptly, or the jet contains states with lifetimes comparable to the detector scale, allowing a portion of them to decay in the detector and leave displaced vertices. The former case is the prototypical semivisible jets scenario, while the latter is closer to an emerging jets signal. With our dimension-6 glueball decays through a Higgs portal, these two cases overlap. Depending on the region in parameter space, the $0^{++}$ may decay promptly or have cm to m lifetimes while the heavier states are relatively long-lived. Alternatively, the $0^{++}$ may decay promptly while only a subset of the heavier species leave displaced vertices.  We leave the interesting task of developing an optimal search strategy that takes advantage of this class of signals to future work.

\section{Conclusions}
\label{section:conclusion}

Confining dark sectors appear as a component of a large class of possible BSM scenarios with a wide range of possible signatures and very broad theoretical motivation.
The case of a pure Yang-Mills dark sector, corresponding to $N_f = 0$ QCD, is an important representative of this class, but one whose study has been hampered by our ignorance of pure-glue hadronization into dark glueballs. 
While first steps to constrain the possible range of hadronization outcomes based on the perturbative gluon shower, energy conservation, and the large $m_0/\LD$ mass gap were taken in~\cite{Curtin:2022tou}, the absence of a realistic hadronization model left large theoretical uncertainties, especially for exclusive production rates of individual glueball species, which determine most collider observables.
This motivates developing a more sophisticated phenomenological parameterization of the non-perturbative physics, which allows us to make more quantitative predictions for the final state from dark glue sector showers.

In this work, we present the first implementation of a color string dynamics inspired hadronization model for dark glueball fragmentation. 
We borrow from the Lund string model to parameterize how color-singlet flux tubes produced by the dark gluon shower fragment into dark glueballs. 
The specific algorithm is quite simple, and local in the sense that it assumes each combination of fragmenting string pieces chooses democratically from all kinematically available glueball species. This makes one of our main results all the more remarkable: 
the relative multiplicities of different glueball species approximately follows the theoretically expected thermal distribution with $T_\mathrm{had} \simeq T_c \simeq \LD$, 
independent of the chosen fragmentation function and with a weak dependence on the shower cutoff scale. 
Also, our algorithm's intuitive geometric picture of self-intersecting color flux rings suggests our approach may capture the relevant dynamics of glueball fragmentation.

Our shower and hadronization model has a handful of parameters beyond the physical glueball mass and initial center of mass energy: the shower cutoff scale, the choice of fragmentation function, and that function's two parameters. 
We suggest three sets of benchmarks that span a broad space of our model's physically reasonable predictions for different possible values of these nuisance parameters, see \cref{table:benchmarks}. 

We apply our new simulation framework to explore the potential reach of emerging and semivisble jet searches, finding both search strategies would probe distinct but overlapping regions of dark glueball parameter space. We then focus on two important phenomenological demonstrations: the study of glueball decays in the proposed MATHUSLA detector and a realistic benchmark model for semivisible/emerging jets.

For dark glueballs decaying in MATHUSLA, we focus on the simplified glueball parameter space motivated by theories of neutral naturalness like the Fraternal Twin Higgs or Folded Supersymmetry, where the dark QCD sector is coupled to the SM via the Higgs portal.
We find that including the multiplicity-enhancing effects of the shower and the realistic full spectrum of glueball species produced from our hadronization model dramatically increase the projected sensitivity of MATHUSLA
compared to earlier simplified estimates, see \cref{figure:MATH}.

We also considered a simplified model of a $Z'$ coupled to a dark QCD in the quirk-like regime that lead to the production of dark glueballs via resonant $Z'$ production.  This model has broad and potentially overlapping regions of parameter space that can possibly have  emerging jet and semivisible jet signatures. 
The main result we obtained using our simulation of the pure-glue shower and hadronization in the $Z^\prime$ model was finding the distributions of $r_{\text{inv}}$ that arise due to the pure-glue theory's multiplicity of glueball states with potentially widely separated lifetimes, see \cref{figure:rinv}. This demonstrates how a  semivisible jet search can yield realistic constraints for pure-glue dark sectors.

We hope that some version of our approach can be incorporated into the \textsc{Pythia} Hidden Valley Module. 
For future studies, we suggest further lattice calculations of glueball decay matrix elements, which will reduce systematic uncertainties in glueball lifetimes relevant to the long-lived particle regime. 
Our hadronization model could also be improved or generalized in various ways, such as accounting for parity and charge selection rules, and generalizing the number of dark colors beyond 3 (though fully characterizing the glueball mass spectrum for $N_c\neq3$ would again require additional lattice studies). 
A more sophisticated implementation might explicitly simulate a closed oscillating classical string in position space.
This would improve the IR/collinear unsafety 
inherent in our approach, which relies on discretizing rings into string pieces in a way that explicitly depends on the number of gluon splittings during the shower.

A reliable glueball Monte Carlo also enables new cosmological studies of these dark sectors. 
Previous analyses derived constraints from avoiding overproduction of surviving stable glueball states~\cite{Faraggi:2000pv,Feng:2011ik,Boddy:2014yra,Forestell:2016qhc,Acharya:2017szw,Soni:2017nlm} or late-time decays modifying big-bang nucleosynthesis and the cosmic microwave background~\cite{Forestell:2017wov}. These analyses assumed that the glueball relic densities originated in the dark QCD phase transition, which itself is not well understood. However, it is possible for glueball densities to receive important contributions from late-time decays, annihilations, or other entropy injections. For these scenarios, our shower and hadronization model could supply new predictions. The same can be said for models where glueball production plays an important role in the production of cosmic rays~\cite{Winkler:2022zdu, Curtin:2022oec}.

The most immediate application of our work will be to enable many new detailed collider studies of $N_f = 0$ QCD dark sectors. For example, including factors such as detector effects is required in order to quantitatively establish the distinction between the emerging and semivisible jet regimes, as well as the realistic sensitivity of the ATLAS and CMS main detectors to dark glueball decays.
This work also provides us with a tool we can use to further develop search strategies that are insensitive to dark hadronization uncertainties~\cite{Cohen:2023mya}, perhaps relying on some aspects of jet substructure and that could even incorporate machine learning, see e.g.~\cite{Bernreuther:2020vhm, Buss:2022lxw, Barron:2021btf, Canelli:2021aps, Dillon:2022mkq, Faucett:2022zie,  Lu:2023gjk, Bardhan:2023mia, Anzalone:2023ugq, Pedro:2023sdp}.
In general, understanding the detailed phenomenology of these dark sectors will help us design the searches that could lead to the next discovery beyond the Standard Model.

\acknowledgments
We thank Tom Bouley for many insightful conversations.
The research of AB, TC, and GDK is supported by the U.S.~Department of Energy under grant number DE-SC0011640.
The research of DC and CG was supported in
part by Discovery Grants from the Natural Sciences
and Engineering Research Council of Canada and the
Canada Research Chair program. The research of DC
was also supported by the Alfred P. Sloan Foundation,
the Ontario Early Researcher Award, and the University of Toronto McLean Award. The work of CG was
also supported by the University of Toronto Connaught
International Scholarship.

\clearpage

\appendix

\section*{Appendices}
\section{Glueball Hadronization in Detail}
\label{appendix:algorithm}

Here, we describe how we implement our algorithm by modifying the public version of
\textsc{Pythia} version 8.307 
to simulate a perturbative gluon shower, group the final gluons into color singlets, and produce glueballs. Operationally, we collide $e^+e^-$ beams set to produce only a single at-rest Higgs boson of arbitrary mass, which is decayed to two gluons. 
Using Pythia's standard showering functionality, the gluons undergo a pure-glue $p_T$-ordered shower with the running of the strong coupling
given by the 3-loop $N_f=0$ beta function presented in \cite{Prosperi:2006hx}.
The shower is cut off once the $p_T$ scale reaches $p_{T\text{min}} = c \,\LD$, with $c$ being a dimensionless nuisance parameter. Then, we execute color reconnection as described in \cref{section:lund} with $m_\text{ref} = m_0$ in \cref{equation:stringlength} (though the exact value of $m_\mathrm{ref}$ has a negligible effect on the outcome), but color connection swaps are only allowed if the invariant mass of the resulting color-singlet rings of Lund strings is at least $2\s m_0$. This guarantees each ring is allowed to decay to glueballs with no additional momentum transfer. 

So far, the implementation is very similar to how the standard Lund string model would handle a pure-gluon shower that is about to hadronize. In SM QCD, or Hidden Valley QCD with light dark quark flavors, \textsc{Pythia} would determine how the strings break (including breaking any closed loops of strings) and group the remaining strings into yo-yo modes with two ends for mesons and junction topologies with three ends for baryons \cite{Bierlich:2022pfr,Andersson:1983ia}. In the absence of quarks, our Lund strings are incapable of breaking in this way. Instead, we imagine that our rings of Lund strings iteratively pinch off smaller rings as in color reconnection, but now the smaller rings must be on-shell glueballs selected from the twelve possible species. 

We now describe our algorithm for taking rings of Lund string pieces and having them radiate glueballs.   We must decide on the glueball's species, its momentum, which pieces of the ring get converted into the glueball, and how to distribute momentum throughout the ring that remains after the glueball's emission.  Our choices follow the following principles:

\begin{enumerate}
    \item Enforce strict conservation of energy and momentum at every step.
    \item Maintain Lorentz invariance wherever possible.
    \item Borrow techniques from the Lund String model wherever they apply.
\end{enumerate}

We will first describe how we decide which pieces of the ring get converted into a glueball. Each ring is made of string pieces that are color-connected to two nearest-neighbors. These connections serve as our analog to breakup vertices in the standard procedure. Again for the purpose of minimizing free energy, the color-connection vertex that attaches the two string pieces with the largest string-length $\lambda$ is the seed for glueball emission.\footnote{As discussed in \cref{section:fits}, we can interpret this as a color-string polygon most rapidly decreasing its perimeter. We could instead choose a random vertex to begin fragmentation with minimal impact on the final results, so this choice can be taken as arbitrary.} 
The two string pieces connected by this vertex are (at least) the pieces that will be converted or ``fragmented" into a glueball. If the two pieces have an invariant mass less than $m_0$, then one of their nearest neighbors (selected arbitrarily) will also be added to the list of fragmenting pieces. If this is still not enough invariant mass, the nearest neighbor in the other direction (still in color-connection space) will fragment as well, and so on until we have a selection of string pieces with invariant mass at least $m_0$.

Next, we select the glueball's species. The invariant mass of the fragmenting string pieces is an upper bound on the mass $m_G$ of the selected species. We perform a weighted random selection from the kinematically available species, where the weight $2J+1$ accounts for the higher multiplicity of a state with spin $J$.  A similar approach is used  for SM meson species selection in \textsc{Pythia}. 
For example, a $\text{u}\bar{\text{d}}$ mode could be either a pion or a $\rho$ meson.  The choice is determined in \textsc{Pythia} by incorporating a spin-dependent weight into the random selection. As a correction to better fit the data, \textsc{Pythia} additionally includes suppression of heavier species beyond the na\"ive 3:1 expectation for the vector-to-scalar weight.\footnote{One could add additional mass suppression to our glueball hadronization model at the expense of more nuisance parameters, but the modest mass splitting of the glueball species suggests this may not be necessary. Our code includes options for selecting a species non-democratically, which could be further explored in future works.} As an alternative approach, one could plausibly select the species before the fragmenting string pieces and gather string pieces with invariant mass at least $m_G$. However, that approach causes the species distribution to acquire a strong dependence on the ratio of the invariant mass of the shower, which is distinctly different behavior from what we expect in QCD as discussed in \cref{section:fits}. 

After determining the glueball species, we must specify its momentum. First, we choose the direction of the three momentum $\hat{p}_G$ to be along the momentum of the fragmenting string pieces in the ring's rest frame. This decision is admittedly not Lorentz covariant, but this $\hat{p}_G$ is the best indication of a preferred direction of the fragmenting system. If all of the string pieces in the ring are fragmenting, then $\hat{p}_G$ is selected randomly and isotropically in the ring's rest frame. Then, again inspired by the Lund string model, the glueball takes a fraction $z$ of the fragmenting pieces' light-cone momentum $p_{\pm\text{pieces}} = E_{\text{pieces}} \pm \left| \vec{p}_{\text{pieces}} \right|$, so that the glueball's light-cone momentum is

\begin{equation}
p_{\pm G} = z \, p_{\pm\text{pieces}}\,,
\end{equation}

\noindent which is a relation that is invariant under boosts along $\hat{p}_G$. With the glueball's direction and light-cone momentum fixed, and imposing the on-shell condition, the glueball's four momentum is fully specified. 

The momentum fraction $z$ follows a probability distribution called the fragmentation function. \textsc{Pythia} uses the LSFF in \cref{equation:LSFF}. This fragmentation function was derived by enforcing consistency between whether the Lund string starts fragmenting from one end or the other (this is what is meant by ``symmetric" in the fragmentation function's name). However, our Lund strings lack endpoints, so it is unclear that this fragmentation function is a valid assumption. For this reason, we also consider a beta distribution with the particular parametrization shown in \cref{equation:fbeta}. We chose this parametrization because it has a qualitatively similar shape to the LSFF which deforms similarly as one varies the hadron mass. With these two functions, we can parametrize a broad space of plausible unimodal probability distributions for describing the fraction of momentum the glueball takes away from the fragmenting string pieces.

Once the glueball's momentum $p_G^\mu$ is fixed, there is some leftover recoiling momentum

\begin{equation}
p_{\text{rec}}^\mu = p_{\text{pieces}}^\mu - p_G^\mu \,.
\end{equation}
In sampling the fragmentation function, we impose $z > z_{\text{min}} = m_G^2/m_{\text{pieces}}^2$ to avoid negative invariant mass-squared $p_\text{rec}^2$.
The remaining decision required to execute the algorithm is how to distribute $p_{\text{rec}}^\mu$ among the rest of the ring of Lund strings before it radiates the next glueball. We simply choose to make two string pieces, each with half of $p_{\text{rec}}^\mu$, to replace the string pieces that fragmented. This way, whether a ring has many string pieces or only two, it can radiate several glueballs according to the above algorithm until it runs out of invariant mass.

The stopping condition for the above iterative glueball radiation algorithm is when the invariant mass of the leftover ring is less than $2\s m_0$, so fragmenting it into two on-shell glueballs would be kinematically forbidden. When this happens, we must fragment the ring into two glueballs rather than a glueball plus a ring. We keep the species and direction the original glueball would have had, but then select a species for the second glueball using the same method of sampling from the kinematically accessible options weighted by spin multiplicity. Then, the magnitude of the glueballs' three momentum is fixed by momentum conservation.  This leads to a final state where all string segments have been converted into glueballs.

\clearpage

\begin{figure}[b!]
\centering
\includegraphics[width=0.49\textwidth]{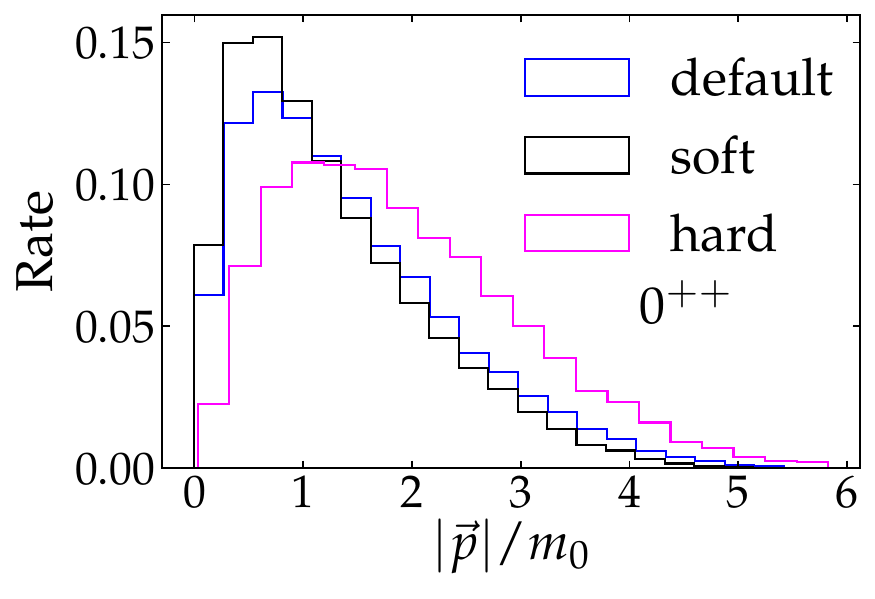}
\includegraphics[width=0.49\textwidth]{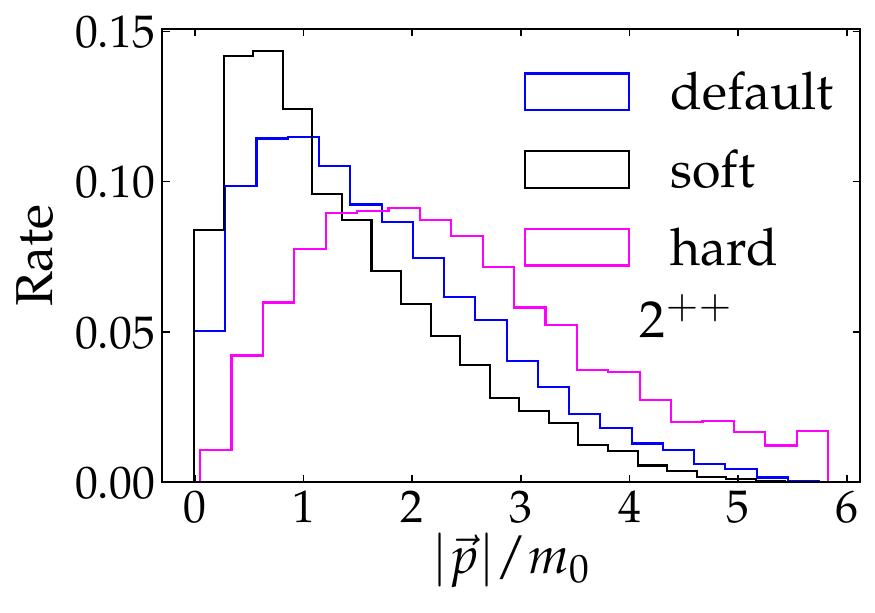}
\includegraphics[width=0.49\textwidth]{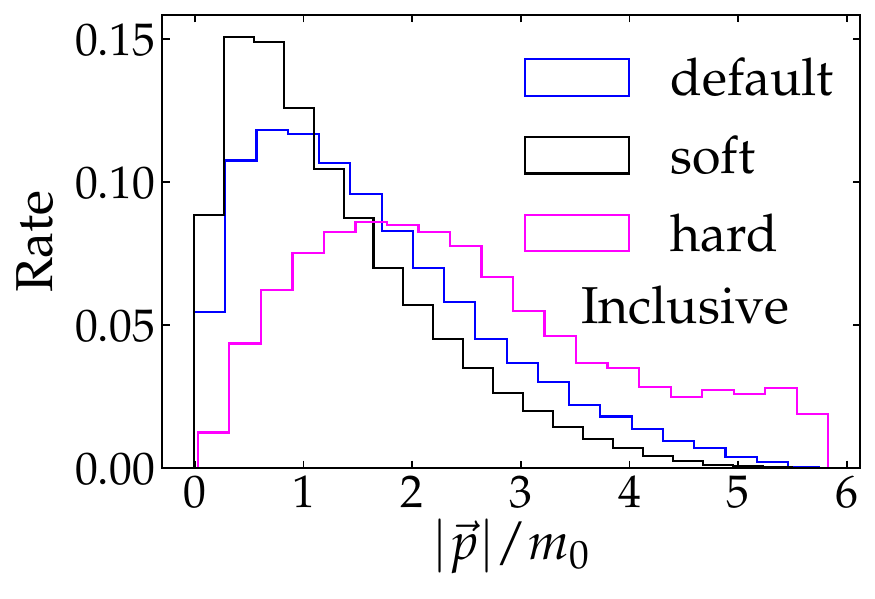}
\caption{Distributions of $|\vec{p}\,|/m_0$ for the three sets of benchmark parameters listed in \cref{table:benchmarks}, measured in the rest frame of the dark gluon shower. Exclusive distributions of the two lightest species are shown, as well as the inclusive distribution. As expected, glueballs from ``harder" parameter variations tend to have larger momentum. \label{figure:pdists}}
\end{figure}

\section{Benchmark Parameter Variations}
\label{appendix:benchmarks}

Here, we show variations in the species production rates and glueball momentum distributions resulting from our different benchmark parameters. The momentum distributions in \cref{figure:pdists} provide an intuitive demonstration of what we mean by glueball ``hardness." As discussed in \cref{section:benchmarks}, there is a straightforward relation between our hadronization algorithm's nuisance parameters and the glueballs' tendency to be produced with smaller or larger momentum. In this sense, the soft and hard variations of our suggested benchmarks are meant to provide the extremes of our algorithm's possible sensible outputs.

\begin{figure}[t!]
\centering
\includegraphics[width=0.8\textwidth]{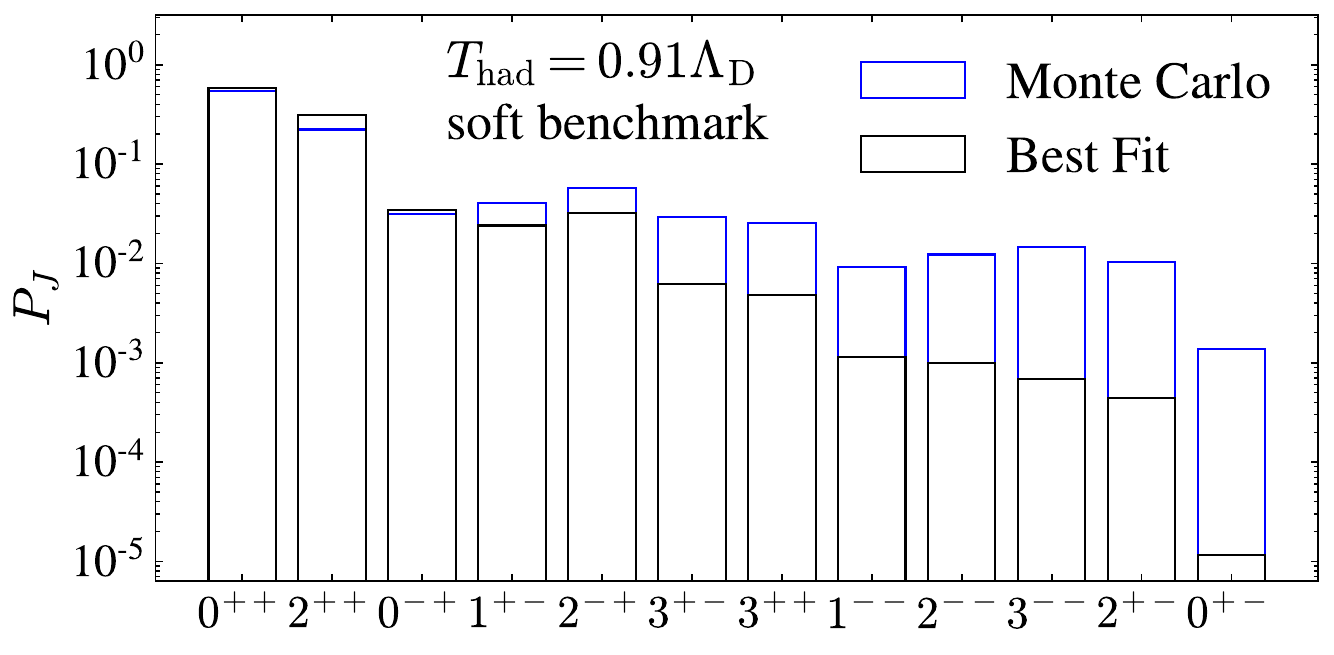}
\includegraphics[width=0.8\textwidth]{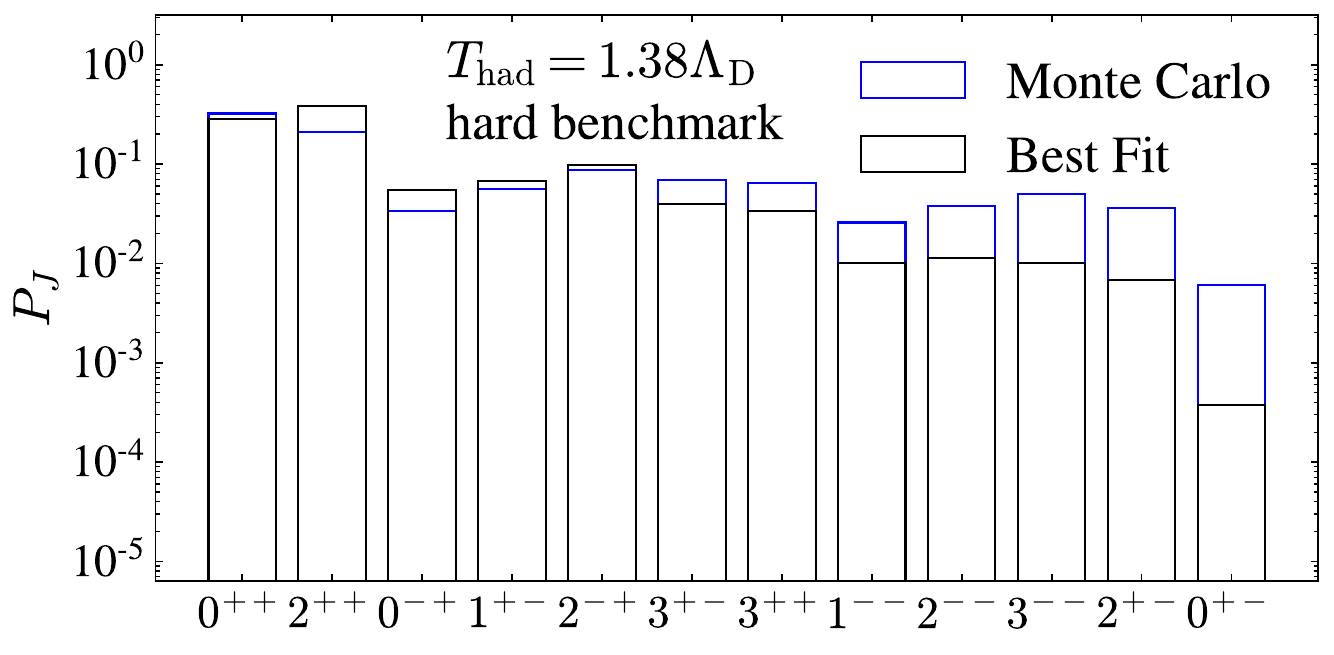}
\caption{Relative species production rates for non-default benchmark parameters. The benchmark with the largest discrepancy between the Monte Carlo output and the fit to the thermal distribution in \cref{equation:thermal} is the hard benchmark, with $R^2=0.60$. For comparison, the default and soft benchmarks yield an $R^2$ of 0.88 and 0.96, respectively. The analogous plot for the default benchmark is in \cref{figure:defaultfit}.\label{figure:otherPJ}}
\end{figure}

As seen in \cref{figure:otherPJ}, as the shower cutoff scale (parametrized by $c$) increases, the best-fit $T_\text{had}$ also increases. This behavior is as expected because a higher shower cutoff scale means the dark gluons have fewer opportunities to branch, leading to higher-mass string pieces during fragmentation and weaker suppression of heavy species production. As in \cref{figure:defaultfit,figure:fits}, \cref{figure:otherPJ,figure:pdists} were generated with $m_0=10\,\text{GeV}$ and dark shower center of mass energy of $125\,\text{GeV}$.

\clearpage

\section{Additional Semivisible Jet Distributions}
\label{appenidx:jetVars}

\begin{figure}[t!]
\centering

\includegraphics[width=\textwidth]{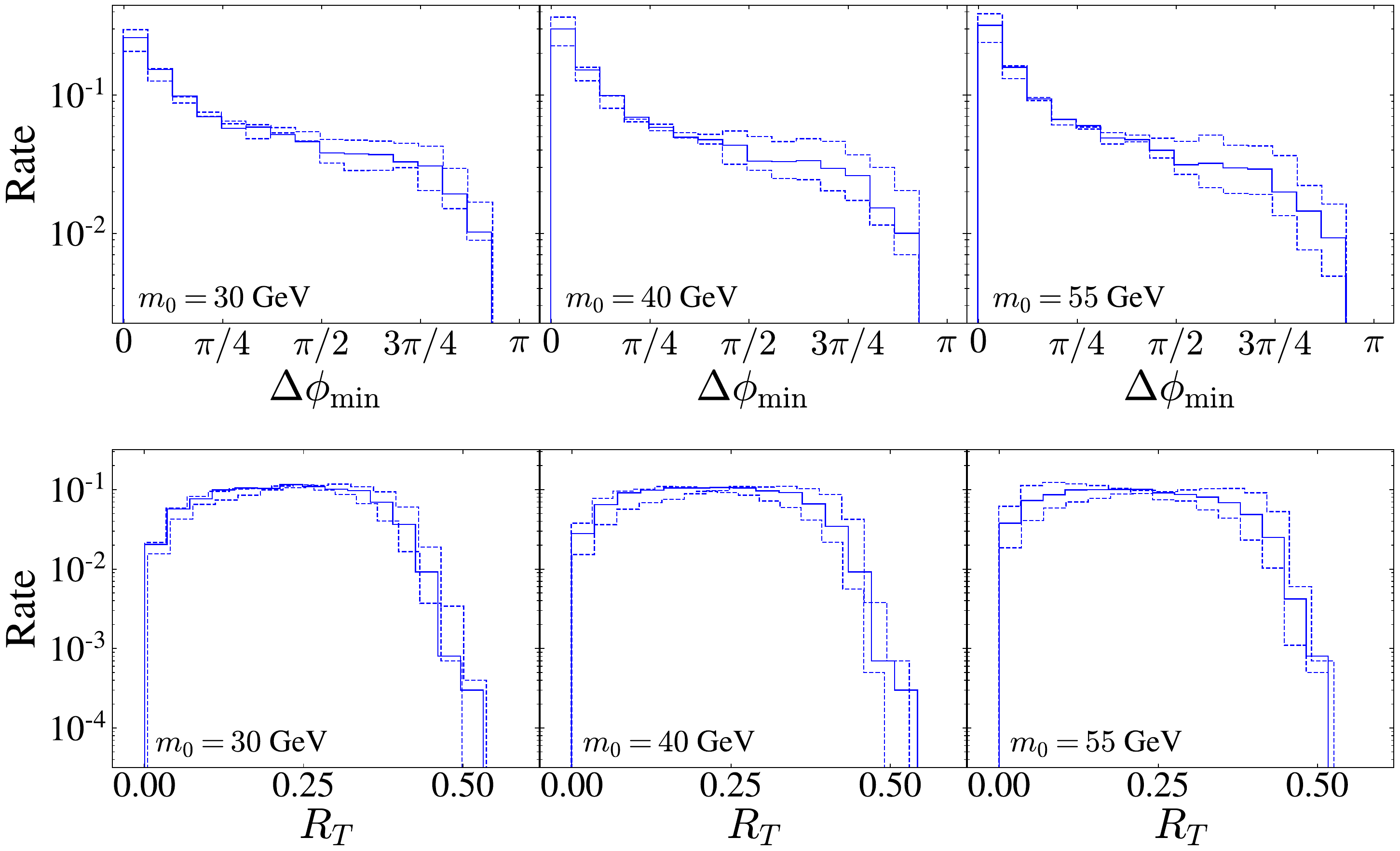}

\caption{Distributions of $\Delta \phi_{\text{min}}$ and $R_T$ for various values of the lightest glueball mass $m_0$ for the semivisible jets scenario. Solid histograms come from using the default hadronization benchmark, and dashed histograms come from the soft and hard variations. \label{figure:phiRT}}
\end{figure}

\begin{figure}[b!]
\centering

\includegraphics[width=\textwidth]{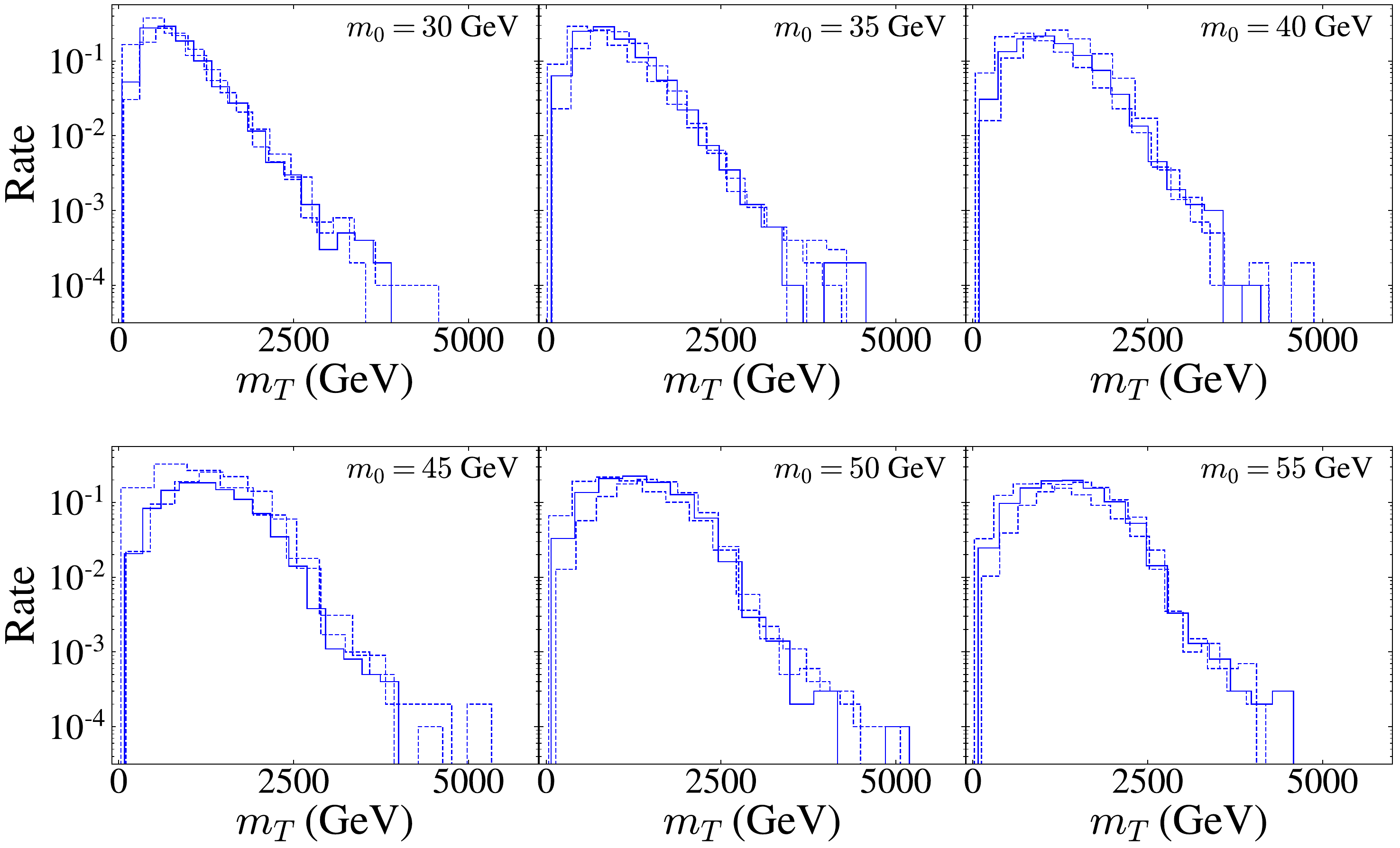}

\caption{Distributions of $m_T$ for various values of the lightest glueball mass $m_0$ for the semivisible jets scenario. Solid histograms come from using the default hadronization benchmark, and dashed histograms come from the soft and hard variations. \label{figure:mT}}
\end{figure}

In addition to $r_{\text{inv}}$ and $r_\text{dec}$, we computed three observables that \cite{CMS:2021dzg} used in their search. The transverse mass $m_T$ is given by

\begin{equation}
\label{equation:mT} m_T^2 = m_{JJ}^2 + 2 |\vec{p}_{T}^{\,\,\text{miss}}|\left( \sqrt{m_{JJ}^2 + |\vec{p}_{T,JJ}|^2} - |\vec{p}_{T,JJ}|\cos (\phi_{JJ}^{\text{miss}})\right),
\end{equation}

\noindent where the two highest-$p_T$ jets have total momentum $p_{JJ}$ with corresponding invariant mass $m_{JJ}$, and $\phi_{JJ}^{\text{miss}}$ is the azimuthal angle between $\vec{p}_{T,JJ}$ and $\vec{p}_T^{\,\,\text{miss}}$. The other two observables are $R_T = |\vec{p}_{T}^{\,\,\text{miss}}|/m_T$ and the minimum azimuthal angle $\Delta \phi_{\text{min}}$ between $\vec{p}_T^{\,\,\text{miss}}$ and the two highest-$p_T$ jets. The $m_T$ distribution is essentially cutoff by the mass of the $Z^\prime$, and $R_T$ and $\Delta \phi_{\text{min}}$ can be used to help cut out background. We found that the $\Delta \phi_{\text{min}}$ and $R_T$ distributions were fairly consistent across the model parameters we simulated, so we show a few representative examples in \cref{figure:phiRT}. The $m_T$ distribution changed more noticeably, so we show more model parameter variations in \cref{figure:mT}. Since $R_T$ depends on $m_T$ but has significantly weaker dependence on the model, there must be a compensating change in $|\vec{p}_{T}^{\,\,\text{miss}}|$ as $m_T$ changes. All of these plots were generated with $M/y=4.5\,\text{TeV}$.

\clearpage

 \addcontentsline{toc}{section}{\protect\numberline{}References}%

\bibliographystyle{JHEP}
\bibliography{biblio.bib}

\providecommand{\href}[2]{#2}\begingroup\raggedright\begin{thebibliography}{100}

\bibitem{Strassler:2006im}
M.J.~Strassler and K.M.~Zurek, \emph{{Echoes of a hidden valley at hadron
  colliders}},
  \href{https://doi.org/10.1016/j.physletb.2007.06.055}{\emph{Phys. Lett. B}
  {\bfseries 651} (2007) 374}
  [\href{https://arxiv.org/abs/hep-ph/0604261}{{\ttfamily hep-ph/0604261}}].

\bibitem{Strassler:2006ri}
M.J.~Strassler and K.M.~Zurek, \emph{{Discovering the Higgs through
  highly-displaced vertices}},
  \href{https://doi.org/10.1016/j.physletb.2008.02.008}{\emph{Phys. Lett. B}
  {\bfseries 661} (2008) 263}
  [\href{https://arxiv.org/abs/hep-ph/0605193}{{\ttfamily hep-ph/0605193}}].

\bibitem{Strassler:2006qa}
M.J.~Strassler, \emph{{Possible effects of a hidden valley on supersymmetric
  phenomenology}},  \href{https://arxiv.org/abs/hep-ph/0607160}{{\ttfamily
  hep-ph/0607160}}.

\bibitem{Han:2007ae}
T.~Han, Z.~Si, K.M.~Zurek and M.J.~Strassler, \emph{{Phenomenology of hidden
  valleys at hadron colliders}},
  \href{https://doi.org/10.1088/1126-6708/2008/07/008}{\emph{JHEP} {\bfseries
  07} (2008) 008} [\href{https://arxiv.org/abs/0712.2041}{{\ttfamily
  0712.2041}}].

\bibitem{Strassler:2008bv}
M.J.~Strassler, \emph{{Why Unparticle Models with Mass Gaps are Examples of
  Hidden Valleys}},  \href{https://arxiv.org/abs/0801.0629}{{\ttfamily
  0801.0629}}.

\bibitem{Strassler:2008fv}
M.J.~Strassler, \emph{{On the Phenomenology of Hidden Valleys with Heavy
  Flavor}},  \href{https://arxiv.org/abs/0806.2385}{{\ttfamily 0806.2385}}.

\bibitem{Luo:2009kf}
M.~Luo and S.~Zheng, \emph{{Gauge Extensions of Supersymmetric Models and
  Hidden Valleys}},
  \href{https://doi.org/10.1088/1126-6708/2009/04/122}{\emph{JHEP} {\bfseries
  04} (2009) 122} [\href{https://arxiv.org/abs/0901.2613}{{\ttfamily
  0901.2613}}].

\bibitem{Cvetic:2012kj}
M.~Cvetic, J.~Halverson and H.~Piragua, \emph{{Stringy Hidden Valleys}},
  \href{https://doi.org/10.1007/JHEP02(2013)005}{\emph{JHEP} {\bfseries 02}
  (2013) 005} [\href{https://arxiv.org/abs/1210.5245}{{\ttfamily 1210.5245}}].

\bibitem{Patt:2006fw}
B.~Patt and F.~Wilczek, \emph{{Higgs-field portal into hidden sectors}},
  \href{https://arxiv.org/abs/hep-ph/0605188}{{\ttfamily hep-ph/0605188}}.

\bibitem{March-Russell:2008lng}
J.~March-Russell, S.M.~West, D.~Cumberbatch and D.~Hooper, \emph{{Heavy Dark
  Matter Through the Higgs Portal}},
  \href{https://doi.org/10.1088/1126-6708/2008/07/058}{\emph{JHEP} {\bfseries
  07} (2008) 058} [\href{https://arxiv.org/abs/0801.3440}{{\ttfamily
  0801.3440}}].

\bibitem{Delgado:2008rq}
A.~Delgado, J.R.~Espinosa, J.M.~No and M.~Quiros, \emph{{The Higgs as a Portal
  to Plasmon-like Unparticle Excitations}},
  \href{https://doi.org/10.1088/1126-6708/2008/04/028}{\emph{JHEP} {\bfseries
  04} (2008) 028} [\href{https://arxiv.org/abs/0802.2680}{{\ttfamily
  0802.2680}}].

\bibitem{Krolikowski:2008qa}
W.~Krolikowski, \emph{{A Hidden Valley model of cold dark matter with photonic
  portal}},  \href{https://arxiv.org/abs/0803.2977}{{\ttfamily 0803.2977}}.

\bibitem{Hambye:2008bq}
T.~Hambye, \emph{{Hidden vector dark matter}},
  \href{https://doi.org/10.1088/1126-6708/2009/01/028}{\emph{JHEP} {\bfseries
  01} (2009) 028} [\href{https://arxiv.org/abs/0811.0172}{{\ttfamily
  0811.0172}}].

\bibitem{Falkowski:2009yz}
A.~Falkowski, J.~Juknevich and J.~Shelton, \emph{{Dark Matter Through the
  Neutrino Portal}},  \href{https://arxiv.org/abs/0908.1790}{{\ttfamily
  0908.1790}}.

\bibitem{Cohen:2015toa}
T.~Cohen, M.~Lisanti and H.K.~Lou, \emph{{Semivisible Jets: Dark Matter
  Undercover at the LHC}},
  \href{https://doi.org/10.1103/PhysRevLett.115.171804}{\emph{Phys. Rev. Lett.}
  {\bfseries 115} (2015) 171804}
  [\href{https://arxiv.org/abs/1503.00009}{{\ttfamily 1503.00009}}].

\bibitem{Cohen:2017pzm}
T.~Cohen, M.~Lisanti, H.K.~Lou and S.~Mishra-Sharma, \emph{{LHC Searches for
  Dark Sector Showers}},
  \href{https://doi.org/10.1007/JHEP11(2017)196}{\emph{JHEP} {\bfseries 11}
  (2017) 196} [\href{https://arxiv.org/abs/1707.05326}{{\ttfamily
  1707.05326}}].

\bibitem{Beauchesne:2017yhh}
H.~Beauchesne, E.~Bertuzzo, G.~Grilli Di~Cortona and Z.~Tabrizi,
  \emph{{Collider phenomenology of Hidden Valley mediators of spin 0 or 1/2
  with semivisible jets}},
  \href{https://doi.org/10.1007/JHEP08(2018)030}{\emph{JHEP} {\bfseries 08}
  (2018) 030} [\href{https://arxiv.org/abs/1712.07160}{{\ttfamily
  1712.07160}}].

\bibitem{Bernreuther:2020vhm}
E.~Bernreuther, T.~Finke, F.~Kahlhoefer, M.~Kr\"amer and A.~M\"uck,
  \emph{{Casting a graph net to catch dark showers}},
  \href{https://doi.org/10.21468/SciPostPhys.10.2.046}{\emph{SciPost Phys.}
  {\bfseries 10} (2021) 046}
  [\href{https://arxiv.org/abs/2006.08639}{{\ttfamily 2006.08639}}].

\bibitem{CMS:2021dzg}
{\scshape CMS} collaboration, \emph{{Search for resonant production of strongly
  coupled dark matter in proton-proton collisions at 13 TeV}},
  \href{https://doi.org/10.1007/JHEP06(2022)156}{\emph{JHEP} {\bfseries 06}
  (2022) 156} [\href{https://arxiv.org/abs/2112.11125}{{\ttfamily
  2112.11125}}].

\bibitem{ATLAS:2023swa}
{\scshape ATLAS} collaboration, \emph{{Search for non-resonant production of
  semi-visible jets using Run\textasciitilde{}2 data in ATLAS}},
  \href{https://arxiv.org/abs/2305.18037}{{\ttfamily 2305.18037}}.

\bibitem{Schwaller:2015gea}
P.~Schwaller, D.~Stolarski and A.~Weiler, \emph{{Emerging Jets}},
  \href{https://doi.org/10.1007/JHEP05(2015)059}{\emph{JHEP} {\bfseries 05}
  (2015) 059} [\href{https://arxiv.org/abs/1502.05409}{{\ttfamily
  1502.05409}}].

\bibitem{CMS:2018bvr}
{\scshape CMS} collaboration, \emph{{Search for new particles decaying to a jet
  and an emerging jet}},
  \href{https://doi.org/10.1007/JHEP02(2019)179}{\emph{JHEP} {\bfseries 02}
  (2019) 179} [\href{https://arxiv.org/abs/1810.10069}{{\ttfamily
  1810.10069}}].

\bibitem{Treado:2748283}
C.J.~Treado, \emph{{A Search for Emerging Jets at $\sqrt{s}$ = 13 TeV at ATLAS
  Run 2}}, Ph.D. thesis, New York University, 2020.

\bibitem{Archer-Smith:2021ntx}
P.~Archer-Smith, D.~Linthorne and D.~Stolarski, \emph{{Emerging jets displaced
  into the future}}, \href{https://doi.org/10.1007/JHEP02(2022)027}{\emph{JHEP}
  {\bfseries 02} (2022) 027}
  [\href{https://arxiv.org/abs/2112.05690}{{\ttfamily 2112.05690}}].

\bibitem{Carrasco:2023loy}
J.~Carrasco and J.~Zurita, \emph{{Emerging jet probes of strongly interacting
  dark sectors}},  \href{https://arxiv.org/abs/2307.04847}{{\ttfamily
  2307.04847}}.

\bibitem{Harnik:2008ax}
R.~Harnik and T.~Wizansky, \emph{{Signals of New Physics in the Underlying
  Event}}, \href{https://doi.org/10.1103/PhysRevD.80.075015}{\emph{Phys. Rev.
  D} {\bfseries 80} (2009) 075015}
  [\href{https://arxiv.org/abs/0810.3948}{{\ttfamily 0810.3948}}].

\bibitem{Knapen:2016hky}
S.~Knapen, S.~Pagan~Griso, M.~Papucci and D.J.~Robinson, \emph{{Triggering Soft
  Bombs at the LHC}},
  \href{https://doi.org/10.1007/JHEP08(2017)076}{\emph{JHEP} {\bfseries 08}
  (2017) 076} [\href{https://arxiv.org/abs/1612.00850}{{\ttfamily
  1612.00850}}].

\bibitem{Barron:2021btf}
J.~Barron, D.~Curtin, G.~Kasieczka, T.~Plehn and A.~Spourdalakis,
  \emph{{Unsupervised hadronic SUEP at the LHC}},
  \href{https://doi.org/10.1007/JHEP12(2021)129}{\emph{JHEP} {\bfseries 12}
  (2021) 129} [\href{https://arxiv.org/abs/2107.12379}{{\ttfamily
  2107.12379}}].

\bibitem{Lory:2022upc}
A.M.~Lory, \emph{{Search for new physics in signatures of soft unclustered
  energy patterns with the ATLAS detector}}, Ph.D. thesis, Munich U., 2022.
\newblock 10.5282/edoc.30891.

\bibitem{Baumgart:2009tn}
M.~Baumgart, C.~Cheung, J.T.~Ruderman, L.-T.~Wang and I.~Yavin,
  \emph{{Non-Abelian Dark Sectors and Their Collider Signatures}},
  \href{https://doi.org/10.1088/1126-6708/2009/04/014}{\emph{JHEP} {\bfseries
  04} (2009) 014} [\href{https://arxiv.org/abs/0901.0283}{{\ttfamily
  0901.0283}}].

\bibitem{Cheng:2015buv}
H.-C.~Cheng, S.~Jung, E.~Salvioni and Y.~Tsai, \emph{{Exotic Quarks in Twin
  Higgs Models}}, \href{https://doi.org/10.1007/JHEP03(2016)074}{\emph{JHEP}
  {\bfseries 03} (2016) 074}
  [\href{https://arxiv.org/abs/1512.02647}{{\ttfamily 1512.02647}}].

\bibitem{Csaki:2015fba}
C.~Csaki, E.~Kuflik, S.~Lombardo and O.~Slone, \emph{{Searching for displaced
  Higgs boson decays}},
  \href{https://doi.org/10.1103/PhysRevD.92.073008}{\emph{Phys. Rev. D}
  {\bfseries 92} (2015) 073008}
  [\href{https://arxiv.org/abs/1508.01522}{{\ttfamily 1508.01522}}].

\bibitem{Park:2017rfb}
M.~Park and M.~Zhang, \emph{{Tagging a jet from a dark sector with
  Jet-substructures at colliders}},
  \href{https://doi.org/10.1103/PhysRevD.100.115009}{\emph{Phys. Rev. D}
  {\bfseries 100} (2019) 115009}
  [\href{https://arxiv.org/abs/1712.09279}{{\ttfamily 1712.09279}}].

\bibitem{Kribs:2018ilo}
G.D.~Kribs, A.~Martin, B.~Ostdiek and T.~Tong, \emph{{Dark Mesons at the LHC}},
  \href{https://doi.org/10.1007/JHEP07(2019)133}{\emph{JHEP} {\bfseries 07}
  (2019) 133} [\href{https://arxiv.org/abs/1809.10184}{{\ttfamily
  1809.10184}}].

\bibitem{Kribs:2018oad}
G.D.~Kribs, A.~Martin and T.~Tong, \emph{{Effective Theories of Dark Mesons
  with Custodial Symmetry}},
  \href{https://doi.org/10.1007/JHEP08(2019)020}{\emph{JHEP} {\bfseries 08}
  (2019) 020} [\href{https://arxiv.org/abs/1809.10183}{{\ttfamily
  1809.10183}}].

\bibitem{Costantino:2020msc}
A.~Costantino, S.~Fichet and P.~Tanedo, \emph{{Effective Field Theory in AdS:
  Continuum Regime, Soft Bombs, and IR Emergence}},
  \href{https://doi.org/10.1103/PhysRevD.102.115038}{\emph{Phys. Rev. D}
  {\bfseries 102} (2020) 115038}
  [\href{https://arxiv.org/abs/2002.12335}{{\ttfamily 2002.12335}}].

\bibitem{Cohen:2020afv}
T.~Cohen, J.~Doss and M.~Freytsis, \emph{{Jet Substructure from Dark Sector
  Showers}}, \href{https://doi.org/10.1007/JHEP09(2020)118}{\emph{JHEP}
  {\bfseries 09} (2020) 118}
  [\href{https://arxiv.org/abs/2004.00631}{{\ttfamily 2004.00631}}].

\bibitem{Knapen:2021eip}
S.~Knapen, J.~Shelton and D.~Xu, \emph{{Perturbative benchmark models for a
  dark shower search program}},
  \href{https://doi.org/10.1103/PhysRevD.103.115013}{\emph{Phys. Rev. D}
  {\bfseries 103} (2021) 115013}
  [\href{https://arxiv.org/abs/2103.01238}{{\ttfamily 2103.01238}}].

\bibitem{Albouy:2022cin}
G.~Albouy et~al., \emph{{Theory, phenomenology, and experimental avenues for
  dark showers: a Snowmass 2021 report}},
  \href{https://doi.org/10.1140/epjc/s10052-022-11048-8}{\emph{Eur. Phys. J. C}
  {\bfseries 82} (2022) 1132}
  [\href{https://arxiv.org/abs/2203.09503}{{\ttfamily 2203.09503}}].

\bibitem{Chacko:2005vw}
Z.~Chacko, Y.~Nomura, M.~Papucci and G.~Perez, \emph{{Natural little hierarchy
  from a partially goldstone twin Higgs}},
  \href{https://doi.org/10.1088/1126-6708/2006/01/126}{\emph{JHEP} {\bfseries
  01} (2006) 126} [\href{https://arxiv.org/abs/hep-ph/0510273}{{\ttfamily
  hep-ph/0510273}}].

\bibitem{Burdman:2006tz}
G.~Burdman, Z.~Chacko, H.-S.~Goh and R.~Harnik, \emph{{Folded supersymmetry and
  the LEP paradox}},
  \href{https://doi.org/10.1088/1126-6708/2007/02/009}{\emph{JHEP} {\bfseries
  02} (2007) 009} [\href{https://arxiv.org/abs/hep-ph/0609152}{{\ttfamily
  hep-ph/0609152}}].

\bibitem{Cai:2008au}
H.~Cai, H.-C.~Cheng and J.~Terning, \emph{{A Quirky Little Higgs Model}},
  \href{https://doi.org/10.1088/1126-6708/2009/05/045}{\emph{JHEP} {\bfseries
  05} (2009) 045} [\href{https://arxiv.org/abs/0812.0843}{{\ttfamily
  0812.0843}}].

\bibitem{Poland:2008ev}
D.~Poland and J.~Thaler, \emph{{The Dark Top}},
  \href{https://doi.org/10.1088/1126-6708/2008/11/083}{\emph{JHEP} {\bfseries
  11} (2008) 083} [\href{https://arxiv.org/abs/0808.1290}{{\ttfamily
  0808.1290}}].

\bibitem{Curtin:2015fna}
D.~Curtin and C.B.~Verhaaren, \emph{{Discovering Uncolored Naturalness in
  Exotic Higgs Decays}},
  \href{https://doi.org/10.1007/JHEP12(2015)072}{\emph{JHEP} {\bfseries 12}
  (2015) 072} [\href{https://arxiv.org/abs/1506.06141}{{\ttfamily
  1506.06141}}].

\bibitem{Craig:2015pha}
N.~Craig, A.~Katz, M.~Strassler and R.~Sundrum, \emph{{Naturalness in the Dark
  at the LHC}}, \href{https://doi.org/10.1007/JHEP07(2015)105}{\emph{JHEP}
  {\bfseries 07} (2015) 105}
  [\href{https://arxiv.org/abs/1501.05310}{{\ttfamily 1501.05310}}].

\bibitem{Cohen:2015gaa}
T.~Cohen, N.~Craig, H.K.~Lou and D.~Pinner, \emph{{Folded Supersymmetry with a
  Twist}}, \href{https://doi.org/10.1007/JHEP03(2016)196}{\emph{JHEP}
  {\bfseries 03} (2016) 196}
  [\href{https://arxiv.org/abs/1508.05396}{{\ttfamily 1508.05396}}].

\bibitem{Craig:2016kue}
N.~Craig, S.~Knapen, P.~Longhi and M.~Strassler, \emph{{The Vector-like Twin
  Higgs}}, \href{https://doi.org/10.1007/JHEP07(2016)002}{\emph{JHEP}
  {\bfseries 07} (2016) 002}
  [\href{https://arxiv.org/abs/1601.07181}{{\ttfamily 1601.07181}}].

\bibitem{Cohen:2018mgv}
T.~Cohen, N.~Craig, G.F.~Giudice and M.~Mccullough, \emph{{The Hyperbolic
  Higgs}}, \href{https://doi.org/10.1007/JHEP05(2018)091}{\emph{JHEP}
  {\bfseries 05} (2018) 091}
  [\href{https://arxiv.org/abs/1803.03647}{{\ttfamily 1803.03647}}].

\bibitem{Cheng:2018gvu}
H.-C.~Cheng, L.~Li, E.~Salvioni and C.B.~Verhaaren, \emph{{Singlet Scalar Top
  Partners from Accidental Supersymmetry}},
  \href{https://doi.org/10.1007/JHEP05(2018)057}{\emph{JHEP} {\bfseries 05}
  (2018) 057} [\href{https://arxiv.org/abs/1803.03651}{{\ttfamily
  1803.03651}}].

\bibitem{Hur:2007uz}
T.~Hur, D.-W.~Jung, P.~Ko and J.Y.~Lee, \emph{{Electroweak symmetry breaking
  and cold dark matter from strongly interacting hidden sector}},
  \href{https://doi.org/10.1016/j.physletb.2010.12.047}{\emph{Phys. Lett. B}
  {\bfseries 696} (2011) 262}
  [\href{https://arxiv.org/abs/0709.1218}{{\ttfamily 0709.1218}}].

\bibitem{Kribs:2009fy}
G.D.~Kribs, T.S.~Roy, J.~Terning and K.M.~Zurek, \emph{{Quirky Composite Dark
  Matter}}, \href{https://doi.org/10.1103/PhysRevD.81.095001}{\emph{Phys. Rev.
  D} {\bfseries 81} (2010) 095001}
  [\href{https://arxiv.org/abs/0909.2034}{{\ttfamily 0909.2034}}].

\bibitem{Bai:2013xga}
Y.~Bai and P.~Schwaller, \emph{{Scale of dark QCD}},
  \href{https://doi.org/10.1103/PhysRevD.89.063522}{\emph{Phys. Rev. D}
  {\bfseries 89} (2014) 063522}
  [\href{https://arxiv.org/abs/1306.4676}{{\ttfamily 1306.4676}}].

\bibitem{Appelquist:2015yfa}
T.~Appelquist et~al., \emph{{Stealth Dark Matter: Dark scalar baryons through
  the Higgs portal}},
  \href{https://doi.org/10.1103/PhysRevD.92.075030}{\emph{Phys. Rev. D}
  {\bfseries 92} (2015) 075030}
  [\href{https://arxiv.org/abs/1503.04203}{{\ttfamily 1503.04203}}].

\bibitem{Appelquist:2015zfa}
T.~Appelquist et~al., \emph{{Detecting Stealth Dark Matter Directly through
  Electromagnetic Polarizability}},
  \href{https://doi.org/10.1103/PhysRevLett.115.171803}{\emph{Phys. Rev. Lett.}
  {\bfseries 115} (2015) 171803}
  [\href{https://arxiv.org/abs/1503.04205}{{\ttfamily 1503.04205}}].

\bibitem{Antipin:2015xia}
O.~Antipin, M.~Redi, A.~Strumia and E.~Vigiani, \emph{{Accidental Composite
  Dark Matter}}, \href{https://doi.org/10.1007/JHEP07(2015)039}{\emph{JHEP}
  {\bfseries 07} (2015) 039}
  [\href{https://arxiv.org/abs/1503.08749}{{\ttfamily 1503.08749}}].

\bibitem{Freytsis:2016dgf}
M.~Freytsis, S.~Knapen, D.J.~Robinson and Y.~Tsai, \emph{{Gamma-rays from Dark
  Showers with Twin Higgs Models}},
  \href{https://doi.org/10.1007/JHEP05(2016)018}{\emph{JHEP} {\bfseries 05}
  (2016) 018} [\href{https://arxiv.org/abs/1601.07556}{{\ttfamily
  1601.07556}}].

\bibitem{Kribs:2016cew}
G.D.~Kribs and E.T.~Neil, \emph{{Review of strongly-coupled composite dark
  matter models and lattice simulations}},
  \href{https://doi.org/10.1142/S0217751X16430041}{\emph{Int. J. Mod. Phys. A}
  {\bfseries 31} (2016) 1643004}
  [\href{https://arxiv.org/abs/1604.04627}{{\ttfamily 1604.04627}}].

\bibitem{Mitridate:2017oky}
A.~Mitridate, M.~Redi, J.~Smirnov and A.~Strumia, \emph{{Dark Matter as a
  weakly coupled Dark Baryon}},
  \href{https://doi.org/10.1007/JHEP10(2017)210}{\emph{JHEP} {\bfseries 10}
  (2017) 210} [\href{https://arxiv.org/abs/1707.05380}{{\ttfamily
  1707.05380}}].

\bibitem{Beauchesne:2018myj}
H.~Beauchesne, E.~Bertuzzo and G.~Grilli Di~Cortona, \emph{{Dark matter in
  Hidden Valley models with stable and unstable light dark mesons}},
  \href{https://doi.org/10.1007/JHEP04(2019)118}{\emph{JHEP} {\bfseries 04}
  (2019) 118} [\href{https://arxiv.org/abs/1809.10152}{{\ttfamily
  1809.10152}}].

\bibitem{Francis:2018xjd}
A.~Francis, R.J.~Hudspith, R.~Lewis and S.~Tulin, \emph{{Dark Matter from
  Strong Dynamics: The Minimal Theory of Dark Baryons}},
  \href{https://doi.org/10.1007/JHEP12(2018)118}{\emph{JHEP} {\bfseries 12}
  (2018) 118} [\href{https://arxiv.org/abs/1809.09117}{{\ttfamily
  1809.09117}}].

\bibitem{Chacko:2015fbc}
Z.~Chacko, D.~Curtin and C.B.~Verhaaren, \emph{{A Quirky Probe of Neutral
  Naturalness}}, \href{https://doi.org/10.1103/PhysRevD.94.011504}{\emph{Phys.
  Rev. D} {\bfseries 94} (2016) 011504}
  [\href{https://arxiv.org/abs/1512.05782}{{\ttfamily 1512.05782}}].

\bibitem{Burdman:2018ehe}
G.~Burdman and G.~Lichtenstein, \emph{{Displaced Vertices from Hidden Glue}},
  \href{https://doi.org/10.1007/JHEP08(2018)146}{\emph{JHEP} {\bfseries 08}
  (2018) 146} [\href{https://arxiv.org/abs/1807.03801}{{\ttfamily
  1807.03801}}].

\bibitem{Kilic:2018sew}
C.~Kilic, S.~Najjari and C.B.~Verhaaren, \emph{{Discovering the Twin Higgs
  Boson with Displaced Decays}},
  \href{https://doi.org/10.1103/PhysRevD.99.075029}{\emph{Phys. Rev. D}
  {\bfseries 99} (2019) 075029}
  [\href{https://arxiv.org/abs/1812.08173}{{\ttfamily 1812.08173}}].

\bibitem{Faraggi:2000pv}
A.E.~Faraggi and M.~Pospelov, \emph{{Self-interacting dark matter from the
  hidden heterotic string sector}},
  \href{https://doi.org/10.1016/S0927-6505(01)00121-9}{\emph{Astropart. Phys.}
  {\bfseries 16} (2002) 451}
  [\href{https://arxiv.org/abs/hep-ph/0008223}{{\ttfamily hep-ph/0008223}}].

\bibitem{Boddy:2014yra}
K.K.~Boddy, J.L.~Feng, M.~Kaplinghat and T.M.P.~Tait, \emph{{Self-Interacting
  Dark Matter from a Non-Abelian Hidden Sector}},
  \href{https://doi.org/10.1103/PhysRevD.89.115017}{\emph{Phys. Rev. D}
  {\bfseries 89} (2014) 115017}
  [\href{https://arxiv.org/abs/1402.3629}{{\ttfamily 1402.3629}}].

\bibitem{Boddy_2014_1}
K.K.~Boddy, J.L.~Feng, M.~Kaplinghat, Y.~Shadmi and T.M.~Tait, \emph{Strongly
  interacting dark matter: Self-interactions and {keV} lines},
  \href{https://doi.org/10.1103/physrevd.90.095016}{\emph{Physical Review D}
  {\bfseries 90} (2014) }.

\bibitem{Garc_a_Garc_a_2015}
I.G.~Garc{\'{\i}}a, R.~Lasenby and J.~March-Russell, \emph{Twin higgs {WIMP}
  dark matter},
  \href{https://doi.org/10.1103/physrevd.92.055034}{\emph{Physical Review D}
  {\bfseries 92} (2015) }.

\bibitem{Soni_2016}
A.~Soni and Y.~Zhang, \emph{Hidden \textbf{SU(N)} glueball dark matter},
  \href{https://doi.org/10.1103/physrevd.93.115025}{\emph{Physical Review D}
  {\bfseries 93} (2016) }.

\bibitem{Yamanaka:2019aeq}
N.~Yamanaka, H.~Iida, A.~Nakamura and M.~Wakayama, \emph{{Dark matter
  scattering cross section and dynamics in dark Yang-Mills theory}},
  \href{https://doi.org/10.1016/j.physletb.2020.136056}{\emph{Phys. Lett. B}
  {\bfseries 813} (2021) 136056}
  [\href{https://arxiv.org/abs/1910.01440}{{\ttfamily 1910.01440}}].

\bibitem{Yamanaka:2019yek}
N.~Yamanaka, H.~Iida, A.~Nakamura and M.~Wakayama, \emph{{Glueball scattering
  cross section in lattice SU(2) Yang-Mills theory}},
  \href{https://doi.org/10.1103/PhysRevD.102.054507}{\emph{Phys. Rev. D}
  {\bfseries 102} (2020) 054507}
  [\href{https://arxiv.org/abs/1910.07756}{{\ttfamily 1910.07756}}].

\bibitem{Curtin:2022oec}
D.~Curtin and C.~Gemmell, \emph{{Indirect Detection of Dark Matter Annihilating
  into Dark Glueballs}},  \href{https://arxiv.org/abs/2211.05794}{{\ttfamily
  2211.05794}}.

\bibitem{Asadi:2022vkc}
P.~Asadi, E.D.~Kramer, E.~Kuflik, T.R.~Slatyer and J.~Smirnov, \emph{{Glueballs
  in a thermal squeezeout model}},
  \href{https://doi.org/10.1007/JHEP07(2022)006}{\emph{JHEP} {\bfseries 07}
  (2022) 006} [\href{https://arxiv.org/abs/2203.15813}{{\ttfamily
  2203.15813}}].

\bibitem{Forestell:2016qhc}
L.~Forestell, D.E.~Morrissey and K.~Sigurdson, \emph{{Non-Abelian Dark Forces
  and the Relic Densities of Dark Glueballs}},
  \href{https://doi.org/10.1103/PhysRevD.95.015032}{\emph{Phys. Rev. D}
  {\bfseries 95} (2017) 015032}
  [\href{https://arxiv.org/abs/1605.08048}{{\ttfamily 1605.08048}}].

\bibitem{Forestell:2017wov}
L.~Forestell, D.E.~Morrissey and K.~Sigurdson, \emph{{Cosmological Bounds on
  Non-Abelian Dark Forces}},
  \href{https://doi.org/10.1103/PhysRevD.97.075029}{\emph{Phys. Rev. D}
  {\bfseries 97} (2018) 075029}
  [\href{https://arxiv.org/abs/1710.06447}{{\ttfamily 1710.06447}}].

\bibitem{Soni:2017nlm}
A.~Soni, H.~Xiao and Y.~Zhang, \emph{{Cosmic selection rule for the glueball
  dark matter relic density}},
  \href{https://doi.org/10.1103/PhysRevD.96.083514}{\emph{Phys. Rev. D}
  {\bfseries 96} (2017) 083514}
  [\href{https://arxiv.org/abs/1704.02347}{{\ttfamily 1704.02347}}].

\bibitem{Buen-Abad:2018mas}
M.A.~Buen-Abad, R.~Emami and M.~Schmaltz, \emph{{Cannibal Dark Matter and Large
  Scale Structure}},
  \href{https://doi.org/10.1103/PhysRevD.98.083517}{\emph{Phys. Rev. D}
  {\bfseries 98} (2018) 083517}
  [\href{https://arxiv.org/abs/1803.08062}{{\ttfamily 1803.08062}}].

\bibitem{Jo:2020ggs}
B.~Jo, H.~Kim, H.D.~Kim and C.S.~Shin, \emph{{Exploring the Universe with dark
  light scalars}},
  \href{https://doi.org/10.1103/PhysRevD.103.083528}{\emph{Phys. Rev. D}
  {\bfseries 103} (2021) 083528}
  [\href{https://arxiv.org/abs/2010.10880}{{\ttfamily 2010.10880}}].

\bibitem{Carenza:2022pjd}
P.~Carenza, R.~Pasechnik, G.~Salinas and Z.-W.~Wang, \emph{{Glueball Dark
  Matter Revisited}},
  \href{https://doi.org/10.1103/PhysRevLett.129.261302}{\emph{Phys. Rev. Lett.}
  {\bfseries 129} (2022) 261302}
  [\href{https://arxiv.org/abs/2207.13716}{{\ttfamily 2207.13716}}].

\bibitem{Carloni:2010tw}
L.~Carloni and T.~Sjostrand, \emph{{Visible Effects of Invisible Hidden Valley
  Radiation}}, \href{https://doi.org/10.1007/JHEP09(2010)105}{\emph{JHEP}
  {\bfseries 09} (2010) 105} [\href{https://arxiv.org/abs/1006.2911}{{\ttfamily
  1006.2911}}].

\bibitem{Carloni:2011kk}
L.~Carloni, J.~Rathsman and T.~Sjostrand, \emph{{Discerning Secluded Sector
  gauge structures}},
  \href{https://doi.org/10.1007/JHEP04(2011)091}{\emph{JHEP} {\bfseries 04}
  (2011) 091} [\href{https://arxiv.org/abs/1102.3795}{{\ttfamily 1102.3795}}].

\bibitem{Curtin:2022tou}
D.~Curtin, C.~Gemmell and C.B.~Verhaaren, \emph{{Simulating glueball production
  in Nf=0 QCD}}, \href{https://doi.org/10.1103/PhysRevD.106.075015}{\emph{Phys.
  Rev. D} {\bfseries 106} (2022) 075015}
  [\href{https://arxiv.org/abs/2202.12899}{{\ttfamily 2202.12899}}].

\bibitem{Andersson:1983ia}
B.~Andersson, G.~Gustafson, G.~Ingelman and T.~Sjostrand, \emph{{Parton
  Fragmentation and String Dynamics}},
  \href{https://doi.org/10.1016/0370-1573(83)90080-7}{\emph{Phys. Rept.}
  {\bfseries 97} (1983) 31}.

\bibitem{Bierlich:2022pfr}
C.~Bierlich et~al., \emph{{A comprehensive guide to the physics and usage of
  PYTHIA 8.3}},  \href{https://arxiv.org/abs/2203.11601}{{\ttfamily
  2203.11601}}.

\bibitem{Chou:2016lxi}
J.P.~Chou, D.~Curtin and H.J.~Lubatti, \emph{{New Detectors to Explore the
  Lifetime Frontier}},
  \href{https://doi.org/10.1016/j.physletb.2017.01.043}{\emph{Phys. Lett. B}
  {\bfseries 767} (2017) 29}
  [\href{https://arxiv.org/abs/1606.06298}{{\ttfamily 1606.06298}}].

\bibitem{MATHUSLA:2018bqv}
{\scshape MATHUSLA} collaboration, \emph{{A Letter of Intent for MATHUSLA: A
  Dedicated Displaced Vertex Detector above ATLAS or CMS.}},
  \href{https://arxiv.org/abs/1811.00927}{{\ttfamily 1811.00927}}.

\bibitem{MATHUSLA:2020uve}
{\scshape MATHUSLA} collaboration, \emph{{An Update to the Letter of Intent for
  MATHUSLA: Search for Long-Lived Particles at the HL-LHC}},
  \href{https://arxiv.org/abs/2009.01693}{{\ttfamily 2009.01693}}.

\bibitem{Curtin:2018mvb}
D.~Curtin et~al., \emph{{Long-Lived Particles at the Energy Frontier: The
  MATHUSLA Physics Case}},
  \href{https://doi.org/10.1088/1361-6633/ab28d6}{\emph{Rept. Prog. Phys.}
  {\bfseries 82} (2019) 116201}
  [\href{https://arxiv.org/abs/1806.07396}{{\ttfamily 1806.07396}}].

\bibitem{FIELD19781}
R.~Field and R.~Feynman, \emph{A parametrization of the properties of quark
  jets},
  \href{https://doi.org/https://doi.org/10.1016/0550-3213(78)90015-9}{\emph{Nuclear
  Physics B} {\bfseries 136} (1978) 1}.

\bibitem{Bahr:2008pv}
M.~Bahr et~al., \emph{{Herwig++ Physics and Manual}},
  \href{https://doi.org/10.1140/epjc/s10052-008-0798-9}{\emph{Eur. Phys. J. C}
  {\bfseries 58} (2008) 639} [\href{https://arxiv.org/abs/0803.0883}{{\ttfamily
  0803.0883}}].

\bibitem{Gleisberg:2008ta}
T.~Gleisberg, S.~Hoeche, F.~Krauss, M.~Schonherr, S.~Schumann, F.~Siegert
  et~al., \emph{{Event generation with SHERPA 1.1}},
  \href{https://doi.org/10.1088/1126-6708/2009/02/007}{\emph{JHEP} {\bfseries
  02} (2009) 007} [\href{https://arxiv.org/abs/0811.4622}{{\ttfamily
  0811.4622}}].

\bibitem{Webber:1983if}
B.R.~Webber, \emph{{A QCD Model for Jet Fragmentation Including Soft Gluon
  Interference}},
  \href{https://doi.org/10.1016/0550-3213(84)90333-X}{\emph{Nucl. Phys. B}
  {\bfseries 238} (1984) 492}.

\bibitem{Kupco:1998fx}
A.~Kupco, \emph{{Cluster hadronization in HERWIG 5.9}},  in \emph{{Workshop on
  Monte Carlo Generators for HERA Physics (Plenary Starting Meeting)}},
  pp.~292--300, 4, 1998 [\href{https://arxiv.org/abs/hep-ph/9906412}{{\ttfamily
  hep-ph/9906412}}].

\bibitem{Cohen:2023mya}
T.~Cohen, J.~Roloff and C.~Scherb, \emph{{Dark sector showers in the Lund jet
  plane}}, \href{https://doi.org/10.1103/PhysRevD.108.L031501}{\emph{Phys. Rev.
  D} {\bfseries 108} (2023) L031501}
  [\href{https://arxiv.org/abs/2301.07732}{{\ttfamily 2301.07732}}].

\bibitem{Elder:2017bkd}
B.T.~Elder, M.~Procura, J.~Thaler, W.J.~Waalewijn and K.~Zhou,
  \emph{{Generalized Fragmentation Functions for Fractal Jet Observables}},
  \href{https://doi.org/10.1007/JHEP06(2017)085}{\emph{JHEP} {\bfseries 06}
  (2017) 085} [\href{https://arxiv.org/abs/1704.05456}{{\ttfamily
  1704.05456}}].

\bibitem{Morningstar:1999rf}
C.J.~Morningstar and M.J.~Peardon, \emph{{The Glueball spectrum from an
  anisotropic lattice study}},
  \href{https://doi.org/10.1103/PhysRevD.60.034509}{\emph{Phys. Rev. D}
  {\bfseries 60} (1999) 034509}
  [\href{https://arxiv.org/abs/hep-lat/9901004}{{\ttfamily hep-lat/9901004}}].

\bibitem{Chen:2005mg}
Y.~Chen et~al., \emph{{Glueball spectrum and matrix elements on anisotropic
  lattices}}, \href{https://doi.org/10.1103/PhysRevD.73.014516}{\emph{Phys.
  Rev. D} {\bfseries 73} (2006) 014516}
  [\href{https://arxiv.org/abs/hep-lat/0510074}{{\ttfamily hep-lat/0510074}}].

\bibitem{Athenodorou:2021qvs}
A.~Athenodorou and M.~Teper, \emph{{SU(N) gauge theories in 3+1 dimensions:
  glueball spectrum, string tensions and topology}},
  \href{https://doi.org/10.1007/JHEP12(2021)082}{\emph{JHEP} {\bfseries 12}
  (2021) 082} [\href{https://arxiv.org/abs/2106.00364}{{\ttfamily
  2106.00364}}].

\bibitem{Jaffe:1985qp}
R.L.~Jaffe, K.~Johnson and Z.~Ryzak, \emph{{Qualitative Features of the
  Glueball Spectrum}},
  \href{https://doi.org/10.1016/0003-4916(86)90035-7}{\emph{Annals Phys.}
  {\bfseries 168} (1986) 344}.

\bibitem{VanApeldoorn:1981gx}
G.W.~Van~Apeldoorn et~al., \emph{{Thermal Emission of Pions in Anti-proton
  Proton Annihilation at 12-{GeV}/c}},
  \href{https://doi.org/10.1007/BF01436312}{\emph{Z. Phys. C} {\bfseries 7}
  (1981) 235}.

\bibitem{Manes:2001cs}
J.L.~Manes, \emph{{Emission spectrum of fundamental strings: An Algebraic
  approach}}, \href{https://doi.org/10.1016/S0550-3213(01)00578-8}{\emph{Nucl.
  Phys. B} {\bfseries 621} (2002) 37}
  [\href{https://arxiv.org/abs/hep-th/0109196}{{\ttfamily hep-th/0109196}}].

\bibitem{Blanchard:2004du}
P.~Blanchard, S.~Fortunato and H.~Satz, \emph{{The Hagedorn temperature and
  partition thermodynamics}},
  \href{https://doi.org/10.1140/epjc/s2004-01673-0}{\emph{Eur. Phys. J. C}
  {\bfseries 34} (2004) 361}
  [\href{https://arxiv.org/abs/hep-ph/0401103}{{\ttfamily hep-ph/0401103}}].

\bibitem{Noronha-Hostler:2010nut}
J.~Noronha-Hostler, J.~Noronha and C.~Greiner, \emph{{Particle Ratios and the
  QCD Critical Temperature}},
  \href{https://doi.org/10.1088/0954-3899/37/9/094062}{\emph{J. Phys. G}
  {\bfseries 37} (2010) 094062}
  [\href{https://arxiv.org/abs/1001.2610}{{\ttfamily 1001.2610}}].

\bibitem{Lucini:2012wq}
B.~Lucini, A.~Rago and E.~Rinaldi, \emph{{SU($N_c$) gauge theories at
  deconfinement}},
  \href{https://doi.org/10.1016/j.physletb.2012.04.070}{\emph{Phys. Lett. B}
  {\bfseries 712} (2012) 279}
  [\href{https://arxiv.org/abs/1202.6684}{{\ttfamily 1202.6684}}].

\bibitem{Ellis:1991qj}
R.K.~Ellis, W.J.~Stirling and B.R.~Webber, \emph{{QCD and collider physics}},
  vol.~8, Cambridge University Press (2, 2011).

\bibitem{Juknevich:2009gg}
J.E.~Juknevich, \emph{{Pure-glue hidden valleys through the Higgs portal}},
  \href{https://doi.org/10.1007/JHEP08(2010)121}{\emph{JHEP} {\bfseries 08}
  (2010) 121} [\href{https://arxiv.org/abs/0911.5616}{{\ttfamily 0911.5616}}].

\bibitem{Djouadi:2018xqq}
{\scshape HDECAY} collaboration, \emph{{HDECAY: Twenty$_{++}$ years after}},
  \href{https://doi.org/10.1016/j.cpc.2018.12.010}{\emph{Comput. Phys. Commun.}
  {\bfseries 238} (2019) 214}
  [\href{https://arxiv.org/abs/1801.09506}{{\ttfamily 1801.09506}}].

\bibitem{Winkler:2018qyg}
M.W.~Winkler, \emph{{Decay and detection of a light scalar boson mixing with
  the Higgs boson}},
  \href{https://doi.org/10.1103/PhysRevD.99.015018}{\emph{Phys. Rev. D}
  {\bfseries 99} (2019) 015018}
  [\href{https://arxiv.org/abs/1809.01876}{{\ttfamily 1809.01876}}].

\bibitem{Alwall:2014hca}
J.~Alwall, R.~Frederix, S.~Frixione, V.~Hirschi, F.~Maltoni, O.~Mattelaer
  et~al., \emph{{The automated computation of tree-level and next-to-leading
  order differential cross sections, and their matching to parton shower
  simulations}}, \href{https://doi.org/10.1007/JHEP07(2014)079}{\emph{JHEP}
  {\bfseries 07} (2014) 079} [\href{https://arxiv.org/abs/1405.0301}{{\ttfamily
  1405.0301}}].

\bibitem{Bozzi:2005wk}
G.~Bozzi, S.~Catani, D.~de~Florian and M.~Grazzini, \emph{{Transverse-momentum
  resummation and the spectrum of the Higgs boson at the LHC}},
  \href{https://doi.org/10.1016/j.nuclphysb.2005.12.022}{\emph{Nucl. Phys. B}
  {\bfseries 737} (2006) 73}
  [\href{https://arxiv.org/abs/hep-ph/0508068}{{\ttfamily hep-ph/0508068}}].

\bibitem{deFlorian:2011xf}
D.~de~Florian, G.~Ferrera, M.~Grazzini and D.~Tommasini,
  \emph{{Transverse-momentum resummation: Higgs boson production at the
  Tevatron and the LHC}},
  \href{https://doi.org/10.1007/JHEP11(2011)064}{\emph{JHEP} {\bfseries 11}
  (2011) 064} [\href{https://arxiv.org/abs/1109.2109}{{\ttfamily 1109.2109}}].

\bibitem{ATLAS:2015egz}
{\scshape ATLAS} collaboration, \emph{{Measurements of the Higgs boson
  production and decay rates and coupling strengths using pp collision data at
  $\sqrt{s}=7$ and 8 TeV in the ATLAS experiment}},
  \href{https://doi.org/10.1140/epjc/s10052-015-3769-y}{\emph{Eur. Phys. J. C}
  {\bfseries 76} (2016) 6} [\href{https://arxiv.org/abs/1507.04548}{{\ttfamily
  1507.04548}}].

\bibitem{Workman:2022ynf}
{\scshape Particle Data Group} collaboration, \emph{{Review of Particle
  Physics}}, \href{https://doi.org/10.1093/ptep/ptac097}{\emph{PTEP} {\bfseries
  2022} (2022) 083C01}.

\bibitem{CMS:2022qva}
{\scshape CMS} collaboration, \emph{{Search for invisible decays of the Higgs
  boson produced via vector boson fusion in proton-proton collisions at
  s=13\,\,TeV}}, \href{https://doi.org/10.1103/PhysRevD.105.092007}{\emph{Phys.
  Rev. D} {\bfseries 105} (2022) 092007}
  [\href{https://arxiv.org/abs/2201.11585}{{\ttfamily 2201.11585}}].

\bibitem{CMS:2020iwv}
{\scshape CMS} collaboration, \emph{{Search for long-lived particles using
  displaced jets in proton-proton collisions at $\sqrt{s} = $ 13 TeV}},
  \href{https://doi.org/10.1103/PhysRevD.104.012015}{\emph{Phys. Rev. D}
  {\bfseries 104} (2021) 012015}
  [\href{https://arxiv.org/abs/2012.01581}{{\ttfamily 2012.01581}}].

\bibitem{Barron:2020kfo}
J.~Barron and D.~Curtin, \emph{{On the Origin of Long-Lived Particles}},
  \href{https://doi.org/10.1007/JHEP12(2020)061}{\emph{JHEP} {\bfseries 12}
  (2020) 061} [\href{https://arxiv.org/abs/2007.05538}{{\ttfamily
  2007.05538}}].

\bibitem{Kang:2008ea}
J.~Kang and M.A.~Luty, \emph{{Macroscopic Strings and 'Quirks' at Colliders}},
  \href{https://doi.org/10.1088/1126-6708/2009/11/065}{\emph{JHEP} {\bfseries
  11} (2009) 065} [\href{https://arxiv.org/abs/0805.4642}{{\ttfamily
  0805.4642}}].

\bibitem{Altarelli:1989ff}
G.~Altarelli, B.~Mele and M.~Ruiz-Altaba, \emph{{Searching for New Heavy Vector
  Bosons in $p \bar{p}$ Colliders}},
  \href{https://doi.org/10.1007/BF01556677}{\emph{Z. Phys. C} {\bfseries 45}
  (1989) 109}.

\bibitem{Fuks:2017vtl}
B.~Fuks and R.~Ruiz, \emph{{A comprehensive framework for studying $W'$ and
  $Z'$ bosons at hadron colliders with automated jet veto resummation}},
  \href{https://doi.org/10.1007/JHEP05(2017)032}{\emph{JHEP} {\bfseries 05}
  (2017) 032} [\href{https://arxiv.org/abs/1701.05263}{{\ttfamily
  1701.05263}}].

\bibitem{Cacciari:2011ma}
M.~Cacciari, G.P.~Salam and G.~Soyez, \emph{{FastJet User Manual}},
  \href{https://doi.org/10.1140/epjc/s10052-012-1896-2}{\emph{Eur. Phys. J. C}
  {\bfseries 72} (2012) 1896}
  [\href{https://arxiv.org/abs/1111.6097}{{\ttfamily 1111.6097}}].

\bibitem{Cacciari:2008gp}
M.~Cacciari, G.P.~Salam and G.~Soyez, \emph{{The anti-$k_t$ jet clustering
  algorithm}}, \href{https://doi.org/10.1088/1126-6708/2008/04/063}{\emph{JHEP}
  {\bfseries 04} (2008) 063} [\href{https://arxiv.org/abs/0802.1189}{{\ttfamily
  0802.1189}}].

\bibitem{CMS:1997ema}
{\scshape CMS} collaboration, \emph{{CMS: The electromagnetic calorimeter.
  Technical design report}},  CERN-LHCC-97-33, CMS-TDR-4.

\bibitem{Feng:2011ik}
J.L.~Feng and Y.~Shadmi, \emph{{WIMPless Dark Matter from Non-Abelian Hidden
  Sectors with Anomaly-Mediated Supersymmetry Breaking}},
  \href{https://doi.org/10.1103/PhysRevD.83.095011}{\emph{Phys. Rev. D}
  {\bfseries 83} (2011) 095011}
  [\href{https://arxiv.org/abs/1102.0282}{{\ttfamily 1102.0282}}].

\bibitem{Acharya:2017szw}
B.S.~Acharya, M.~Fairbairn and E.~Hardy, \emph{{Glueball dark matter in
  non-standard cosmologies}},
  \href{https://doi.org/10.1007/JHEP07(2017)100}{\emph{JHEP} {\bfseries 07}
  (2017) 100} [\href{https://arxiv.org/abs/1704.01804}{{\ttfamily
  1704.01804}}].

\bibitem{Winkler:2022zdu}
M.W.~Winkler, P.~De~La Torre~Luque and T.~Linden, \emph{{Cosmic ray antihelium
  from a strongly coupled dark sector}},
  \href{https://doi.org/10.1103/PhysRevD.107.123035}{\emph{Phys. Rev. D}
  {\bfseries 107} (2023) 123035}
  [\href{https://arxiv.org/abs/2211.00025}{{\ttfamily 2211.00025}}].

\bibitem{Buss:2022lxw}
T.~Buss, B.M.~Dillon, T.~Finke, M.~Kr\"amer, A.~Morandini, A.~M\"uck et~al.,
  \emph{{What's Anomalous in LHC Jets?}},
  \href{https://arxiv.org/abs/2202.00686}{{\ttfamily 2202.00686}}.

\bibitem{Canelli:2021aps}
F.~Canelli, A.~de~Cosa, L.L.~Pottier, J.~Niedziela, K.~Pedro and M.~Pierini,
  \emph{{Autoencoders for semivisible jet detection}},
  \href{https://doi.org/10.1007/JHEP02(2022)074}{\emph{JHEP} {\bfseries 02}
  (2022) 074} [\href{https://arxiv.org/abs/2112.02864}{{\ttfamily
  2112.02864}}].

\bibitem{Dillon:2022mkq}
B.M.~Dillon, L.~Favaro, T.~Plehn, P.~Sorrenson and M.~Kr\"amer, \emph{{A
  Normalized Autoencoder for LHC Triggers}},
  \href{https://arxiv.org/abs/2206.14225}{{\ttfamily 2206.14225}}.

\bibitem{Faucett:2022zie}
T.~Faucett, S.-C.~Hsu and D.~Whiteson, \emph{{Learning to identify semi-visible
  jets}}, \href{https://doi.org/10.1007/JHEP12(2022)132}{\emph{JHEP} {\bfseries
  12} (2022) 132} [\href{https://arxiv.org/abs/2208.10062}{{\ttfamily
  2208.10062}}].

\bibitem{Lu:2023gjk}
C.-T.~Lu, H.~Lv, W.~Shen, L.~Wu and J.~Zhang, \emph{{Probing dark QCD sector
  through the Higgs portal with machine learning at the LHC}},
  \href{https://doi.org/10.1007/JHEP08(2023)187}{\emph{JHEP} {\bfseries 08}
  (2023) 187} [\href{https://arxiv.org/abs/2304.03237}{{\ttfamily
  2304.03237}}].

\bibitem{Bardhan:2023mia}
D.~Bardhan, Y.~Kats and N.~Wunch, \emph{{Searching for dark jets with displaced
  vertices using weakly supervised machine learning}},
  \href{https://doi.org/10.1103/PhysRevD.108.035036}{\emph{Phys. Rev. D}
  {\bfseries 108} (2023) 035036}
  [\href{https://arxiv.org/abs/2305.04372}{{\ttfamily 2305.04372}}].

\bibitem{Anzalone:2023ugq}
L.~Anzalone, S.S.~Chhibra, B.~Maier, N.~Chernyavskaya and M.~Pierini,
  \emph{{Triggering Dark Showers with Conditional Dual Auto-Encoders}},
  \href{https://arxiv.org/abs/2306.12955}{{\ttfamily 2306.12955}}.

\bibitem{Pedro:2023sdp}
K.~Pedro and P.~Shyamsundar, \emph{{Optimal Mass Variables for Semivisible
  Jets}},  \href{https://arxiv.org/abs/2303.16253}{{\ttfamily 2303.16253}}.

\bibitem{Prosperi:2006hx}
G.M.~Prosperi, M.~Raciti and C.~Simolo, \emph{{On the running coupling constant
  in QCD}}, \href{https://doi.org/10.1016/j.ppnp.2006.09.001}{\emph{Prog. Part.
  Nucl. Phys.} {\bfseries 58} (2007) 387}
  [\href{https://arxiv.org/abs/hep-ph/0607209}{{\ttfamily hep-ph/0607209}}].

\end{thebibliography}\endgroup

\end{document}